\documentclass[prd,12pt,letterpaper,showpacs,groupedaddress,superscriptaddress,nofootinbib,floatfix,preprintnumbers,tightenlines]{revtex4-2}
\usepackage{hyperref,amssymb,amsmath,graphicx,xcolor,cancel,mathrsfs,multirow,bm,mathtools,tikz,placeins,bbm,tikz-feynman,appendix}
\usepackage[caption=false]{subfig}
\usepackage{bm}

\DeclareRobustCommand\circled[1]{\tikz[baseline=(char.base)]{\node[shape=circle,draw,inner sep=2pt] (char) {#1};}}

\begin{document}
\begin{figure}[!t]
	
	\vskip -1.5cm
	\leftline{\includegraphics[width=0.25\textwidth]{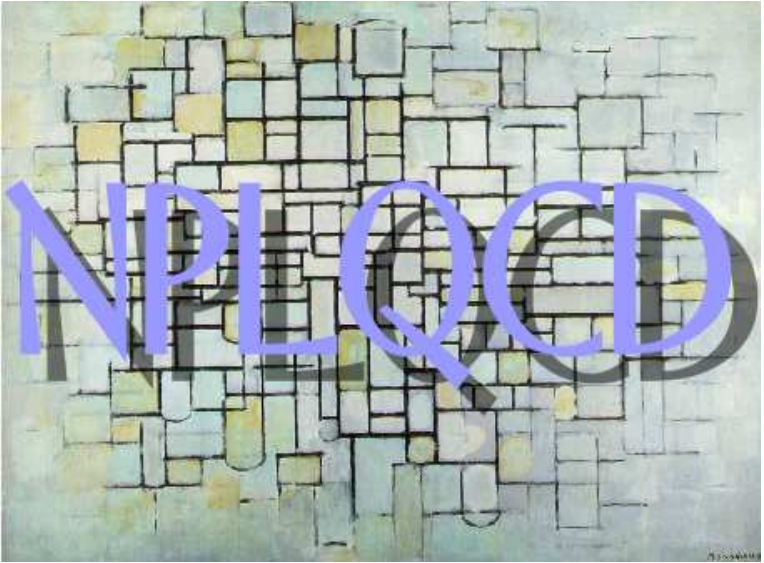}}
\end{figure}

\title{Neutrinoless Double Beta Decay from Lattice QCD: The Long-Distance $\pi^{-} \rightarrow \pi^{+} e^{-} e^{-}$ Amplitude}
\author{W.~Detmold}
\affiliation{Center for Theoretical Physics, Massachusetts Institute of Technology, Boston, MA 02139, USA}
\author{D.J.~Murphy}
\affiliation{Center for Theoretical Physics, Massachusetts Institute of Technology, Boston, MA 02139, USA}
\collaboration{NPLQCD Collaboration}

\date{\today}

\preprint{MIT-CTP/5196}



\begin{abstract}
Neutrinoless double beta decay (\( 0 \nu \beta \beta \)) is a hypothetical nuclear decay mode with important implications. In particular, observation of this decay would demonstrate that the neutrino is a Majorana particle and that lepton number conservation is violated in nature. Relating experimental constraints on \(0 \nu \beta \beta\) decay rates to the neutrino masses requires theoretical input in the form of non-perturbative nuclear matrix elements which remain difficult to calculate reliably. This work marks a first step toward providing a general lattice QCD framework for computing long-distance \(0 \nu \beta \beta\) matrix elements in the case where the decay is mediated by a light Majorana neutrino. The relevant formalism is developed and then tested by computing the simplest such matrix element describing an unphysical \( \pi^{-} \rightarrow \pi^{+} e^{-} e^{-} \) transition on a series of domain wall fermion ensembles. The resulting lattice data is then fit to next-to-leading-order chiral perturbation theory, allowing a fully-controlled extraction of the low energy constant governing the transition rate, \(g_{\nu}^{\pi \pi}(\mu = 770 \,\, \mathrm{MeV}) = -10.78(12)_{\rm stat}(51)_{\rm sys}\). Finally, future prospects for calculations of more complicated processes, such as the phenomenologically important \(n^{0} n^{0} \rightarrow p^{+} p^{+} e^{-} e^{-}\) decay, are discussed.

\end{abstract}

\maketitle

\section{Introduction}
\label{sec:intro}
\begin{figure}[!h]
	\centering
	\subfloat[$2 \nu \beta \beta$]{
		\resizebox{0.44\linewidth}{!}{
			\begin{tikzpicture}
			\begin{feynman}
			\vertex (a1) {\(u\)};
			\vertex[right=8.8cm of a1] (a2) {\(u\)};
			\vertex[below=0.3cm of a1] (b1) {\(d\)};
			\vertex[right=8.8cm of b1] (b2) {\(d\)};
			\vertex[below=0.6cm of a1] (c1) {\(d\)};
			\vertex[right=2.2cm of c1] (c2);
			\vertex[right=4.4cm of c1] (c3);
			\vertex[right=6.6cm of c1] (c4);
			\vertex[right=8.8cm of c1] (c5) {\(u\)};
			\vertex[below=1.1cm of a1] (d1);
			\vertex[right=2.2cm of d1] (d2);
			\vertex[right=4.4cm of d1] (d3);
			\vertex[right=6.6cm of d1] (d4);
			\vertex[right=8.8cm of d1] (d5) {\(\overline{\nu}_{e}\)};
			\vertex[below=1.6cm of a1] (e1);
			\vertex[right=2.2cm of e1] (e2);
			\vertex[right=4.4cm of e1] (e3);
			\vertex[right=6.6cm of e1] (e4);
			\vertex[right=8.8cm of e1] (e5) {\(e^{-}\)};
			\vertex[below=2.1cm of a1] (f1);
			\vertex[right=2.2cm of f1] (f2);
			\vertex[right=4.4cm of f1] (f3);
			\vertex[right=6.6cm of f1] (f4);
			\vertex[right=8.8cm of f1] (f5);
			\vertex[below=2.6cm of a1] (g1);
			\vertex[right=2.2cm of g1] (g2);
			\vertex[right=4.4cm of g1] (g3);
			\vertex[right=6.6cm of g1] (g4);
			\vertex[right=8.8cm of g1] (g5) {\(e^{-}\)};
			\vertex[below=3.1cm of a1] (h1);
			\vertex[right=2.2cm of h1] (h2);
			\vertex[right=4.4cm of h1] (h3);
			\vertex[right=6.6cm of h1] (h4);
			\vertex[right=8.8cm of h1] (h5) {\(\overline{\nu}_{e}\)};
			\vertex[below=3.6cm of a1] (i1) {\(d\)};
			\vertex[right=2.2cm of i1] (i2);
			\vertex[right=4.4cm of i1] (i3);
			\vertex[right=6.6cm of i1] (i4);
			\vertex[right=8.8cm of i1] (i5) {\(u\)};
			\vertex[below=3.9cm of a1] (j1) {\(d\)};
			\vertex[right=8.8cm of j1] (j2) {\(d\)};
			\vertex[below=4.2cm of a1] (k1) {\(u\)};
			\vertex[right=8.8cm of k1] (k2) {\(u\)};
			\diagram* {
				(a1) -- [fermion] (a2),
				(b1) -- [fermion] (b2),
				(c1) -- [fermion] (c5),
        (c1) -- (c2) -- [boson, edge label'=\(W^{-}\)] (e3) -- [fermion] (d5),
				(c1) -- (c2) -- [boson] (e3) -- [fermion] (e5),
        (i1) -- (i2) -- [boson, edge label=\(W^{-}\)] (g3) -- [fermion] (g5),
				(i1) -- (i2) -- [boson] (g3) -- [fermion] (h5),
				(i1) -- [fermion] (i5),
				(j1) -- [fermion] (j2),
				(k1) -- [fermion] (k2),
			};
		\draw[decoration={brace}, decorate] (c1.south west) -- (a1.north west)
        node[pos=0.5, left] { \(n^{0}\) };
		\draw[decoration={brace}, decorate] (k1.south west) -- (i1.north west)
        node[pos=0.5, left] { \(n^{0}\) };
		\draw[decoration={brace}, decorate] (a2.north east) -- (c5.south east)
    node[pos=0.5, right] { \(p^{+}\) };
		\draw[decoration={brace}, decorate] (i5.north east) -- (k2.south east)
    node[pos=0.5, right] { \(p^{+}\) };
			\end{feynman}
			\end{tikzpicture}
		}
	}
	\subfloat[$0 \nu \beta \beta$]{
		\resizebox{0.44\linewidth}{!}{
			\begin{tikzpicture}
			\begin{feynman}
			\vertex (a1) {\(u\)};
			\vertex[right=8.8cm of a1] (a2) {\(u\)};
			\vertex[below=0.3cm of a1] (b1) {\(d\)};
			\vertex[right=8.8cm of b1] (b2) {\(d\)};
			\vertex[below=0.6cm of a1] (c1) {\(d\)};
			\vertex[right=2.2cm of c1] (c2);
			\vertex[right=4.4cm of c1] (c3);
			\vertex[right=6.6cm of c1] (c4);
			\vertex[right=8.8cm of c1] (c5) {\(u\)};
			\vertex[below=1.1cm of a1] (d1);
			\vertex[right=2.2cm of d1] (d2);
			\vertex[right=4.4cm of d1] (d3);
			\vertex[right=6.6cm of d1] (d4);
			\vertex[right=8.8cm of d1] (d5);
			\vertex[below=1.6cm of a1] (e1);
			\vertex[right=2.2cm of e1] (e2);
			\vertex[right=4.4cm of e1] (e3);
			\vertex[right=6.6cm of e1] (e4);
			\vertex[right=8.8cm of e1] (e5) {\(e^{-}\)};
			\vertex[below=2.1cm of a1] (f1);
			\vertex[right=2.2cm of f1] (f2);
			\vertex[right=4.4cm of f1] (f3);
			\vertex[right=6.6cm of f1] (f4);
			\vertex[right=8.8cm of f1] (f5);
			\vertex[below=2.6cm of a1] (g1);
			\vertex[right=2.2cm of g1] (g2);
			\vertex[right=4.4cm of g1] (g3);
			\vertex[right=6.6cm of g1] (g4);
			\vertex[right=8.8cm of g1] (g5) {\(e^{-}\)};
			\vertex[below=3.1cm of a1] (h1);
			\vertex[right=2.2cm of h1] (h2);
			\vertex[right=4.4cm of h1] (h3);
			\vertex[right=6.6cm of h1] (h4);
			\vertex[right=8.8cm of h1] (h5);
			\vertex[below=3.6cm of a1] (i1) {\(d\)};
			\vertex[right=2.2cm of i1] (i2);
			\vertex[right=4.4cm of i1] (i3);
			\vertex[right=6.6cm of i1] (i4);
			\vertex[right=8.8cm of i1] (i5) {\(u\)};
			\vertex[below=3.9cm of a1] (j1) {\(d\)};
			\vertex[right=8.8cm of j1] (j2) {\(d\)};
			\vertex[below=4.2cm of a1] (k1) {\(u\)};
			\vertex[right=8.8cm of k1] (k2) {\(u\)};
			\diagram* {
				(a1) -- [fermion] (a2),
				(b1) -- [fermion] (b2),
				(c1) -- [fermion] (c5),
        (c1) -- (c2) -- [boson, edge label'=\(W^{-}\)] (e3) -- [fermion] (e5),
				(e3) -- [majorana, edge label'=\(\nu_{e}\)] (g3),
        (i1) -- (i2) -- [boson, edge label=\(W^{-}\)] (g3) -- [fermion] (g5),
				(i1) -- [fermion] (i5),
				(j1) -- [fermion] (j2),
				(k1) -- [fermion] (k2),
			};
			\draw[decoration={brace}, decorate] (c1.south west) -- (a1.north west)
        node[pos=0.5, left] { \(n^{0}\) };
			\draw[decoration={brace}, decorate] (k1.south west) -- (i1.north west)
        node[pos=0.5, left] { \(n^{0}\) };
			\draw[decoration={brace}, decorate] (a2.north east) -- (c5.south east)
        node[pos=0.5, right] { \(p^{+}\) };
			\draw[decoration={brace}, decorate] (i5.north east) -- (k2.south east)
      node[pos=0.5, right] { \(p^{+}\) };
			\end{feynman}
			\end{tikzpicture}
		}
	} \\
\caption{Quark-level Standard Model processes responsible for neutrinoful (left) and neutrinoless (right) double beta decay.}
\label{fig:dbd_sm}
\end{figure}
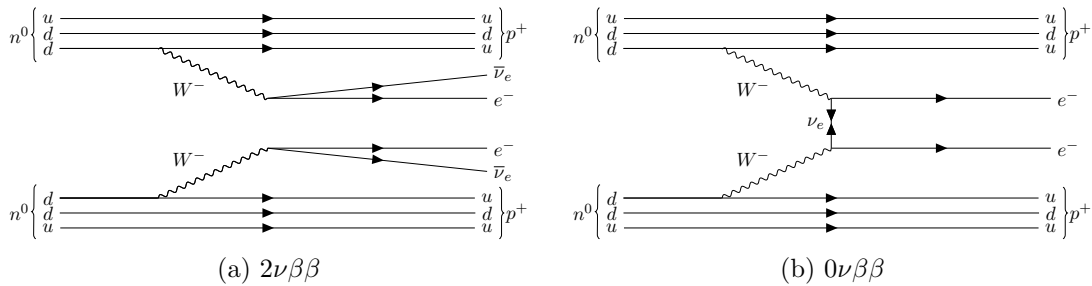

Neutrinoless double beta decay (\(0 \nu \beta \beta \)), depicted in Figure \ref{fig:dbd_sm}, is a hypothetical nuclear decay process which, if observed, would provide a wealth of information about the properties of neutrinos. In particular, it is the only known experimentally viable method for resolving the long-standing question of whether neutrinos are Majorana or Dirac particles. In addition, it would also provide a first example of a lepton-number violating process, which may help to explain baryogenesis in the early universe, as well as provide additional constraints on the parameters describing the neutrino sector in the Standard Model of particle physics. While \( 0 \nu \beta \beta \) has not been observed, it is the subject of a large and active experimental search effort, with bounds on the half-lives of relevant nuclei at the level of \( T_{1/2}^{0 \nu} \gtrsim 10^{25} - 10^{26} \) yrs \cite{KamLAND-Zen:2016pfg}. Next-generation experiments currently underway are aiming to probe half-lives that are an additional one to two orders of magnitude larger in the near future \cite{Dolinski:2019nrj}.

Relating a future experimental measurement of a \( 0 \nu \beta \beta \) decay rate \( T_{1/2}^{0 \nu} \) for a particular nucleus to the effective Majorana neutrino  mass \( m_{\beta \beta} = \vert \sum_{k} U_{ek}^{2} m_{k} \vert \) --- where \( \{ m_{k} \} \) are the neutrino eigenstate masses and \( U_{e k} \) are elements of the Pontecorvo-Maki-Nakagawa-Sakata (PMNS) neutrino mixing matrix --- requires theoretical input in the form of a nuclear matrix element, \( M^{0 \nu} \), describing the non-perturbative, hadronic part of the decay. These quantities are related by
\begin{equation}
\left( T_{1/2}^{0 \nu} \right)^{-1} \propto \left\vert m_{\beta \beta} \right\vert^{2} G^{0 \nu} \left\vert M^{0 \nu} \right\vert^{2},
\end{equation}
where \(G^{0 \nu}\) is a known kinematic factor. Reliably calculating \( M^{0 \nu} \) for nuclear systems relevant to experimental searches has proven to be a difficult challenge. A variety of phenomenological nuclear models have been used to perform these calculations \cite{Engel:2016xgb,Vergados:2012xy}, with predictions for a given nucleus from different models typically varying by 100\% or more \cite{Giuliani_Poves}, and with no principled method for assigning systematic uncertainties. Improving this situation will be crucial for interpreting experimental results from \( 0 \nu \beta \beta \) searches as constraints on the parameters of particular models of neutrinoless double beta decay moving forward.

In principle, lattice QCD and the electroweak theory jointly provide an entirely \textit{ab-initio} method for determining \( M^{0 \nu} \). However, in practice, computing matrix elements of the large nuclei relevant to \( 0 \nu \beta \beta \) searches is well beyond the computational and algorithmic limits of lattice QCD for the forseeable future. More realistically, one could hope to compute QCD matrix elements of sub-processes such as the \( n^{0} n^{0} \rightarrow p^{+} p^{+} e^{-} e^{-} \) decay, and then relate these to matrix elements of many-body systems within an effective field theory framework \cite{Cirigliano:2018yza}. Another possibility is to compute matrix elements of small nuclei which could then be used to probe the systematics of nuclear model calculations by directly comparing lattice and model predictions.

First calculations of the long-distance contributions to the neutrinoful double beta decay process \( n^{0} n^{0} \rightarrow p^{+} p^{+} e^{-} e^{-} \overline{\nu}_{e} \overline{\nu}_{e} \), and of the leading order short-distance contributions to neutrinoless double beta decay arising from new physics beyond the electroweak scale, were reported in Refs.~\cite{Tiburzi:2017iux} and \cite{PhysRevLett.121.172501}, respectively. More recently, first calculations of the simplest long-distance \(0 \nu \beta \beta\) amplitude describing an unphysical \( \pi^{-} \rightarrow \pi^{+} e^{-} e^{-} \) transition have appeared in the literature \cite{Detmold:2018zan,Tuo:2019bue}, as well as a calculation of the related \(\pi^{-} \pi^{-} \rightarrow e^{-} e^{-}\) decay amplitude \cite{Feng:2018pdq}. 

This work presents a complete calculation of the long-distance \(\pi^{-} \rightarrow \pi^{+} e^{-} e^{-} \) amplitude using a series of domain wall fermion ensembles. The paper is organized as follows: Section \ref{sec:methods} and Appendix \ref{appendix:formalism} develop the necessary formalism, including a novel treatment of the light Majorana neutrino on the lattice using a regulated form of the continuum, infinite volume scalar propagator. Section \ref{sec:calc} describes the lattice ensembles and numerical implementations of the two- and four-point correlation functions needed to extract the \(\pi^{-} \rightarrow \pi^{+} e^{-} e^{-}\) matrix element, as well as a series of fits to next-to-leading-order chiral perturbation theory (\(\chi\)PT) used to extrapolate the lattice data to the physical mass, infinite volume, and continuum limit, as well as determine the relevant \(\chi\)PT low energy constant \(g_{\nu}^{\pi \pi}(\mu)\) and assign a full statistical and systematic error budget. Finally, Sections \ref{sec:discussion} and \ref{sec:conclusions} discuss the results of this calculation in the context of other calculations in the literature, and lay out the prospects for future work.

\section{Methodology}
\label{sec:methods}
It is assumed throughout this work that neutrinoless double beta decay is mediated by the long-distance, light Majorana neutrino exchange mechanism. At low energies, and after integrating out the \(W\) boson, the underlying Standard Model interaction responsible for beta decay is described by the effective electroweak Hamiltonian
\begin{equation}
  \label{eqn:weak_hamiltonian}
  H_{W} = 2 \sqrt{2} G_{F} V_{ud} \left( \overline{u}_{L} \gamma_{\mu} d_{L} \right) \left( \overline{e}_{L} \gamma_{\mu} \nu_{e L} \right),
\end{equation}
where \(G_{F}\) is the Fermi constant and \(V_{ud}\) is the Cabibbo-Kobayashi-Maskawa (CKM) matrix element describing the strength of the \( d \rightarrow u \) transition arising from the flavor-changing weak interaction. \(0 \nu \beta \beta\) is induced at second order in electroweak perturbation theory, leading to the bilocal matrix element \cite{doi:10.1142/S0217751X1530001X}
\begin{equation}
  \int d^{4} x \, d^{4} y \, \big\langle f e e \big\vert \mathcal{T} \left\{ H_{W}(x) H_{W}(y) \right\} \big\vert i \big\rangle = 4 m_{\beta \beta} G_{F}^{2} V_{ud}^{2} \int d^{4} x \, d^{4} y \, H_{\alpha \beta}(x,y) L_{\alpha \beta}(x,y),
  \label{eqn:dbd_me}
\end{equation}
which can be factorized into tensors
\begin{equation}
  L_{\alpha \beta} \equiv \overline{e}_{L}(p_{1}) \gamma_{\alpha} S_{\nu}(x,y) \gamma_{\beta} e_{L}^{C}(p_{2}) e^{- i p_{1} \cdot x} e^{- i p_{2} \cdot y} \equiv \Gamma_{\alpha \beta}^{\rm lept.} S(x,y) e^{-i p_{1} \cdot x} e^{-i p_{2} \cdot y},
\label{eqn:leptonic_tensor}
\end{equation}
describing the leptonic part of the decay and
\begin{equation}
  H_{\alpha \beta} \equiv \big\langle f \big\vert \mathcal{T} \left\{ \overline{u}_{L}(x) \gamma_{\alpha} d_{L}(x) \overline{u}_{L}(y) \gamma_{\beta} d_{L}(y) \right\} \big\vert i \big\rangle \equiv \big\langle f \big\vert \mathcal{T} \left\{ j_{\alpha}(x) j_{\beta}(y) \right\} \big\vert i \big\rangle,
\label{eqn:hadronic_tensor}
\end{equation}
describing the hadronic part of the decay, respectively. In addition, \( S(x,y) \) is the neutrino propagator, \( e^{C}_{L} \equiv C \overline{e}_{L}^{\top} \) denotes charge conjugation, and \(\mathcal{T}\{\cdots\}\) denotes the time-ordering operation. Since current constraints from oscillation experiments \cite{PhysRevD.98.030001} suggest that \(m_{\beta \beta}\) is very small compared to typical scales relevant to QCD or nuclear physics, it is also assumed throughout this work that the massless scalar propagator
\begin{equation}
\label{eqn:continuum_neutrino_prop}
  S(x,y) = \int \frac{d^{4} q}{(2 \pi)^{4}} \frac{1}{q^{2}} e^{i q \cdot (x-y)} 
\end{equation}
is sufficient to describe the neutrino up to corrections which are much smaller than the percent-scale statistical and systematic errors of the lattice calculations\footnote{Previous, exploratory work in Ref.~\cite{Detmold:2018zan} examined the neutrino mass dependence of the \(\pi^{-} \rightarrow \pi^{+} e^{-} e^{-}\) amplitude and found that the computed signals were indeed indisinguishable within statistical uncertainties when \(m_{\beta \beta} \ll m_{\pi}\).}. 

To develop methodology, it is instructive to begin by considering the simplest \(0 \nu \beta \beta\) process from the perspective of lattice field theory: an unphysical \(\pi^{-} \rightarrow \pi^{+} e^{-} e^{-}\) transition for pions at rest. While this decay does not occur in nature, it is a well-defined amplitude in quantum field theory, and serves as a natural starting point for lattice calculations since systems of pions are free of the well-known signal-to-noise issue plaguing calculations of nucleon and nuclear systems \cite{Parisi:1983ae,Lepage:1989hd}. In addition, since this transition has only single hadron initial and final states, the volume dependence of the hadronic matrix element is expected to be mild and exponentially suppressed. The \(\pi^{-} \rightarrow \pi^{+} e^{-} e^{-}\) amplitude has been computed at next-to-leading-order in chiral perturbation theory \cite{Cirigliano:2017ymo,Cirigliano:2017djv,Cirigliano:2017tvr,Cirigliano:2018hja,Cirigliano:2018yza,Cirigliano:2019vdj}, allowing for a simple case study of matching lattice results to \(\chi\)PT in the context of \( 0 \nu \beta \beta \) amplitudes. 

The desired \(0 \nu \beta \beta\) matrix element can be extracted in lattice QCD using methods which have been successfully applied to other second-order electroweak processes, including the neutrinoful double beta decay (\(2 \nu \beta \beta\)) amplitude for the process \(n^{0} n^{0} \rightarrow p^{+} p^{+} e^{-} e^{-} \overline{\nu}_{e} \overline{\nu}_{e}\) \cite{Tiburzi:2017iux}, as well as various kaon decays \cite{PhysRevLett.113.112003,Bai:2016gzv,Bai:2018hqu}. The key observation underlying these calculations is that after integrating the Euclidean-space four-point function
\begin{equation}
  \label{eqn:lattice_bilocal_4pt}
  C_{\pi \rightarrow \pi e e}(t_{-}, t_{x}, t_{y}, t_{+}) = \sum_{\vec{x},\vec{y}} \int \frac{d^{4} q}{(2 \pi)^{4}} \frac{1}{q^{2}} e^{i q \cdot (x-y)} \Gamma_{\alpha \beta}^{\rm lept.}\big\langle \mathscr{O}_{\pi^{+}}(t_{+}) \mathcal{T} \left\{ j_{\alpha}(x) j_{\beta}(y) \right\} \mathscr{O}^{\dagger}_{\pi^{-}}(t_{-}) \big\rangle
\end{equation}
over \(t_{x}\) and \(t_{y}\), the required matrix element
\begin{equation}
  M^{0 \nu} = \sum_{n=0}^{\infty} \sum_{\vec{x},\vec{y}} \int \frac{d^{3} q}{(2 \pi)^{3}} \frac{ \Gamma_{\alpha \beta}^{\rm lept.} \big\langle \pi e e \big\vert j_{\alpha}(\vec{x}) \big\vert n \big\rangle \big\langle n \big\vert j_{\beta}(\vec{y}) \big\vert \pi \big\rangle}{2 E_{n} \vert\vec{q}\vert \left( \vert\vec{q}\vert + E_{n} - m_{\pi} \right)} e^{i \vec{q} \cdot (\vec{x}-\vec{y})}
  \label{eqn:0vbb_me}
\end{equation}
appears as the slope of the linear contribution in the \(T \rightarrow \infty\) regime:
\begin{equation}
\begin{split}
  \mathbbm{C}_{\pi \rightarrow \pi e e}(T) &= \sum_{t_{x}=\Delta}^{T-\Delta} \sum_{t_{y}=\Delta}^{T-\Delta} C_{\pi \rightarrow \pi e e}(0,t_{x},t_{y},T) \\ 
  &= \sum_{n=0}^{\infty} \sum_{\vec{x},\vec{y}} \int \frac{d^{3} q}{(2 \pi)^{3}} \frac{\Gamma_{\alpha \beta}^{\rm lept.} \big\langle \pi e e \big\vert j_{\alpha}(\vec{x}) \big\vert n \big\rangle \big\langle n \big\vert j_{\beta}(\vec{y}) \big\vert \pi \big\rangle}{2 E_{n} \vert\vec{q}\vert \left( \vert\vec{q}\vert + E_{n} - m_{\pi} \right)} e^{i \vec{q} \cdot (\vec{x}-\vec{y})} \\
  &\hspace{4cm}\times \left( \left( T - 2 \Delta \right) + \frac{e^{-(\vert\vec{q}\vert + E_{n} - m_{\pi}) ( T - 2 \Delta )} - 1}{\vert\vec{q}\vert + E_{n} - m_{\pi}} \right),
\end{split}
\label{eqn:lattice_bilocal_me}
\end{equation}
where \(T\) is the size of the integration window, and \(n\) indexes all possible intermediate states. In practice, \(T\) is taken as large as possible to suppress the additional exponential and constant contributions appearing in Eq.~\eqref{eqn:lattice_bilocal_me}, and the cutoff \( \Delta \ll T \) is chosen sufficiently large to avoid potential coupling to excited initial or final states which may enter if the current insertions are near the pion sources and sinks. At large \(T\) the matrix element \eqref{eqn:0vbb_me} can be extracted from a simple linear fit to the \(T\) dependence of Eq.~\eqref{eqn:lattice_bilocal_me}.

The procedure described above can be spoiled by the appearance of long-distance intermediate states which would introduce exponentially growing, rather than exponentionally suppressed, contamination into Eq.~\eqref{eqn:lattice_bilocal_me}. This is certainly the case for the \( \pi^{-} \rightarrow \pi^{+} e^{-} e^{-} \) transition, for which a pion-to-vacuum transition
\begin{equation}
\big\langle 0 \big\vert j_{\mu} \big\vert \pi(p) \big\rangle = -i p_{\mu} f_{\pi}
\end{equation}
is allowed. A standard procedure for dealing with this contamination is to compute all such transition amplitudes on the lattice, allowing their contributions to be removed from the four-point function \eqref{eqn:lattice_bilocal_4pt} prior to performing the temporal integration, and thus removing the exponential divergence \cite{Bai:2018hqu}. The contributions to the matrix element \eqref{eqn:0vbb_me} from these low-lying states can then be reintroduced \textit{ex post facto}. This particular aspect of the calculation is more difficult for Majorana exchange processes than for purely hadronic decays, since, in general, this subtraction would require the relevant first-order matrix elements to be computed for the full range of momenta needed to saturate the integral over the neutrino momentum \( \vec{q} \). Fortunately, the only relevant long-distance intermediate state for the \( \pi^{-} \rightarrow \pi^{+} e^{-} e^{-} \) decay is the vacuum, for which the integration over \( \vec{q} \) and the hadronic matrix element decouple. For the phenomenologically important \(n^{0} n^{0} \rightarrow p^{+} p^{+} e^{-} e^{-}\) decay, this issue is likely avoided altogether, since the lightest long-distance intermediate state is the deuteron \cite{Tiburzi:2017iux}, and the power-law fall-off of the neutrino propagator at large separations is expected to overwhelm the exponentionally growing hadronic contribution arising from the small energy splitting between the dinucleon and deuteron states.

A second complication in evaluating the four-point function defined by Eq.~\eqref{eqn:lattice_bilocal_4pt} on the lattice is that the continuum neutrino propagator \eqref{eqn:continuum_neutrino_prop} is divergent in the limit \( x \rightarrow y \). In the context of a lattice calculation, this divergence must be explicitly regulated, and a variety of choices of this regulator have been explored in the literature. The approach taken in the exploratory long-distance \( \pi^{-} \rightarrow \pi^{+} e^{-} e^{-} \) calculation preceding this work \cite{Detmold:2018zan}, as well as in some lattice QCD+QED calculations which implement photon-exchange processes \cite{Endres:2015gda} and suffer from a similar divergence, is to use a lattice-regularized propagator with a non-zero bare mass, which can ultimately be extrapolated to zero. Another possibility explored extensively in the lattice QCD+QED literature is to work directly with a massless lattice propagator after removing the divergent zero-mode contribution --- one such example is the first-principles determination of the neutron-proton mass difference reported in Ref.~\cite{Borsanyi:2014jba} --- which is known to introduce power-law finite volume effects \cite{Duncan:1996xy,10.1143/PTP.120.413}. Yet another regularization scheme is the infinite volume reconstruction method introduced by Feng and Jin \cite{Feng:2018qpx} and applied to neutrinoless double beta decay in Ref.~\cite{Tuo:2019bue}. In the present work an alternative to these methods is explored: the neutrino propagator is implemented with a Gaussian-regulated form of the continuum, infinite volume, massless scalar propagator, 
\begin{equation}
  S_{\Lambda}(x,y) = \int \frac{d^{4} q}{(2 \pi)^{4}} \frac{1}{q^{2}} e^{i \cdot q \left( x - y \right)} e^{-q^{2}/\Lambda^{2}},
\label{eqn:regulated_neutrino_prop}
\end{equation}
which reduces to Eq.~\eqref{eqn:continuum_neutrino_prop} in the limit \(\Lambda \rightarrow \infty\). This approach has a number of advantages: Eq.~\eqref{eqn:regulated_neutrino_prop} is computationally cheap and easily implemented, and does not introduce power-law finite volume effects. In addition, there is a natural choice of the regulator cutoff on the lattice --- \( \Lambda = \pi / a\), where \( a \) is the lattice spacing --- which ensures that the regulator is removed in the continuum limit \(a \rightarrow 0\) without introducing an additional parameter extrapolation. This approach does, however, modify the forms of Eqs.~\eqref{eqn:lattice_bilocal_4pt}-\eqref{eqn:lattice_bilocal_me}. A derivation of the appropriate generalizations of these expressions is given in Appendix \ref{appendix:formalism}.

\section{Calculation}
\label{sec:calc}
The calculations detailed in this work make use of a series of \( N_{f} = 2 + 1 \) domain wall fermion gauge field ensembles generated by the RBC/UKQCD collaboration and summarized in Table \ref{tab:ensembles}. These ensembles use the Iwasaki gauge action \cite{Iwasaki:1984cj} and the domain wall fermion action with the Shamir Kernel \cite{Kaplan:1992bt,Shamir:1993zy} for the quarks. Each ensemble incorporates the sea effects of two isospin-symmetric light quark flavors with bare mass \( a m_{l} \) and a single heavy quark flavor with bare mass \( a m_{h} \). While the bare mass of the heavy flavor has been tuned to closely reproduce the physical strange quark mass, the bare masses of the light quarks are somewhat heavier than the physical up and down quark masses, leading to simulated pion masses in the range \( 300 \) MeV \( \lesssim m_{\pi} \lesssim 430 \) MeV. The range of simulated masses, as well as the two independent lattice spacings and physical volumes, allow for the \( 0 \nu \beta \beta \) matrix element \(M^{0 \nu}\) to be matched to its predicted pion mass dependence from \(\chi\)PT, as well as for the results to be extrapolated to the infinite volume and zero lattice spacing limits. Details of the ensemble generation and fits to the low-energy spectrum are described in Refs.~\cite{Allton:2008pn} and \cite{Aoki:2010dy} for the 24I and 32I ensembles, respectively. The scale-setting analysis used to extract the lattice cutoffs in physical units is described in Ref.~\cite{Boyle:2015exm}.

\begin{table}[!ht]
\setlength{\tabcolsep}{5pt}
\centering
\begin{tabular}{c|cccc|ccc}
\hline
\hline
  \rule{0cm}{0.4cm}Ensemble & \( a m_{l} \) & \( a m_{s} \) & \( \beta \) & \( L^{3} \times T \times L_{s} \) & \( m_{\pi} L \) & \( m_{\pi} \) (MeV) & \( a^{-1} \) (GeV) \\
\hline
  \rule{0cm}{0.4cm}24I & 0.01 & \multirow{2}{*}{0.04} & \multirow{2}{*}{2.13} & \multirow{2}{*}{\(24^{3} \times 64 \times 16\)} & 5.81(1) & 432.2(1.4) & \multirow{2}{*}{1.784(5)} \\
  24I & 0.005 & & & & 4.57(1) & 339.6(1.2) & \\
\hline
  32I & 0.008 & & & & 5.53(1) & 410.8(1.5) & \multirow{3}{*}{2.382(8)} \\
  32I & 0.006 & 0.03 & 2.25 & \(32^{3} \times 64 \times 16\) & 4.84(1) & 359.7(1.2) & \\
  32I & 0.004 & & & & 4.06(1) & 302.0(1.1) & \\
\hline
\hline
\end{tabular}
  \caption{Summary of the ensembles and input parameters used in this analysis. Here, \(\beta\) is the gauge coupling, \(L^{3} \times T \times L_{s}\) is the lattice volume decomposed into the length of the spatial (\(L\)), temporal (\(T\)), and fifth (\(L_{s}\)) dimensions, and \(a m_{l}\) and \(a m_{h}\) are the bare, input light and heavy quark masses. Details of the ensemble generation and scale setting can be found in Refs.~\cite{Allton:2008pn,Aoki:2010dy,Boyle:2015exm}.}
\label{tab:ensembles}
\end{table}

The remainder of this section describes the results of the calculations that were performed, as well as the fits that were used to extract physical quantities of interest. Section \ref{subsec:spectrum} describes fits to two-point correlation functions used to extract the pion masses, decay constants, and normalization factors of each simulation. Section \ref{subsec:LD_amplitude} describes fits to the four-point function used to extract \( M^{0 \nu}\). Finally, in Section \ref{subsec:chpt_extrap}, chiral perturbation theory is used to extrapolate the lattice results for \( M^{0 \nu} \) to the physical pion mass, continuum, and infinite volume limit, as well as to extract the relevant low energy constant \( g_{\nu}^{\pi \pi}(\mu) \).

\subsection{Spectrum}
\label{subsec:spectrum}

Extracting \(M^{0 \nu}\) from the lattice four-point function \eqref{eqn:lattice_bilocal_4pt} requires four inputs: the pion mass, the pion decay constant, the renormalizaton factor for the local \(V-A\) electroweak current, and the pion-to-vacuum transition matrix element
\begin{equation}
  \mathcal{N}_{\mathscr{O}}^{s_{1} s_{2}} = \big\langle 0 \big\vert \mathscr{O}_{\pi}^{s_{1} s_{2}} \big\vert \pi \big\rangle,
\end{equation}
where \(\mathscr{O}_{\pi}^{s_{1} s_{2}}\) are the pion interpolating operators with the same source (\(s_{1}\)) and sink (\(s_{2}\)) smearing as used to compute the four-point function. These quantities can be determined entirely from appropriate two-point functions, which, in this analysis, are constructed from Coulomb-gauge fixed wall source lattice propagators computed using a deflated, mixed-precision conjugate gradient solver \cite{Stathopoulos:2007zi} with 1000 low-mode deflation vectors and a stopping tolerance of \(r = 10^{-8}\). One such propagator is computed for each time slice, and the correlation functions are computed using both a local sink (L) and a zero-momentum projected wall sink (W), and ultimately time-translation averaged over the entire lattice to improve the signal. These techniques, as well as the details of the specific correlation functions and fitting procedures described below, have been developed and used previously in Refs.~\cite{Blum:2014tka,Boyle:2015exm}, and will only be briefly discussed here.   

In this analysis six types of two-point functions are computed: the pseudoscalar-pseudoscalar correlator \(\langle P P \rangle\) with the interpolating operator \(P(x) = \overline{q}(x) \gamma_{5} q(x)\) and a local or wall sink, the axial-pseudoscalar correlator \(\langle A P \rangle\) with \(A_{\mu}(x) = \overline{q}(x) \gamma_{\mu} \gamma_{5} q(x)\) and a local or wall sink, and the correlators
\begin{equation}
  C_{\mathscr{A}}(t) \equiv \Big\langle 0 \Big\vert \sum_{\vec{x}} \partial_{\mu} \mathscr{A}_{\mu}(\vec{x},t) \Big\vert \pi \Big\rangle
\end{equation}
and
\begin{equation}
  C_{A}(t) \equiv \Big\langle 0 \Big\vert \sum_{\vec{x}} \partial_{\mu} A_{\mu}(\vec{x},t) \Big\vert \pi \Big\rangle,
\end{equation}
where \(\mathscr{A}_{\mu}\) is the non-local, five-dimensional conserved axial current \cite{Blum:2014tka} and \(A_{\mu}\) is the local, four-dimensional axial current as defined above. The first four correlators can be used to determine \(m_{\pi}\), \(f_{\pi}\), and the overlap factors \(\mathcal{N}_{\mathscr{O}}^{s_{1} s_{2}}\) by fitting the lattice results to the expected time-dependence of the ground states,
\begin{equation}
  \big\langle 0 \big\vert \mathscr{O}_{1}^{s_{1} s_{2}}(t) (\mathscr{O}^{s_{1} s_{2}}_{2})^{\dagger}(0) \big\vert 0 \big\rangle \stackrel{t \gg 1}{\simeq} \frac{ \mathcal{N}_{\mathscr{O}_{1}}^{s_{1} s_{2}} { \mathcal{N}_{\mathscr{O}_{2}}^{s_{1} s_{2}} }^{\dagger} }{ 2 m_{\pi} } \Big( e^{-m_{\pi} t} \pm e^{-m_{\pi} (T-t)} \Big),
\label{eqn:pion_2pt_ansatz}
\end{equation}
where the sign is + (-) for \(\langle PP \rangle\) (\(\langle AP \rangle\)), and extracting \(f_{\pi}\) from the relation
\begin{equation}
  f_{\pi} = \frac{1}{m_{\pi} V} \frac{\vert \mathcal{N}_{A}^{WL} \vert \vert \mathcal{N}_{P}^{WL} \vert}{\vert \mathcal{N}_{P}^{WW} \vert}.
\end{equation}
The final two correlators involving the divergences of the axial currents are used to extract the axial current renormalization coefficient \(Z_{A}\) by fitting a constant to the ratio
\begin{equation}
  \label{eqn:za_ratio}
  \frac{1}{2} \left[ \frac{ C_{\mathscr{A}}(t-1) + C_{\mathscr{A}}(t) }{ 2 C_{A}(t-\frac{1}{2}) } + \frac{ 2 C_{\mathscr{A}}(t) }{ C_{A}(t+\frac{1}{2}) + C_{A}(t-\frac{1}{2}) } \right] \stackrel{t \gg 1}{\simeq} \frac{Z_{A}}{Z_{\mathscr{A}}}.
\end{equation}
In the remainder of this work it is assumed that \(Z_{\mathscr{A}} \approx 1\) and that \(Z_{V} \approx Z_{A}\), so that the renormalization factor for the \(V-A\) electroweak current may also be approximated by \(Z_{A}\). These approximations are valid up to small \(\mathcal{O}(m_{\rm res})\) and \(\mathcal{O}(m_{\rm res}^{2})\) corrections, respectively, where \(m_{\rm res}\) is the domain wall residual mass. Additional detail can be found in Refs.~\cite{Blum:2014tka,Boyle:2015exm}.

The fits are performed simultaneously to the four \(\langle PP \rangle\) and \(\langle AP \rangle\) two-point functions, as well as to the ratio defined in Eq.~\eqref{eqn:za_ratio}, by minimizing the fully correlated \(\chi^{2}\)
\begin{equation}
  \chi^{2} = \sum_{i=1}^{N} \sum_{j=1}^{N} \left( y_{i} - f(x_{i};\vec{\beta}) \right) \Sigma_{ij}^{-1} \left( y_{j} - f(x_{j};\vec{\beta}) \right),
  \label{eqn:correlated_chi2}
\end{equation}
where \(y_{i}\) are the data,
\begin{equation}
  \Sigma_{ij} = \big\langle \left( y_{i} - \langle y \rangle \right) \left( y_{j} - \langle y \rangle \right) \big\rangle
\end{equation}
is the covariance matrix computed from the data, and \(f\) is the assumed fit form depending on independent variables \(x_{i}\) and parameters \(\vec{\beta}\). The extracted parameters and corresponding \(\chi^{2}/\)dof for each ensemble are summarized in Table \ref{tab:2pt_fits}, and are found to be consistent with the previous determinations in Refs.~\cite{Allton:2008pn,Aoki:2010dy}. In addition, plots of the data and the corresponding fits can be found in Appendix \ref{appendix:2pt}.

\begin{table}[!ht]
\setlength{\tabcolsep}{5pt}
  \centering
  \begin{tabular}{ccccccc}
    \hline
    \hline
    Ensemble & \( a m_{l} \) & \( \vert \mathcal{N}_{P}^{WW} \vert \) & \( a m_{\pi} \) & \( a f_{\pi} \) & \( Z_{A} \) & \(\chi^{2} / {\rm dof}\) \\
    \hline
    \multirow{2}{*}{24I} & 0.01 & \( 1.2224(47) \times 10^{6} \) & 0.24160(45) & 0.09177(25) & 0.717766(57) & 1.20 \\
    & 0.005 & \( 1.1997(54) \times 10^{6} \) & 0.19131(51) & 0.08495(25) & 0.717161(59) & 1.70 \\
    \hline
    \multirow{3}{*}{32I} & 0.008 & \( 3.511(18) \times 10^{6} \) & 0.17277(56) & 0.06802(30) & 0.745357(44) & 1.31 \\
    & 0.006 & \( 3.458(16) \times 10^{6} \) & 0.15077(45) & 0.06477(20) & 0.745088(32) & 1.20 \\
    & 0.004 & \( 3.398(18) \times 10^{6} \) & 0.12652(39) & 0.06194(27) & 0.745020(40) & 0.76 \\
    \hline
    \hline
  \end{tabular}
  \caption{Results for the pion mass, pion decay constant, \(\mathcal{N}_{P}^{WW}\) for the pseudoscalar interpolating operator \(P(x) = \overline{q}(x) \gamma_{5} q(x)\) with a Coulomb gauge-fixed wall source and zero-momentum projected wall sink (WW), the axial current renormalization factor \(Z_{A}\), and the correlated \(\chi^{2}/\)dof for each lattice ensemble. The errors are purely statistical and are computed using the jackknife resampling technique.}
  \label{tab:2pt_fits}
\end{table}

\subsection{Long-Distance \(\pi^{-} \rightarrow \pi^{+} e^{-} e^{-}\) Amplitude}
\label{subsec:LD_amplitude}

\begin{figure}[!ht]
\centering
\subfloat[Neutrino block]{
\resizebox{0.4\linewidth}{!}{
\begin{tikzpicture}
\begin{feynman}
\vertex[small] (a1) {};
\vertex[small, below=0.75cm of a1] (a2) {};
\vertex[small, below=1.5cm of a1] (a3) {};
\vertex[small, below=1.5cm of a1] (a3) {};
\vertex[below=1.875cm of a1] (a4);
\vertex[small, dot, below=2.25cm of a1] (a5) {};
\vertex[small, below=3.0cm of a1] (a6) {};
\vertex[crossed dot, right=1.5cm of a2] (b2) {};
\vertex[crossed dot, right=3.0cm of a4] (c4) {};
\vertex[small, right=4.5cm of a1] (d1) {};
\vertex[small, right=4.5cm of a2] (d2) {};
\vertex[small, right=4.5cm of a3] (d3) {};
\vertex[small, dot, right=4.5cm of a5] (d5) {};
\vertex[small, right=4.5cm of a6] (d6) {};
\diagram* {
	(a5) -- [fermion, out=0, in=210, edge label'=\(d\)] (c4)
			 -- [fermion, out=-30, in=180, edge label'=\(u\)] (d5),
	(b2) -- [majorana, edge label'=\(\nu\)] (c4),
	(a1) -- [scalar] (a6),
	(d1) -- [scalar] (d6),
};
\vertex[left=1.5em of a1];
\vertex[left=1.0em of b2] {\(x\)};
\vertex[left=1.5em of a5] {\(\alpha, i\)};
\vertex[left=1.5em of a6] {\(t_{-}\)};
\vertex[above=1.0em of c4] {\(y\)};
\vertex[right=1.5em of d1];
\vertex[right=1.5em of d5] {\(\beta,j\)};
\vertex[right=1.5em of d6] {\(t_{+}\)};
\end{feynman}
\end{tikzpicture}
}} \\
  \hspace{0.2cm}
  \subfloat[Type \circled{1} contraction]{\includegraphics[width=0.4\linewidth]{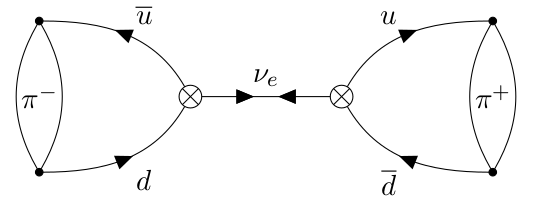}}
\hspace{0.2cm}
  \subfloat[Type \circled{2} contraction]{\includegraphics[width=0.4\linewidth]{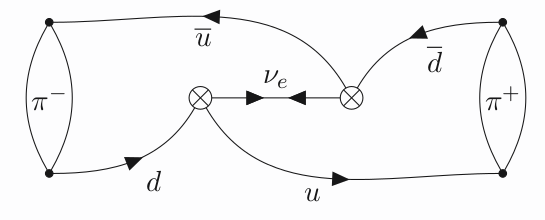}}
\begin{equation}
  \circled{1} = \mathrm{Tr} \left[ S_{u}^{\dagger}(t_{-} \rightarrow x) \gamma_{\alpha} \left( 1 - \gamma_{5} \right) S_{d}(t_{-} \rightarrow x) \right] \cdot \mathrm{Tr} \left[ S_{u}^{\dagger}(t_{+} \rightarrow y) \gamma_{\beta} \left( 1 - \gamma_{5} \right) S_{d}(t_{+} \rightarrow y) \right]
\label{eqn:pion_contraction_1}
\end{equation}
\begin{equation}
  \circled{2} = \mathrm{Tr} \left[ S_{u}^{\dagger}(t_{+} \rightarrow x) \gamma_{\alpha} \left( 1 - \gamma_{5} \right) S_{d}(t_{-} \rightarrow x) S_{u}^{\dagger}(t_{-} \rightarrow y) \gamma_{\beta} \left( 1 - \gamma_{5} \right) S_{d}(t_{+} \rightarrow y) \right]
  \label{eqn:pion_contraction_2}
\end{equation}
  \caption{Top: diagrammatic representation of the neutrino block construction (Eq.~\eqref{eqn:neutrino_block}). The labels \((\alpha,i)\) and \((\beta,j)\) reflect the open spin and color indices at the source and sink, respectively. Bottom: two classes of hadronic contractions for the $\pi^{-} \rightarrow \pi^{+} e^{-} e^{-}$ decay. Crossed circles denote insertions of the electroweak current.}
  \label{fig:pion_contractions}
\end{figure}

Applying Wick's theorem to the hadronic matrix element, Eq.~\eqref{eqn:hadronic_tensor}, results in two classes of diagrams and four total contractions, depicted in Figure \ref{fig:pion_contractions}. 
In practice, computing these contractions by brute force is prohibitively expensive due to the double summation over the spacetime locations of the current insertions. In Ref.~\cite{Tiburzi:2017iux}, this problem was solved for neutrinoful double beta decay amplitudes by computing quark propagators in the presence of an additional background axial field, which can be shown to implicitly induce this summation. Unfortunately, background field techniques do not easily generalize to include the neutrino propagator, requiring the development of other techniques for \(0 \nu \beta \beta\) decays. 

A general method for computing \(0 \nu \beta \beta\) contractions, including the diagrams in Figure \ref{fig:pion_contractions}, with the full integration over the locations of both current insertions was introduced in Ref.~\cite{Detmold:2018zan}, and is briefly reviewed here. This method works by exploiting the convolution theorem and the translational invariance of the neutrino propagator to reduce the cost of the summation from \(\mathcal{O}(V^{2})\), where \(V\) is the lattice volume, to \(\mathcal{O}(V \log V)\) using the fast Fourier transform (FFT). The key idea, depicted in the top panel of Figure \ref{fig:pion_contractions}, is to use the FFT to integrate the leptonic tensor, Eq.~\eqref{eqn:leptonic_tensor}, against the quark lines passing through one of the two weak current insertions. More explicitly, for each fixed time ordering of the operators a \(12 \times 12\) spin-color matrix-valued field\footnote{For the special case of the type 1 contraction \eqref{eqn:pion_contraction_1} this cost can be reduced by a further factor of 144 since this contraction factorizes into two independent spin-color traces. Thus, it suffices to compute a scalar neutrino block rather than the full \(12 \times 12\) spin-color matrix}
\begin{equation}
  \begin{split}
    B_{\alpha}(x; t_{-}, t_{+}) &= \int d^{3} y \, L_{\alpha \beta}(x-y) \Big[ S_{u}^{\dagger}(t_{-} \rightarrow y) \gamma_{\beta} \left( 1 - \gamma_{5} \right) S_{d}(t_{+} \rightarrow y) \Big] \\
    &= \mathscr{F}^{-1} \Big[ \mathscr{F}(L_{\alpha \beta}) \cdot \mathscr{F} \left( S_{u}^{\dagger} \gamma_{\beta} \left( 1 - \gamma_{5} \right) S_{d} \right) \Big]
  \end{split}
  \label{eqn:neutrino_block}
\end{equation}
is computed using the FFT (\(\mathscr{F}\)) and its inverse. The full contractions --- for example, Eqs.~\eqref{eqn:pion_contraction_1} and \eqref{eqn:pion_contraction_2} for the \(\pi^{-} \rightarrow \pi^{+} e^{-} e^{-}\) transition --- are then assembled by integrating this ``neutrino block'' \(B_{\alpha}(x)\) against quark propagators to \(x\) and contracting the remaining open indices in the appropriate combinations, for a total cost scaling as \(\mathcal{O}(V \log V)\). Scaling benchmarks for this algorithm on CPUs and GPUs which demonstrate its efficiency were reported in Ref.~\cite{Detmold:2018zan}. 

\begin{figure}[!ht]
\centering
\subfloat{\includegraphics[width=0.48\linewidth]{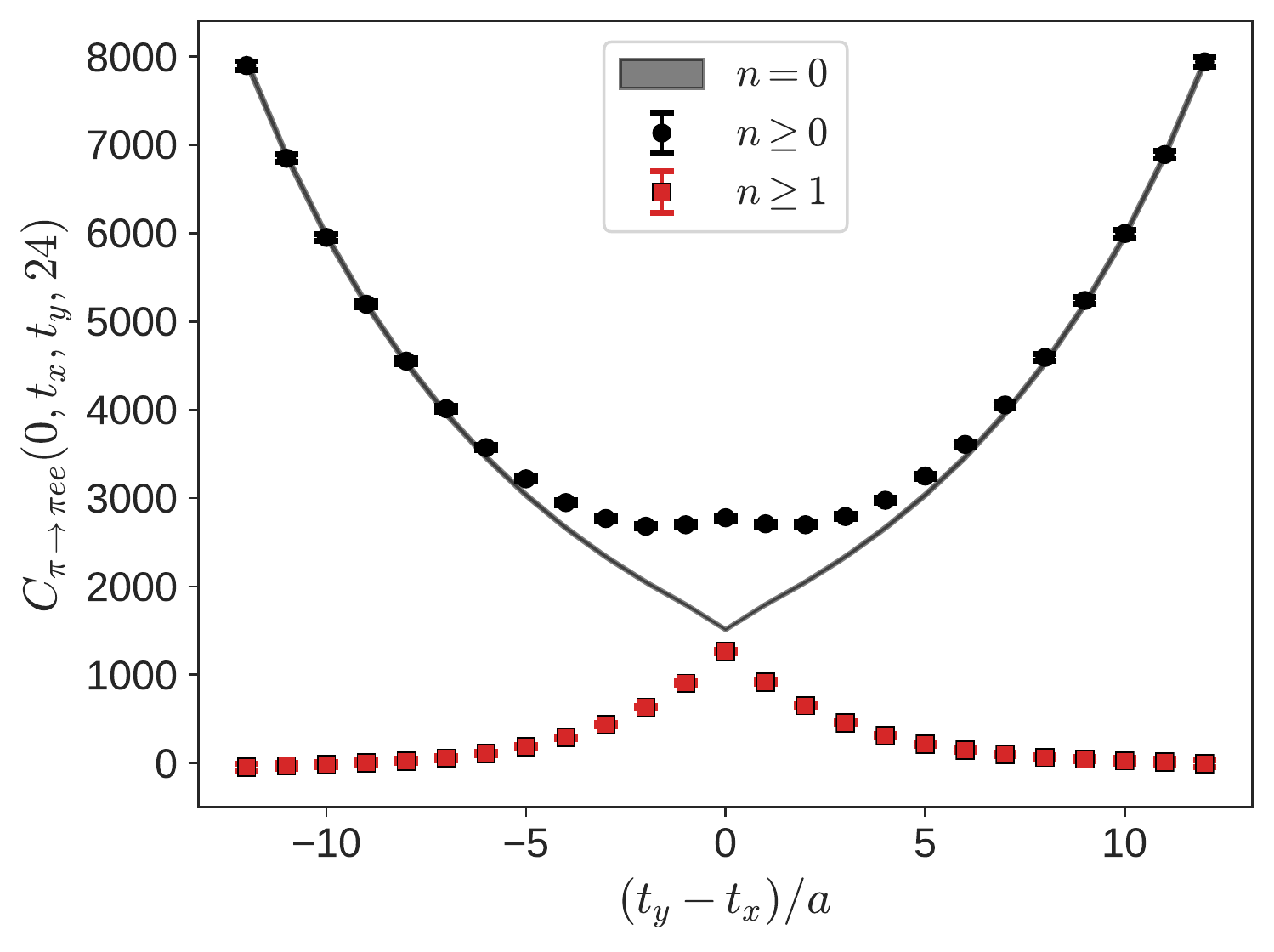}}
\subfloat{\includegraphics[width=0.47\linewidth]{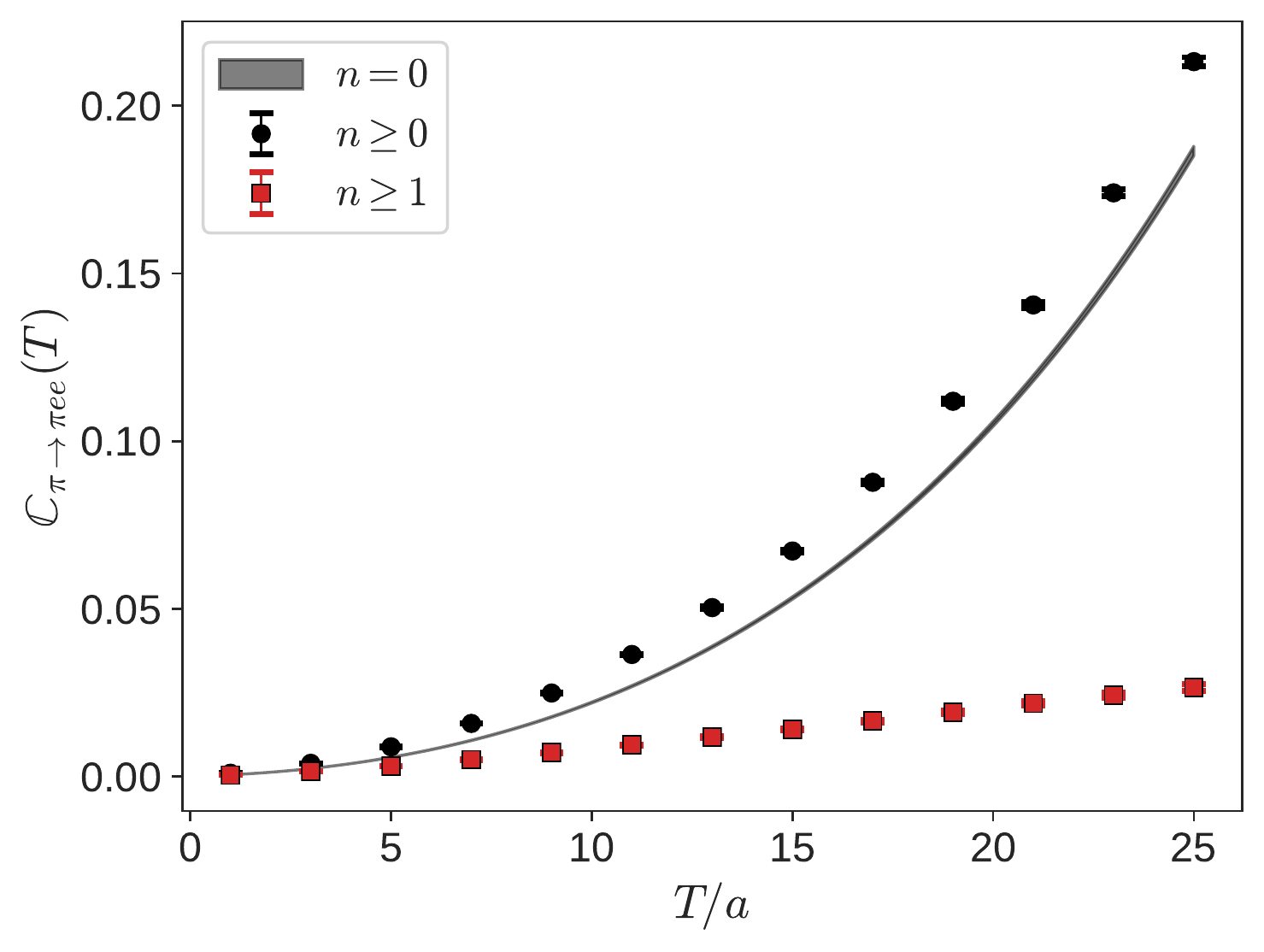}}
  \caption{Left (right): Example signals for the raw (integrated) four-point function, shown in black (red) and defined in Eq.~\eqref{eqn:lattice_bilocal_4pt} (Eq.~\eqref{eqn:lattice_bilocal_me}), before and after subtracting the contribution from the vacuum intermediate state on the 24I \(a m_{l} = 0.01\) ensemble. The gray band is the vacuum intermediate state contribution computed using Eq.~\eqref{eqn:regulated_vacuum_4pt} and the results of Section \ref{subsec:spectrum}.}
\label{fig:24I_vacuum_subtraction}
\end{figure}

In this work, the four-point function, Eq.~\eqref{eqn:lattice_bilocal_4pt}, is computed using the algorithm described above for all \(\pi^{-}\) source and \(\pi^{+}\) sink separations between 12 and 24 lattice units, and for all time orderings of the weak current insertions which are a distance of at least 6 lattice units from the source and sink\footnote{This minimum separation corresponds to a physical distance of 0.7 (0.5) fm on the 24I (32I) ensembles.}. In addition, on each gauge field configuration the type 1 contraction is averaged over all time translations for each fixed time-ordering of the operators, while the more expensive type 2 contraction is time-averaged over four randomly chosen translations. The neutrino propagator is regulated using the UV cutoff imposed by the lattice itself, \(\Lambda = \pi/a\). The procedure for extracting \(M^{0 \nu}\) from this data is as follows: first, the fitted values of \(m_{\pi}\), \(f_{\pi}\), \(Z_{A}\), and \(\mathcal{N}_{\pi}^{WW}\) from Section \ref{subsec:spectrum} are used to remove the exponentionally divergent contribution from the vacuum intermediate state by subtracting Eq.~\eqref{eqn:regulated_vacuum_4pt} from the data. Then, the subtracted four point function is normalized according to Eq.~\eqref{eqn:normalized_4pt} to remove the dependence on the source-sink separation. Next, this normalized four-point function is integrated in the remaining time dependence of the current insertions for each source-sink separation, and a linear fit is performed to this signal at large separation \(T\). The slope of this fit determines \(M^{0 \nu}\) up to the contribution from the vacuum intermediate state (\(M^{0 \nu}_{(0)}\)). Finally, this missing contribution is reintroduced using Eq.~\eqref{eqn:regulated_vacuum_me}. Figure \ref{fig:24I_vacuum_subtraction} illustrates the vacuum subtraction procedure using data computed on the 24I \(a m_{l} = 0.01\) ensemble.

\begin{figure}[!ht]
\centering
\subfloat[24I ensembles]{\includegraphics[width=0.48\linewidth]{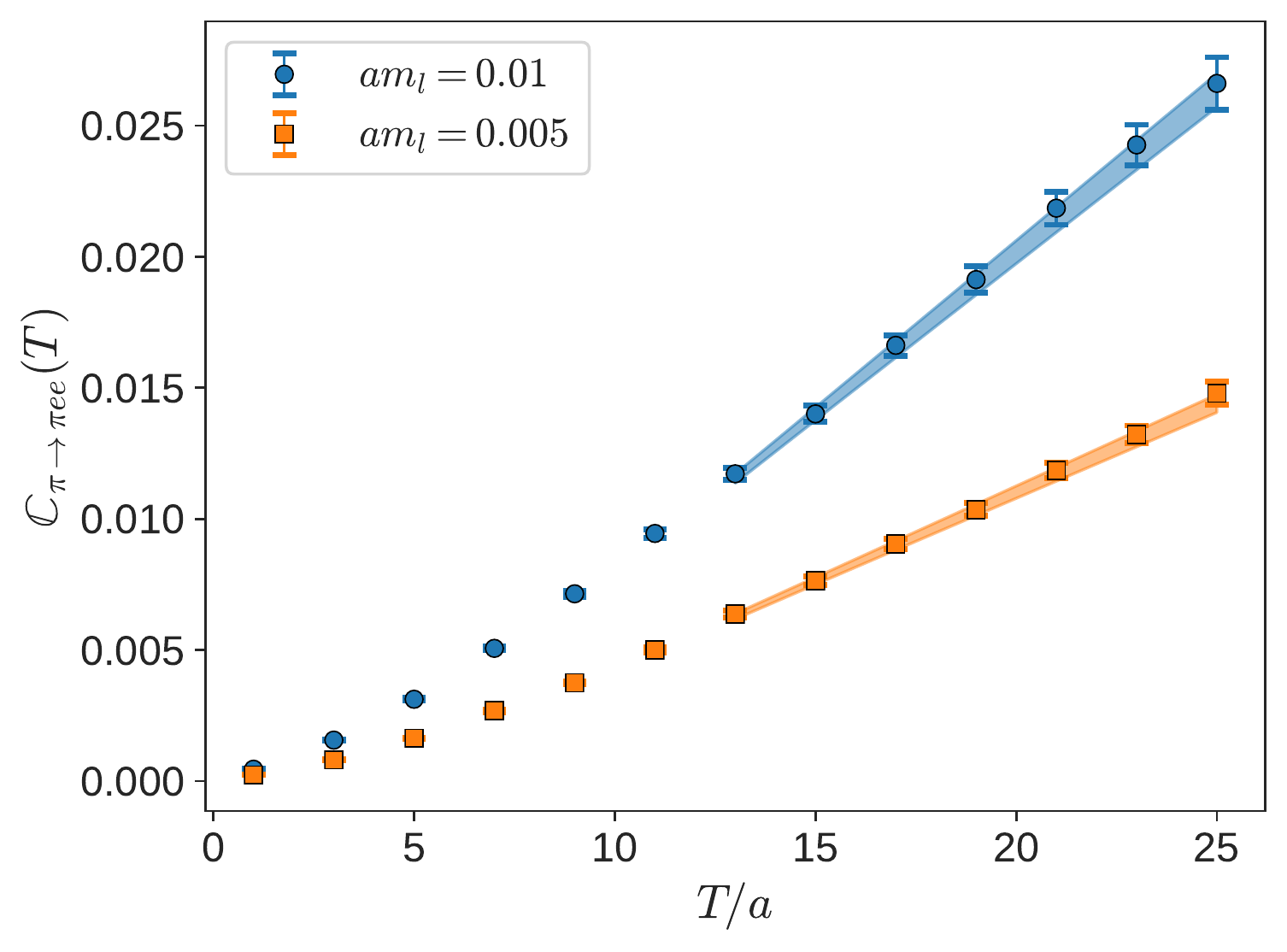}}
\subfloat[32I ensembles]{\includegraphics[width=0.48\linewidth]{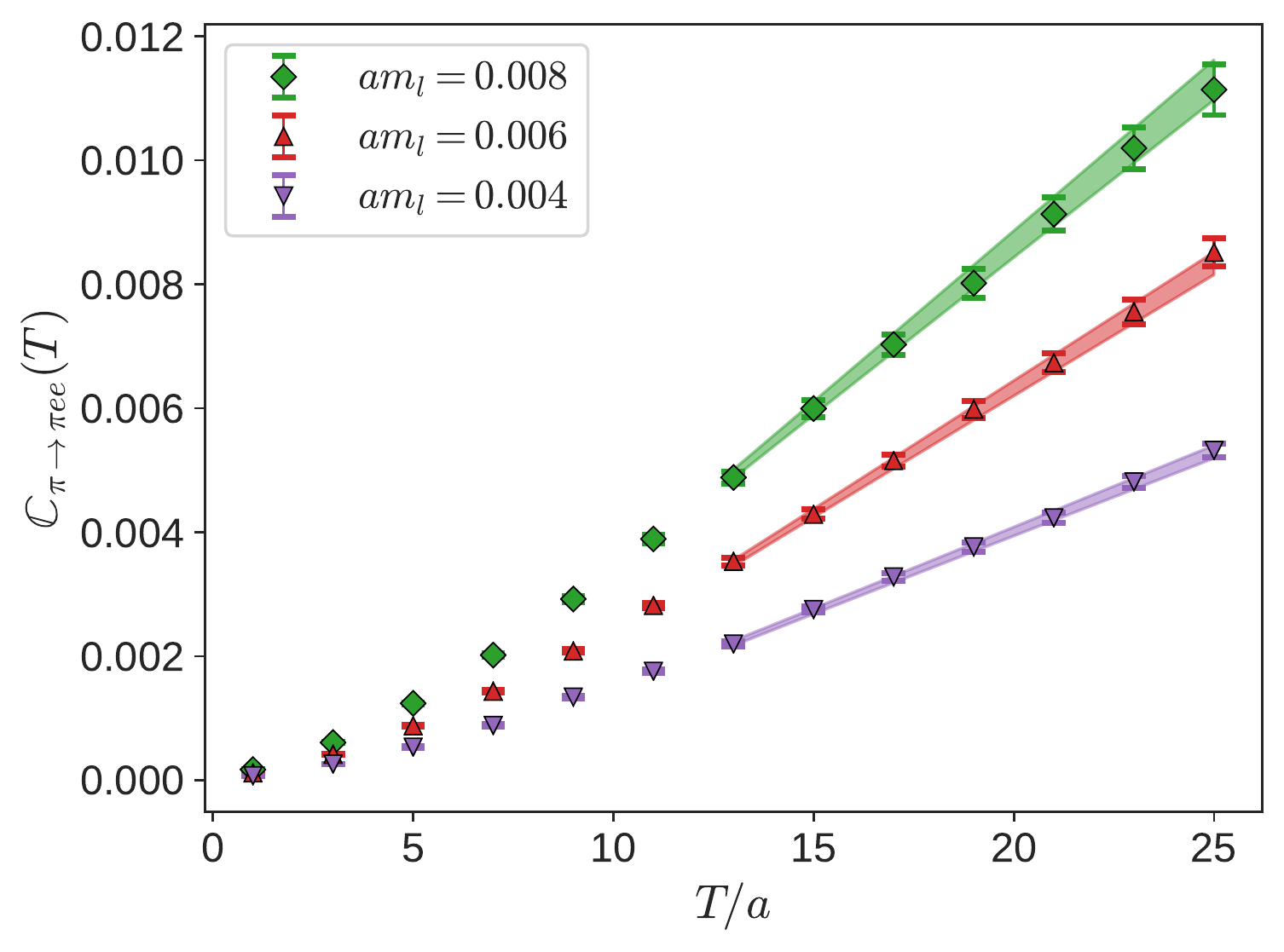}}
  \caption{Lattice signals and example fits to the window \([T_{\rm min}/a, T_{\rm max}/a] = [13,25]\) for the integrated four-point function, Eq.~\eqref{eqn:lattice_bilocal_me}. The data has been processed by first normalizing the raw four-point function, Eq.~\eqref{eqn:lattice_bilocal_4pt}, using Eq.~\eqref{eqn:normalized_4pt}, and then removing the exponentionally divergent contribution from the vacuum intermediate state using Eq.~\eqref{eqn:regulated_vacuum_4pt}.}
\label{fig:lattice_4pt_data}
\end{figure}

In contrast to the fits to the two-point functions --- which, in Appendix \ref{appendix:2pt}, exhibit clear plateau regions where the asymptotic ground state fit forms are valid, and are insensitive to the choice of fit range within this window --- the slopes extracted from linear fits to the integrated four-point functions are observed to be somewhat sensitive to the choice of fit range, while still maintaining acceptable \(\chi^{2}/\mathrm{dof} \simeq 1\). To account for the systematic uncertainty in the choice of fit window, and to avoid potentially introducing a bias into the analysis, the procedure for averaging over fits introduced in Refs.~\cite{Rinaldi:2019thf,
Beane:2020ycc} is adopted. All possible linear fits to the window \([T_{\rm min}/a, T_{\rm max}/a]\) with \( 9 \leq T_{\rm min}/a \leq 19 \), \( 21 \leq T_{\rm max}/a \leq 25 \), and at least four degrees of freedom, are performed by minimizing the correlated \(\chi^{2}\) \eqref{eqn:correlated_chi2}. The parameters \(x = x_{i} \pm \delta x_{i}^{\rm stat.}\) extracted from these fits, labeled by the index \(i\), are then averaged according to
\begin{equation}
  \hat{x} = \frac{\sum_{i} w_{i} x_{i}}{\sum_{i} w_{i}}, \hspace{0.4cm} w_{i} =  \frac{p_{i}}{\left( \delta x_{i}^{\rm stat.} \right)^{2}},
  \label{eqn:fit_avg}
\end{equation}
where \(p_{i}\) is the \(p\)-value corresponding to the \(\chi^{2}\)/dof obtained in fit \(i\), and \(w_{i}\) is its associated weight. This choice of weight is constructed to penalize both poor fits with small \(p_{i}\) and fits with large \(\delta x^{\rm stat.}_{i}\) which poorly constrain the parameters. Uncertainties are assigned by also averaging the statistical uncertainties from each fit
\begin{equation}
\left( \delta \hat{x}^{\rm stat.} \right)^{2} = \frac{\sum_{i} w_{i} \left(\delta x_{i}^{\rm stat.} \right)^{2}}{\sum_{i} w_{i}},
  \label{eqn:fit_stat_err}
\end{equation}
and computing the weighted average deviation from Eq.~\eqref{eqn:fit_avg}
\begin{equation}
  \left( \delta \hat{x}^{\rm sys.} \right)^{2} = \frac{\sum_{i} w_{i} \left( x_{i} - \hat{x}_{i} \right)^{2}}{\sum_{i} w_{i}}
\end{equation}
as an estimate of the systematic uncertainty. Results for \(M^{0 \nu}\) obtained from this procedure, including both statistical and systematic uncertainties, are reported in Table \ref{tab:M0v_fit_results}, along with the dimensionless matrix element
\begin{equation}
  \mathcal{S}_{\pi \pi} = \frac{M^{0 \nu}}{M^{0 \nu}_{(0)}},
  \label{eqn:Spp}
\end{equation}
in anticipation of the fits to \(\chi\)PT discussed in the following section. Example fits to the data for the window \([T_{\rm min}/a, T_{\rm max}/a] = [13,25]\) are shown in Figure \ref{fig:lattice_4pt_data}.

\begin{table}[!ht]
\centering
\begin{tabular}{ccccccc}
\hline
\hline
  Ensemble & \( a m_{{l}} \) & \( a^{2} M_{{(0)}}^{{0 \nu}} \) & \( a^{2} ( M^{0 \nu} - M_{(0)}^{0 \nu} ) \) & \( a^{2} M^{{0 \nu}} \) & \( \mathcal{{S}}_{{\pi \pi}} \) & \( \chi^{{2}} / {{\rm dof}} \) \\
\hline
\rule{0cm}{0.4cm}\multirow{2}{*}{24I} & 0.01 & $0.008422(47)$ & $0.000664(30)(21)$ & $0.009090(51)(14)$ & $1.0788(37)(25)$ & $0.59$ \\
& 0.005 & $0.007217(42)$ & $0.000824(30)(23)$ & $0.008049(54)(22)$ & $1.1140(41)(32)$ & $0.56$ \\
\hline
\rule{0cm}{0.4cm}\multirow{3}{*}{32I} & 0.008 & $0.004627(41)$ & $0.000332(13)(6)$ & $0.004961(40)(6)$ & $1.0717(31)(14)$ & $0.61$ \\
& 0.006 & $0.004196(26)$ & $0.000424(17)(10)$ & $0.004621(34)(9)$ & $1.1012(41)(24)$ & $0.63$ \\
& 0.004 & $0.003837(34)$ & $0.000671(24)(10)$ & $0.004508(46)(11)$ & $1.1749(62)(26)$ & $0.75$ \\
\hline
\hline
\end{tabular}
  \caption{Results for the contribution to the \(0 \nu \beta \beta\) matrix element \(M^{0 \nu}\) from the vacuum intermediate state (\(M^{0 \nu}_{(0)}\)), the correlated difference \(M^{0 \nu} - M^{0 \nu}_{(0)}\) obtained from the slope of a linear fit to the integrated four-point function \(\mathbbm{C}_{\pi \rightarrow \pi e e}(T)\) in the limit \(T \gg 1\), the full matrix element \(M^{0 \nu}\), and the dimensionless matrix element \(S_{\pi \pi} = M^{0 \nu} / M^{0 \nu}_{(0)}\). The quoted uncertainties are statistical and systematic, respectively, as computed by the fit averaging procedure described in the text. \(M^{0 \nu}_{(0)}\) is assigned a systematic error of zero since it is computed from Eq.~\eqref{eqn:regulated_vacuum_me}, and does not depend on the linear fits performed in Section \ref{subsec:LD_amplitude}.}
\label{tab:M0v_fit_results}
\end{table}

\subsection{Chiral/Continuum Extrapolation}
\label{subsec:chpt_extrap}

The final step in the calculation is to extrapolate the lattice data to the combined limits of physical pion mass, zero lattice spacing, and infinite volume. These extrapolations are performed simultaneously using the ansatz
\begin{equation}
  \mathcal{S}_{\pi \pi} = \underbrace{1 + \frac{m_{\pi}^{2}}{8 \pi^{2} f_{\pi}^{2}} \Bigg( 3 \log \left( \frac{\mu^{2}}{m_{\pi}^{2}} \right) + 6 + \frac{5}{6} g_{\nu}^{\pi \pi}(\mu) \Bigg)}_{\mathrm{NLO} \,\, \chi\mathrm{PT}} + \underbrace{c_{FV}^{\rm NLO}\frac{e^{-m_{\pi} L}}{\left( m_{\pi} L \right)^{3/2}}}_{\rm FV} + \underbrace{c_{a} a^{2}}_{\rm Continuum},
  \label{eqn:chiral_ansatz}
\end{equation}
which includes the next-to-leading-order (NLO) pion mass dependence computed in \(\chi\)PT \cite{Cirigliano:2017tvr}, as well as models of the leading order discretization effects and finite volume effects. The term linear in \(a^{2}\) is motivated by the observation that the leading discretization artifacts enter at \(\mathcal{O}(a^{2})\) for domain wall fermions. The finite volume term is motivated by the leading order asymptotic expansions of the NLO \(\chi\)PT finite volume corrections for \(f_{\pi}\) \cite{Allton:2008pn} and the pion vector form factor \cite{Alexandrou:2017blh}, which enter as the \(n = 0\) and \(n = 1\) first-order hadronic matrix elements in Eq.~\eqref{eqn:0vbb_me}, respectively. In both cases \(\chi\)PT predicts 
\begin{equation}
  \Delta_{FV}^{\rm NLO} \propto \frac{e^{-m_{\pi} L}}{\left( m_{\pi} L \right)^{3/2}},
\end{equation}
up to higher order contributions suppressed by additional powers of \(m_{\pi} L\). In principle, the finite volume corrections could be computed self-consistently within the framework of \(\chi\)PT, but this calculation has not been performed for the \(\pi^{-} \rightarrow \pi^{+} e^{-} e^{-}\) amplitude in the literature. We also consider a second, more general fit ansatz
\begin{equation}
  \mathcal{S}_{\pi \pi} = 1 + \frac{m_{\pi}^{2}}{8 \pi^{2} f_{\pi}^{2}} \Bigg( 3 \log \left( \frac{\mu^{2}}{m_{\pi}^{2}} \right) + 6 + \frac{5}{6} g_{\nu}^{\pi \pi}(\mu) \Bigg) + \Bigg( \frac{c_{FV}^{\rm NLO}}{\left( m_{\pi} L \right)^{3/2}} + \frac{c_{FV}^{\rm NNLO}}{\left( m_{\pi} L \right)^{5/2}} \Bigg) e^{-m_{\pi} L} + c_{a} a^{2},
\label{eqn:chiral_ansatz_sys_error}
\end{equation}
which includes a model of the next-to-next-to-leading-order (NNLO) finite volume corrections. This generalized ansatz is only used for the purpose of studying fit systematics associated with the volume dependence.

Table \ref{tab:chiral_fits} summarizes a variety of fits to the ans\"{a}tze Eq.~\eqref{eqn:chiral_ansatz} and Eq.~\eqref{eqn:chiral_ansatz_sys_error} using different subsets of the data, and the superjackknife resampling technique to propagate uncertainties from independent ensembles into a global fit \cite{Bratt:2010jn}. The uncertainties in the inverse lattice spacings used to convert to physical units (Table \ref{tab:ensembles}), as well as the uncertainties in the physical \(m_{\pi^{-}}^{\rm PDG} = 139.5702(4)\) MeV and \(f_{\pi}^{\rm PDG} = 130.4(2)\) MeV reported by the Particle Data Group (PDG) \cite{PhysRevD.98.030001} and used to define the physical point of the extrapolation, are included by generating superjackknife distributions with random fluctuations drawn from an appropriate normal distribution. The renormalization scale for the low energy constant (LEC) \(g_{\nu}^{\pi \pi}\) is fixed at the conventional value \(\mu = 770\) MeV.

\begin{table}[!ht]
\centering
  \resizebox{\linewidth}{!}{
\begin{tabular}{c|cc|cccc|cc|c}
\hline
\hline
  \rule{0cm}{0.4cm}Label & \(m_{\pi}^{\rm min}\) (MeV) & \(m_{\pi}^{\rm max}\) (MeV) & \(g_{\nu}^{\pi \pi}(\mu)\) & \(c_{FV}^{\rm NLO}\) & \(c_{FV}^{\rm NNLO}\) & \(c_{a}\) (fm\(^{-2}\)) & \(\mathcal{S}_{\pi \pi}^{\rm phys.}\) & \(M^{0 \nu}_{\rm phys.}\) (GeV\(^{2}\)) & \(\chi^{2}/\)dof \\ 
\hline
  \rule{0cm}{0.4cm}A1 & 302.0(1.1) & 432.2(1.4) & -10.71(11)(4) & 0.3(3.2)(1.2) & \(\equiv 0\) & 0.80(78)(33) & 1.1062(13)(6) & 0.018813(58)(12) & 24.5 \\
  A2 & 302.0(1.1) & 410.8(1.5) & -10.43(11)(4) & 18.2(3.8)(1.7) & \(\equiv 0\) & -3.8(9)(5) & 1.1096(13)(5) & 0.018871(55)(22) & 3.8 \\
  A3 & 339.6(1.2) & 432.2(1.4) & -10.78(12)(4) & -47.9(5.8)(3.8) & \(\equiv 0\) & 3.6(1.0)(0.4) & 1.1054(14)(6) & 0.018799(57)(17) & 0.2 \\
  A4 & 339.6(1.2) & 410.8(1.5) & -10.69(12)(7) & -34(12)(7) & \(\equiv 0\) & 1.9(1.6)(1.0) & 1.1064(16)(8) & 0.018817(57)(9) & --- \\
\hline
\rule{0cm}{0.4cm}B1 & 339.6(1.2) & 432.2(1.4) & -10.70(10)(6) & \(\equiv 0\) & \(\equiv 0\) & 0.55(67)(44) & 1.1063(12)(8) & 0.018815(58)(17) & 8.9 \\
\hline
  \rule{0cm}{0.4cm}C1 & 302.0(1.1) & 432.2(1.4) & -10.28(11)(6) & -458(48)(29) & 1850(190)(120) & 2.09(79)(43) & 1.1115(13)(8) & 0.018903(58)(5) & 0.5 \\
\hline
\hline
\end{tabular}
  }
  \caption{Summary of fits of the lattice data reported in Table \ref{tab:M0v_fit_results} to the ans\"{a}tze Eq.~\eqref{eqn:chiral_ansatz} and Eq.~\eqref{eqn:chiral_ansatz_sys_error}. Fits (A1)-(A4) use the ansatz Eq.~\eqref{eqn:chiral_ansatz} but vary the subset of the data included in the fit. Fit (B1) uses the same data and ansatz as fit (A3) but discards the infinite volume extrapolation by fixing \(c_{FV}^{\rm NLO} \equiv 0\). Finally, fit (C1) uses all of the available data and the more general ansatz Eq.~\eqref{eqn:chiral_ansatz_sys_error}. The matrix elements \(\mathcal{S}_{\pi \pi}^{\rm phys.}\) and \(M^{0 \nu}_{\rm phys.}\) are obtained by using the fit to extrapolate to the physical \(m_{\pi}\) and \(f_{\pi}\), as well as to zero lattice spacing and infinite volume. The \(\chi\)PT LEC \(g_{\nu}^{\pi \pi}\) is determined at the renormalization scale \(\mu = 770\) MeV.}
  \label{tab:chiral_fits}
\end{table}

\begin{figure}[!ht]
\centering
  \subfloat{\includegraphics[width=0.48\linewidth]{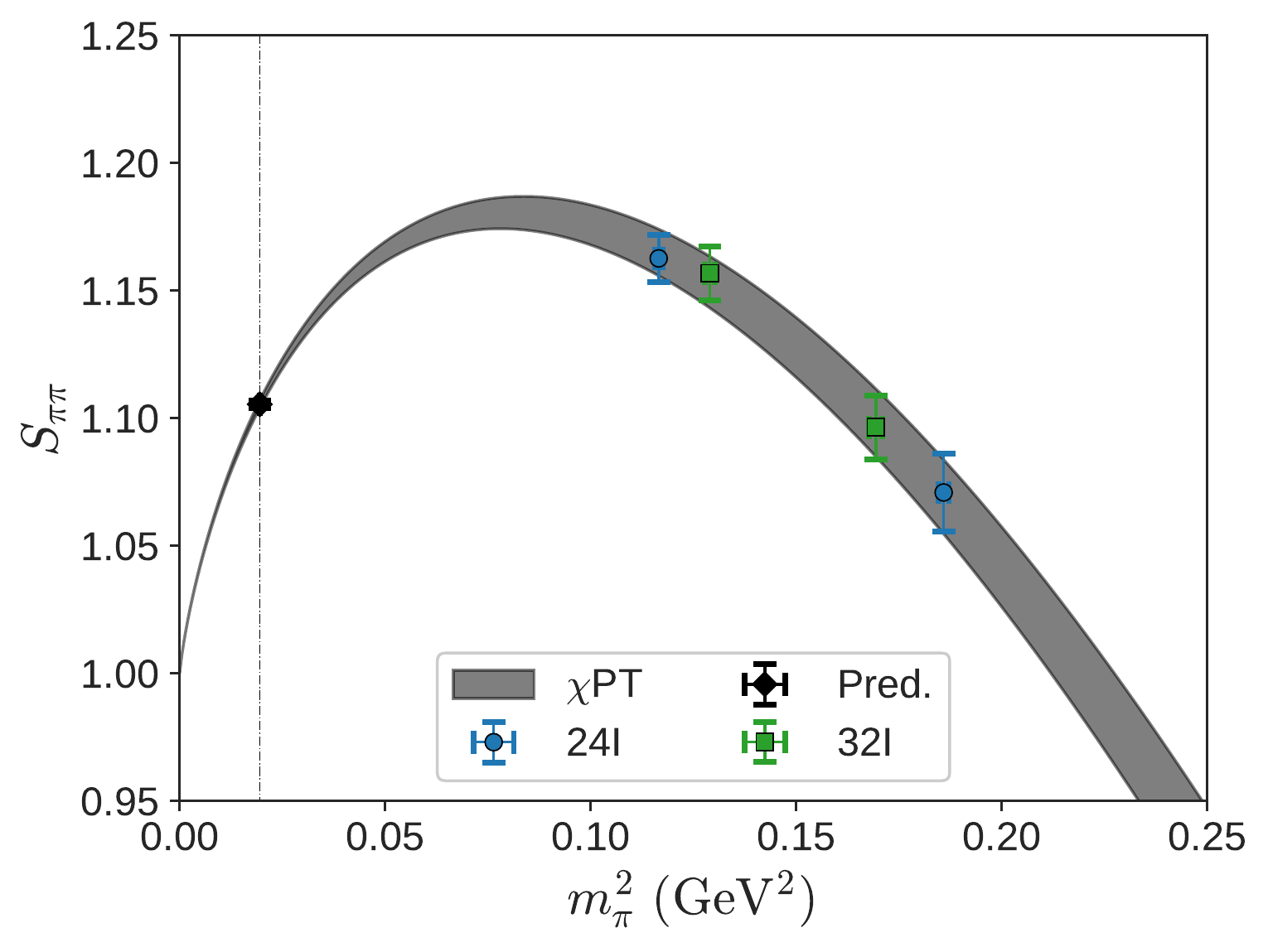}}
  \subfloat{\includegraphics[width=0.48\linewidth]{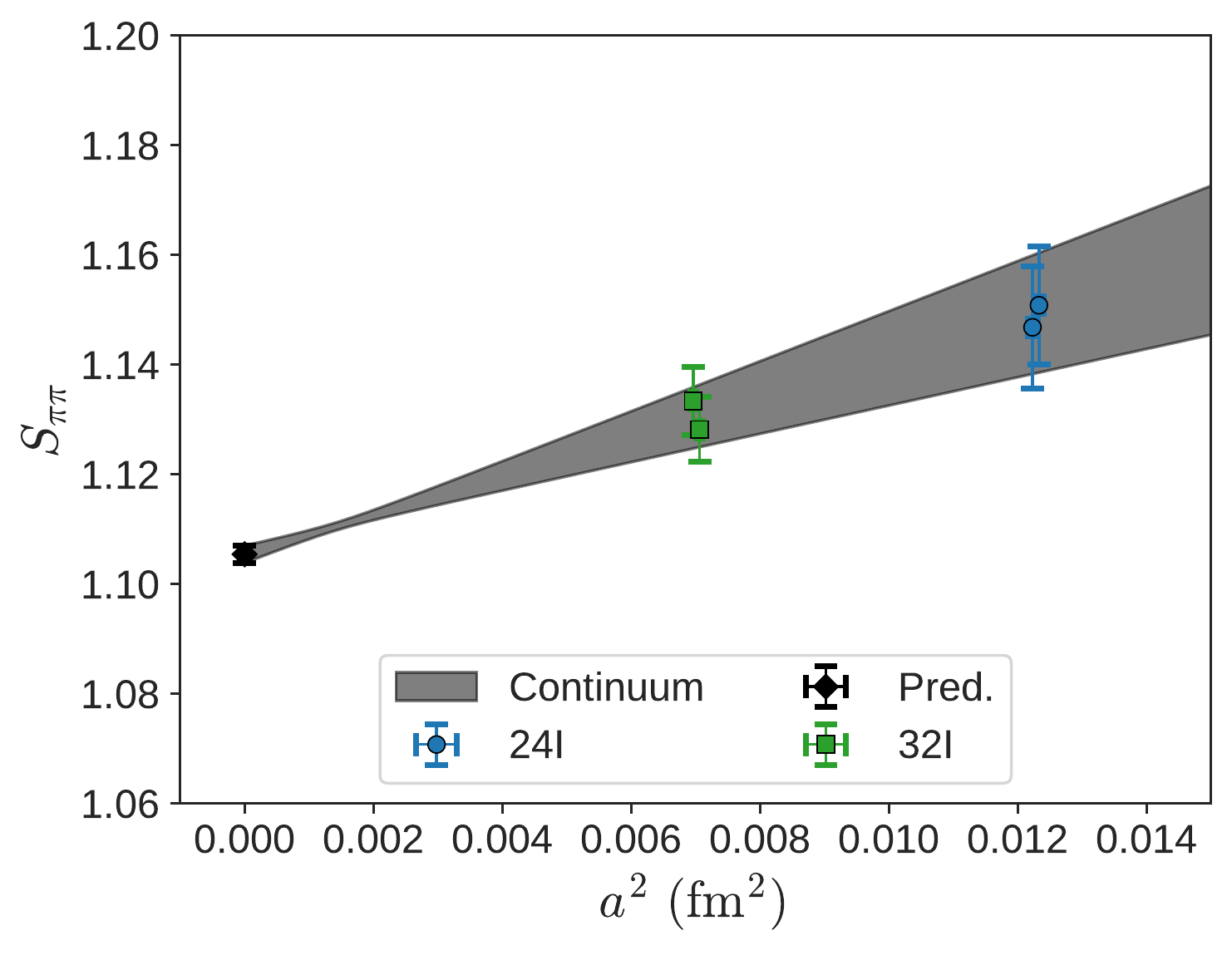}} \\
  \subfloat{\includegraphics[width=0.48\linewidth]{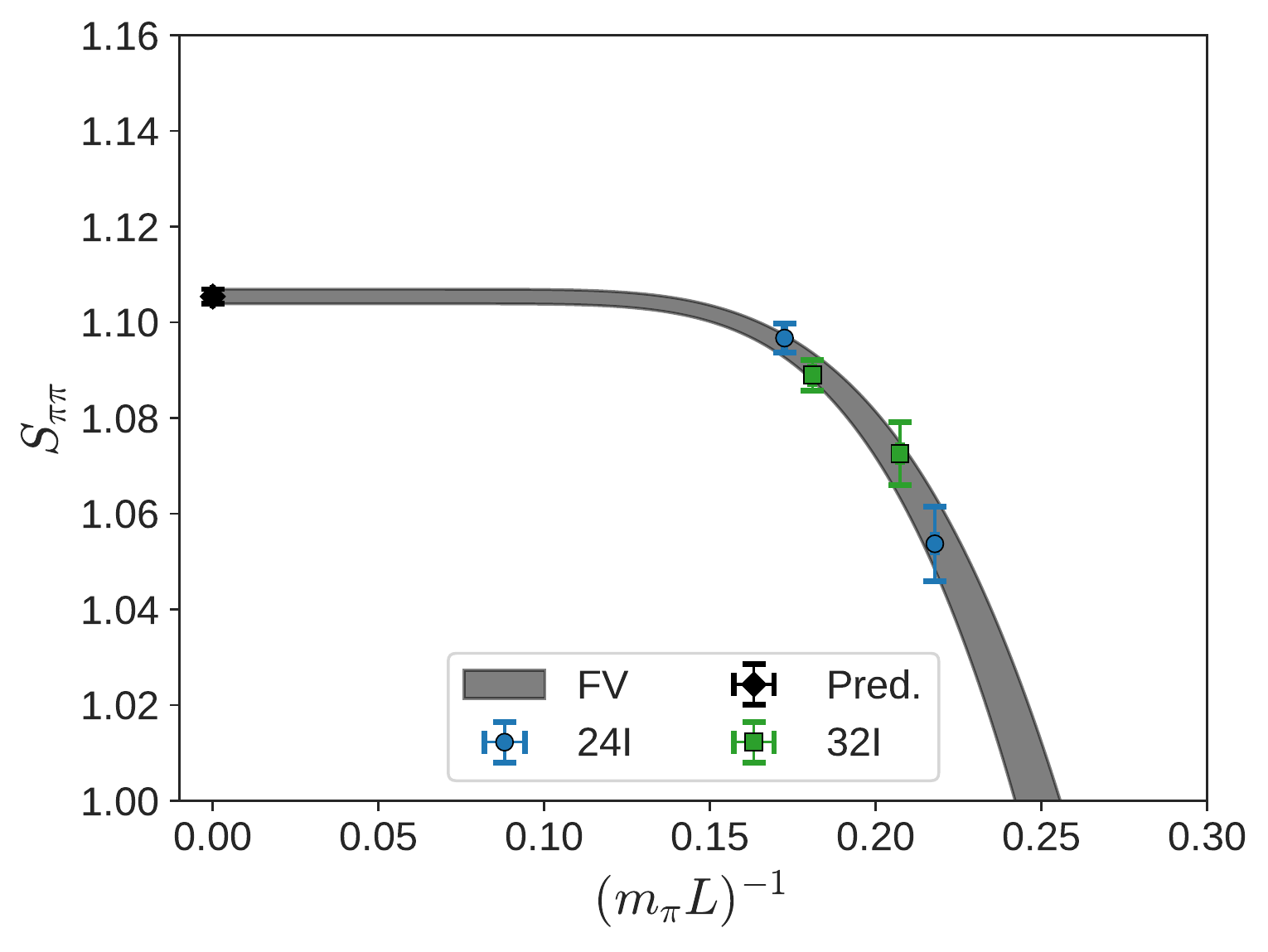}}
  \subfloat{\includegraphics[width=0.48\linewidth]{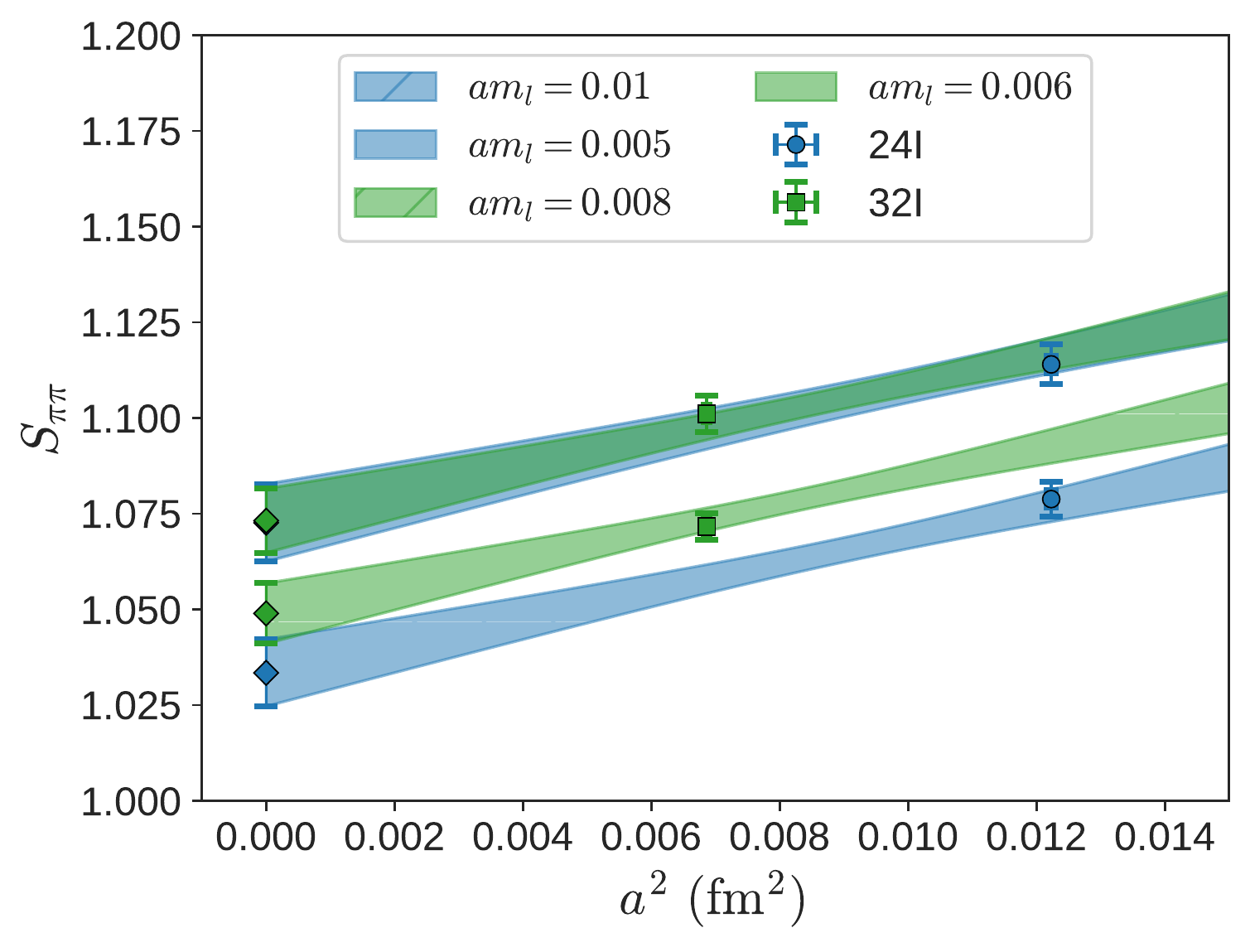}}
  \caption{The chiral (top left), continuum (top/bottom right), and infinite volume (bottom left) extrapolations corresponding to the preferred fit (A3) in Table \ref{tab:chiral_fits}. In all but the bottom right plot the fit has been used to shift the lattice data to the physical point \((m_{\pi} = m_{\pi^{-}}^{\rm PDG}, f_{\pi} = f_{\pi}^{\rm PDG}, a = 0, L = \infty)\), excluding the quantity specified on the horizontal axis. For the continuum extrapolation we also plot the raw data without correcting in \(m_{\pi}\), \(f_{\pi}\), and the lattice volume (bottom right) to illustrate that most of the uncertainty in the top right figure is associated with applying this correction. The vertical dashed line in the upper left plot corresponds to the physical \(m_{\pi^{-}} = 139.5702(4)\) MeV \cite{PhysRevD.98.030001}. In the continuum extrapolation (top right) a slight horizontal shift has been applied successively to each ensemble with the same lattice spacing for clarity.}
  \label{fig:chiral_fits}
\end{figure}

Fits (A1)-(A4) are performed using the ansatz Eq.~\eqref{eqn:chiral_ansatz} and different cuts on the ensembles included in the fit. It is observed that fit (A1) including all data has a poor \(\chi^{2}/\)dof, arising from tension between the data with the lightest pion mass and the data with the heaviest pion mass, but that the \(\chi^{2}/\)dof improves significantly if either of these ensembles is pruned. While some improvement is observed in fit (A2), which prunes the ensemble with the heaviest pion mass and largest value of \(m_{\pi} L\), this is still a relatively poor fit with \(\chi^{2}/\mathrm{dof} = 3.8\). A much more substantial improvement is observed in fit (A3), which instead prunes the lightest ensemble with the smallest value of \(m_{\pi} L\), suggesting that residual finite volume effects drive the observed tension rather than the truncation of the chiral expansion to NLO. Thus, fit (A3) is chosen as the preferred fit determining the central values and statistical errors of \(g_{\nu}^{\pi \pi}\) and the matrix elements \(\mathcal{S}_{\pi \pi}\) and \(M^{0 \nu}\) at the physical point, and is depicted in Figure \ref{fig:chiral_fits}. 

The remaining fits (A4), (B1), and (C1) are variations on fit (A3) used to assign systematic errors. Fit (A4) also prunes the heaviest mass ensemble from fit (A3) and is used to estimate the systematic error associated with truncating the chiral expansion to NLO. This results in a fit with \(N_{\rm dof} = 0\), so no \(\chi^{2}/\)dof can be assigned. Fit (B1) uses the same data as fit (A3) but removes the finite volume correction by fixing \(c_{FV}^{\rm NLO} \equiv 0\). Finally, fit (C1) is performed to all of the data using the more general ansatz Eq.~\eqref{eqn:chiral_ansatz_sys_error}, and includes a second, higher-order finite volume correction term \(\propto (m_{\pi} L)^{-5/2}\). Including this additional term also results in a good fit with \(\chi^{2}/{\rm dof} < 1\), providing further evidence that the tension observed in fit (A1) is driven by residual finite volume effects.

\subsection{Results and Error Budget}
\label{subsec:results}

Based on the arguments presented in the previous section, fit (A3) in Table \ref{tab:chiral_fits} is chosen as the preferred fit to define the central values and statistical uncertainties of the main results of this work. In addition, the following systematic errors are estimated:
\begin{enumerate}
  \item \textit{Sensitivity of the linear fits determining the lattice results for \(M^{0 \nu}\) to the choice of fit range}: In Section \ref{subsec:LD_amplitude}, a systematic uncertainty associated with the variation in the extracted \(0 \nu \beta \beta\) matrix elements as the fit window is varied is computed using a procedure for averaging over possible fits introduced in Refs.~\cite{Rinaldi:2019thf,Beane:2020ycc}. This systematic has been propagated through the chiral extrapolations performed in Section \ref{subsec:chpt_extrap}.  
  \item \textit{Residual finite volume effects}: Since the finite volume term included in the chiral ansatz, Eq.~\eqref{eqn:chiral_ansatz}, is a model rather than a quantity which has been computed self-consistently in \(\chi\)PT, it is possible that the final results still contain residual finite volume errors, and the fit variations studied in Section \ref{subsec:chpt_extrap} suggest that this is indeed the dominant systematic uncertainty. Two procedures for estimating this systematic have been considered: the first, implemented as fit (B1), uses the same data as fit (A3) but drops the finite volume term altogether. The second, implemented as fit (C1), includes all of the available data and adds an additional term \(\propto (m_{\pi} L)^{-5/2}\) modeling the neglected NNLO and higher order finite volume corrections. The larger of the differences in central values between fits (A3) and (B1) or (C1) is used as a conservative estimate of this systematic.
  \item \textit{Truncation of the chiral expansion:} It is possible that higher-order terms in the chiral expansion are needed to accurately describe the lattice simulations over the full range of pion masses reported in this work\footnote{It was found in Ref.~\cite{Boyle:2015exm}, for example, that next-to-next-to-leading-order corrections to the quark mass dependence of \(f_{\pi}\) were needed to obtain a good fit describing a range of lattice data extending from the physical point to the heaviest \(m_{\pi} \approx 430\) MeV 24I ensemble.}. One way to estimate the potential influence of higher order terms is to successively prune the heaviest data from the chiral / continuum / infinite volume extrapolation and examine the resulting variance in the fit parameters. Here the differences in central values between fits (A3) and (A4) are used as an estimate of this systematic.
\end{enumerate}
The main results of this work, extrapolated to the physical pion mass, continuum, and infinite volume limits, and including all sources of statistical and systematic uncertainty discussed in the text, are:
\begin{equation}
\begin{split}
  g_{\nu}^{\pi \pi}(770 \,\, \mathrm{MeV}) &= -10.78(12)_{\rm stat}(4)_{\rm fit}(50)_{\rm FV}(9)_{\chi\mathrm{PT}}, \\
  S_{\pi \pi} &= 1.1054(14)_{\rm stat}(6)_{\rm fit}(61)_{\rm FV}(10)_{\chi\mathrm{PT}}, \\
  M^{0 \nu} &= 0.01880(6)_{\rm stat}(2)_{\rm fit}(10)_{\rm FV}(2)_{\chi\mathrm{PT}} \,\,\,\, \mathrm{GeV}^{2}. \\
\end{split}
\end{equation}

\section{Discussion}
\label{sec:discussion}
The final results, including all sources of error --- \(g_{\nu}^{\pi \pi}(770 \,\, \mathrm{MeV}) = -10.78(12)_{\rm stat}(51)_{\rm sys}\) and \(\mathcal{S}_{\pi \pi} = 1.1054(14)_{\rm stat}(62)_{\rm sys}\) --- are in good agreement with an independent lattice QCD study of the long-distance \(\pi^{-} \rightarrow \pi^{+} e^{-} e^{-}\) amplitude by Tuo, Feng, and Jin \cite{Tuo:2019bue}, which determined \(g_{\nu}^{\pi \pi}(m_{\rho}) = -10.89(28)_{\rm stat}(74)_{\rm sys}\) and \(\mathcal{S}_{\pi \pi} = 1.1045(34)_{\rm stat}(74)_{\rm sys}\). This calculation also used a variant of the domain wall fermion discretization for the quarks, but was performed on a different set of ensembles with near-physical pion masses and coarser \(a \approx 0.2\) fm lattice spacings. In addition, this calculation used a different set of techniques more traditionally associated with lattice QCD+QED calculations to implement the Majorana neutrino in a finite volume, and compared the QED\(_{\rm L}\) \cite{10.1143/PTP.120.413} and infinite volume reconstruction \cite{Feng:2018qpx} techniques for this purpose. Since the calculation was performed at the physical pion mass, \(g_{\nu}^{\pi \pi}(\mu)\) could be extracted directly by inverting 
\begin{equation}
\mathcal{S}_{\pi \pi} = 1 + \frac{m_{\pi}^{2}}{8 \pi^{2} f_{\pi}^{2}} \left( 3 \log \left( \frac{\mu^{2}}{m_{\pi}^{2}} \right) + 6 + \frac{5}{6} g_{\nu}^{\pi \pi}(\mu) \right),
\end{equation}
rather than by performing a chiral fit as in Section \ref{subsec:chpt_extrap} of this work. The same authors also calculated \(g_{\nu}^{\pi \pi}(m_{\rho}) = -11.96(31)\) from the related \(\pi^{-} \pi^{-} \rightarrow e^{-} e^{-}\) decay amplitude in Ref.~\cite{Feng:2018pdq}, which is in \(\approx 4 \sigma\) tension with the determinations from \(\pi^{-} \rightarrow \pi^{+} e^{-} e^{-}\). This latter calculation does not attempt to quantify any sources of systematic error, which, presumably, would help to explain the disagreement. Finally, in Ref.~\cite{Cirigliano:2017tvr} Cirigliano \textit{et al.}~estimate \(g_{\nu}^{\pi \pi}(m_{\rho}) \simeq -7.6\) with an expected uncertainty of 30-50\% by relating this LEC to known LECs describing electromagnetic corrections within \(\chi\)PT \cite{Ananthanarayan:2004qk,Baur:1996ya}, which is also in reasonable agreement with the results presented here.

One advantage of the approach taken in this work is that performing simulations at a range of different pion masses allows for a controlled study of how well NLO \(\chi\)PT describes lattice data. Since connecting first-principles lattice QCD calculations to predictions for the matrix elements of large nuclei used in \(0 \nu \beta \beta\) searches will almost certainly involve an analogous matching to an effective field theory --- allowing for an extrapolation from the few-body systems accessible on the lattice to the many-body systems relevant to experiment --- this study is important to bridge from theory to phenomenology and experiment. Furthermore, lattice calculations of nuclear systems are currently performed at significantly heavier than physical pion masses to ameliorate the signal-to-noise problem, making it crucial to understand how reliably such calculations can be matched to existing effective field theory formalisms.

The chiral fits performed in Section \ref{subsec:chpt_extrap} exhibit a degree of tension between the lattice data --- which spans the range of pion masses \(300 \,\, \mathrm{MeV} \lesssim m_{\pi} \lesssim 430 \,\, \mathrm{MeV}\) and volumes \(4 \lesssim m_{\pi} L \lesssim 6\) --- and the ansatz Eq.~\eqref{eqn:chiral_ansatz}. This ansatz includes the next-to-leading-order continuum \(\chi\)PT amplitude, as well as models of the leading finite volume and discretization artifacts. Dropping the ensemble with the smallest \(m_{\pi} L\) or adding an additional term parametrizing the neglected, higher-order finite volume corrections is observed to dramatically reduce this tension, resulting in good fits with \(\chi^{2}/{\rm dof} < 1\), and suggesting that finite volume artifacts are the dominant systematic uncertainty. In light of this observation, future lattice QCD calculations of long-distance neutrinoless double beta decay amplitudes may benefit from self-consistently addressing the volume dependence within the effective field theory framework used to match to the lattice data, or from performing simulations with sufficiently large volumes that finite volume artifacts are further suppressed. Important formal work in this direction has been performed in Ref.~\cite{Briceno:2019opb}.

\section{Conclusions}
\label{sec:conclusions}
In this work, a novel and general lattice QCD framework for computing four-point functions describing long-distance neutrinoless double beta decay amplitudes mediated by a light Majorana neutrino has been presented and used to compute the amplitude for the unphysical \(\pi^{-} \rightarrow \pi^{+} e^{-} e^{-}\) transition with pions at rest. Data for a series of lattice ensembles with pion masses in the range \(300 \,\, \mathrm{MeV} \lesssim m_{\pi} \lesssim 430 \,\, \mathrm{MeV}\) was fit to the next-to-leading-order chiral perturbation theory amplitude for this decay, and the fit was used to predict the corresponding matrix element in the physical mass, infinite volume, and continuum limit with sub-percent total uncertainty, as well as to determine the \(\chi\)PT low energy constant \(g_{\nu}^{\pi \pi}\) with \(\approx 5\%\) uncertainty. The results were found to be consistent with other estimates of these quantities in the literature. Future work will apply these methods to the phenomenologically important \(n^{0} n^{0} \rightarrow p^{+} p^{+} e^{-} e^{-}\) decay.

\begin{acknowledgements}
The authors wish to thank Z.~Davoudi for pointing out to us in early 2015 the similarity of the \(0 \nu \beta \beta\) matrix element to other matrix elements describing rare kaon decays, and for an earlier lattice formulation of the Euclidean correlation function for this problem. The authors also wish to thank X.~Feng, L.~Jin, E.~Mereghetti, H.~Monge-Camacho, A.~Nicholson, A.~Pochinsky, M.~Savage, P.~Shanahan, M.~Wagman, A.~Walker-Loud and  B.~Wang for useful discussions. The calculations presented in this work were performed using the IBM Blue Gene/Q computers of the RIKEN-BNL Research Center and Brookhaven National Lab and the Stampede2 supercomputer of the Texas Advanced Computing Center (TACC) at the University of Texas at Austin. In addition, computations for this work were carried out in part on facilities of the USQCD Collaboration, which are funded by the Office of Science of the U.S. Department of Energy. WD and DJM are supported in part by the U.S.~Department of Energy, Office of Science, Office of Nuclear Physics under grant Contract Number DE-SC0011090. WD is also supported within the framework of the TMD Topical Collaboration of the U.S.~Department of Energy, Office of Science, Office of Nuclear Physics, and  by the SciDAC4 award DE-SC0018121.

\end{acknowledgements}

\bibliographystyle{apsrev4-2}
\bibliography{LD-pion-0vbb}

\begin{appendices}

  \section{Formalism}
  \label{appendix:formalism}
  In this appendix, a derivation of the formalism describing how to extract the relevant \( 0 \nu \beta \beta \) matrix element from a lattice calculation with a regulated, infinite volume, continuum neutrino propagator, Eq.~\eqref{eqn:regulated_neutrino_prop}, is outlined, beginning from the Euclidean four-point function defined in Eq.~\eqref{eqn:lattice_bilocal_4pt}. While this derivation focuses on the \( \pi^{-} \rightarrow \pi^{+} e^{-} e^{-} \) transition amplitude, the formalism generalizes straightforwardly to other \( 0 \nu \beta \beta \) processes.

The first step in this calculation is to isolate the time dependence of the four-point function arising from the leptonic and hadronic contributions, respectively. The Euclidean time dependence of the neutrino propagator can be extracted by performing the integration over the temporal component of the neutrino's four-momentum, which gives
\begin{equation}
\begin{split}
  \int\limits_{-\infty}^{\infty} &\frac{d q_{4}}{2 \pi} \frac{1}{\vert \vec{q} \vert^{2} + q_{4}^{2}} e^{i q_{4} ( t_{x} - t_{y} )} e^{-q_{4}^{2}/\Lambda^{2}} \\
  &= \frac{1}{4 \vert \vec{q} \vert} e^{\frac{\vert \vec{q} \vert^{2}}{\Lambda^{2}}} \left( e^{-\vert\vec{q}\vert \left\vert t_{x} - t_{y} \right\vert} \mathrm{Erfc} \left[ \frac{\vert\vec{q}\vert}{\Lambda} - \frac{\Lambda}{2} \left\vert t_{x} - t_{y} \right\vert \right] + e^{\vert\vec{q}\vert \left\vert t_{x} - t_{y} \right\vert} \mathrm{Erfc} \left[ \frac{\vert\vec{q}\vert}{\Lambda} + \frac{\Lambda}{2} \left\vert t_{x} - t_{y} \right\vert \right] \right),
\end{split}
\end{equation}
where
\begin{equation}
\mathrm{Erfc}(x) = \frac{2}{\sqrt{\pi}} \int\limits_{x}^{\infty} dt \, e^{-t^{2}}
\end{equation}
is the complementary error function. The time dependence of the hadronic matrix element can be extracted by inserting complete sums over eigenstates of the QCD Hamiltonian,
\begin{equation}
\begin{split}
  \Gamma^{\rm lept.}_{\alpha \beta} \big\langle &\mathscr{O}_{\pi^{+}}(t_{+}) \mathcal{T} \left\{ j_{\alpha}(\vec{x},t_{x}) j_{\beta}(\vec{y},t_{y}) \right\} \mathscr{O}^{\dagger}_{\pi^{-}}(t_{-}) \big\rangle \\
  &= \sum\limits_{l=0}^{\infty} \sum\limits_{m=0}^{\infty} \sum\limits_{n=0}^{\infty} \frac{\Gamma^{\rm lept.}_{\alpha \beta} \langle 0 \vert \mathscr{O}_{\pi^{+}}(t_{+}) \vert l \rangle \langle l \vert j_{\alpha}(\vec{x},t_{x}) \vert m \rangle \langle m \vert j_{\beta}(\vec{y},t_{y}) \vert n \rangle \langle n \vert \mathscr{O}^{\dagger}_{\pi^{-}}(t_{-}) \vert 0 \rangle }{8 E_{l} E_{m} E_{n}} \\
  &\simeq \frac{\vert \mathcal{N}_{\pi} \vert^{2}}{4 m_{\pi}} e^{-m_{\pi} \vert t_{+} - t_{-} \vert} \sum_{m=0}^{\infty} \frac{\Gamma^{\rm lept.}_{\alpha \beta} \langle \pi e e \vert j_{\alpha}(\vec{x}) \vert m \rangle \langle m \vert j_{\beta}(\vec{y}) \vert \pi \rangle }{2 E_{m}} e^{-(E_{m} - m_{\pi}) \vert t_{x} - t_{y} \vert},
\end{split}
\end{equation}
and assuming the current insertion time slices, \( t_{x} \) and \( t_{y} \), are sufficiently separated from the source and sink time slices, \( t_{-} \) and \( t_{+} \), that the sums over \( l \) and \( n \) are saturated by their respective ground states. The dependence on the source-sink separation can be removed by defining a normalized four-point function
\begin{equation}
  \overline{C}_{\pi \rightarrow \pi e e}(t_{x},t_{y}) \equiv \frac{4 m_{\pi}}{\left\vert \mathcal{N}_{\pi} \right\vert^{2}} e^{m_{\pi} \vert t_{+} - t_{-} \vert} C_{\pi \rightarrow \pi e e}(t_{-},t_{x},t_{y},t_{+}),
  \label{eqn:normalized_4pt}
\end{equation}
with \( m_{\pi} \) and \( \mathcal{N}_{\pi} = \big\langle 0 \big\vert \mathscr{O}_{\pi} \big\vert \pi \big\rangle \) determined from the corresponding two-point function. Combining these results and relabling \( m \rightarrow n \) results in
\begin{equation}
  \label{eqn:regulated_4pt}
  \begin{split}
    \overline{C}_{\pi \rightarrow \pi e e}&(t_{x},t_{y}) = \sum_{n=0}^{\infty} \sum_{\vec{x},\vec{y}} \int \frac{d^{3} q}{\left( 2 \pi \right)^{3}} \frac{\Gamma^{\rm lept.}_{\alpha \beta} \langle \pi e e \vert j_{\alpha}(\vec{x}) \vert n \rangle \langle n \vert j_{\beta}(\vec{y}) \vert \pi \rangle }{8 E_{n} \vert\vec{q}\vert} e^{i \vec{q} \cdot (\vec{x} - \vec{y})} e^{-(E_{m} - m_{\pi}) \vert t_{x} - t_{y} \vert} \\
    &\times \left( e^{-\vert\vec{q}\vert \left\vert t_{x} - t_{y} \right\vert} \mathrm{Erfc} \left[ \frac{\vert\vec{q}\vert}{\Lambda} - \frac{\Lambda}{2} \left\vert t_{x} - t_{y} \right\vert \right] + e^{\vert\vec{q}\vert \left\vert t_{x} - t_{y} \right\vert} \mathrm{Erfc} \left[ \frac{\vert\vec{q}\vert}{\Lambda} + \frac{\Lambda}{2} \left\vert t_{x} - t_{y} \right\vert \right] \right).
  \end{split}
\end{equation}
For the vacuum (\(n=0\)) intermediate state, the matrix element
\begin{equation}
  \label{eqn:vacuum_first_order_me}
  \frac{\Gamma^{\rm lept.}_{\alpha \beta} \langle \pi e e \vert j_{\alpha}(0) \vert 0 \rangle \langle 0 \vert j_{\beta}(0) \vert \pi \rangle }{2 E_{0}} = \frac{m_{\pi}^{2} f_{\pi}^{2}}{4 Z_{A}^{2}} \overline{e}_{L} e_{L}^{C}
\end{equation}
decouples from the remaining integration over the neutrino's three-momentum, and the integration can be performed explicitly. The result
\begin{equation}
  \overline{C}^{(0)}_{\pi \rightarrow \pi e e}(t_{x},t_{y}) = \frac{m_{\pi}^{2} f_{\pi}^{2}}{4 Z_{A}^{2}} \sum_{\vec{x},\vec{y}} \frac{1}{4 \pi^{2} \left\vert x - y \right\vert^{2}} \left( 1 - e^{-\frac{\Lambda^{2}}{4} \left\vert x - y \right\vert^{2}} \right) e^{( m_{\pi}-m_{e} ) \vert t_{x} - t_{y} \vert} \overline{e}_{L} e_{L}^{C}
  \label{eqn:regulated_vacuum_4pt}
\end{equation}
can be used to remove the contribution of the vacuum intermediate state to the four-point function Eq.~\eqref{eqn:lattice_bilocal_4pt}, and is manifestly finite in the limit \(x \rightarrow y\) for finite cutoff \( \Lambda \), with \( \overline{C}^{(0)}_{\pi \rightarrow \pi e e} \propto \Lambda^{2} \), and exponentially divergent in the limit \( \vert t_{x} - t_{y} \vert \rightarrow \infty \) for \(m_{e} < m_{\pi}\), as expected.

To derive the analogue of Eq.~\eqref{eqn:0vbb_me}, which is needed to compute the contribution of the vacuum intermediate state to \(M^{0 \nu}\), requires performing the time-ordered integration of Eq.~\eqref{eqn:regulated_4pt} in the operator insertion times. The finite sums over lattice times can be approximated as integrals
\begin{equation}
  \mathbbm{C}_{\pi \rightarrow \pi e e}(T) \approx \frac{1}{2} \int\limits_{0}^{T} d t_{x} \int\limits_{t_{x}}^{T} d t_{y} \, \overline{C}_{\pi \rightarrow \pi e e}(t_{x}, t_{y}),
\end{equation}
and the asymptotic behavior in the limit \(T \rightarrow \infty\) can be isolated using the expansion
\begin{equation}
  \mathrm{Erfc}(x) = \frac{e^{-x^{2}}}{x \sqrt{\pi}} \sum_{n=0}^{\infty} \left( -1 \right)^{n} \frac{\left( 2 n - 1 \right)!!}{\left( 2 x^{2} \right)^{n}}.
\end{equation}
Keeping only the terms proportional to \(T\) results in
\begin{equation}
  \label{eqn:0vbb_me_regulated}
  \begin{split}
    M^{0 \nu} = \sum_{n=0}^{\infty} \sum_{\vec{x},\vec{y}} &\int \frac{d^{3} q}{(2 \pi)^{3}} \frac{\Gamma_{\alpha \beta}^{\rm lept.} \langle \pi e e \vert j_{\alpha}(\vec{x}) \vert n \rangle \langle n \vert j_{\beta}(\vec{y}) \vert \pi \rangle}{4 E_{n} \vert\vec{q}\vert \left[ \vert\vec{q}\vert^{2} - \left( E_{n} - m_{\pi} \right)^{2} \right]} e^{i \vec{q} \cdot (\vec{x}-\vec{y})} \\
    &\times \left( \vert\vec{q}\vert e^{\frac{-\vert\vec{q}\vert^{2} + \left( E_{n} - m_{\pi} \right)^{2}}{\Lambda^{2}}} \mathrm{Erfc} \left[ \frac{E_{n} - m_{\pi}}{\Lambda} \right] - \left( E_{n} - m_{\pi} \right) \mathrm{Erfc} \left[ \frac{\vert\vec{q}\vert}{\Lambda} \right] \right).
  \end{split}
  \end{equation}
From this expression it is easily verified that Eq.~\eqref{eqn:0vbb_me_regulated} reduces to Eq.~\eqref{eqn:0vbb_me} in the limit \(\Lambda \rightarrow \infty\), while also rendering the matrix element finite for finite \( \Lambda \). 

Using this expression, the contribution of any particular long-distance intermediate state to \(M^{0 \nu}\) can be calculated provided one has calculated, or has otherwise modeled using experiment or phenomenology, the corresponding first-order hadronic matrix element as a function of the three-momentum transfer \( \vec{q} \). For the vacuum intermediate state the hadronic matrix element \eqref{eqn:vacuum_first_order_me} and the integration over the momentum again decouple, and the contribution to \(M^{0 \nu}\) may be parametrized as
\begin{equation}
  M^{0 \nu}_{(0)} = \frac{m_{\pi}^{2} f_{\pi}^{2}}{4 Z_{A}^{2}} f \left( m_{\pi}, L, \Lambda \right) \overline{e}_{L} e_{L}^{C},
  \label{eqn:regulated_vacuum_me}
\end{equation}
with
\begin{equation}
  \label{eqn:vacuum_intermediate_state_integral}
  \begin{split}
    f \left( m_{\pi},L,\Lambda \right) = PV &\int\limits_{0}^{\infty} dq \Bigg\{ \sum_{\vec{x},\vec{y}} \frac{1}{4 \pi^{2} \vert \vec{x} - \vec{y} \vert^{2}} \frac{\sin \big( q \vert \vec{x} - \vec{y} \vert \big)}{q^{2} - \left( m_{\pi} - m_{e} \right)^{2}} \\
    &\times \left( q e^{\frac{-q^{2} + \left( m_{\pi} - m_{e} \right)^{2}}{\Lambda^{2}}} \mathrm{Erfc} \left[ \frac{m_{e} - m_{\pi}}{\Lambda} \right] + \left( m_{\pi} - m_{e} \right) \mathrm{Erfc} \left[ \frac{q}{\Lambda} \right] \right) \Bigg\},
  \end{split}
\end{equation}
where \(PV\) denotes the Cauchy principal value\footnote{Formally, the integral is divergent due to the pole at \(q = m_{\pi} - m_{e}\). The integrand has opposite sign depending on the direction from which the pole is approached, however, such that the principal value of the integral is well-defined and finite. Care must be taken in the numerical implementation of Eq.~\eqref{eqn:vacuum_intermediate_state_integral} to address this point.}.

  \section{Two-Point Functions}
  \label{appendix:2pt}
\begin{figure}[!ht]
\centering
\subfloat[24I, \(a m_{l} = 0.01\)]{\includegraphics[width=0.48\linewidth]{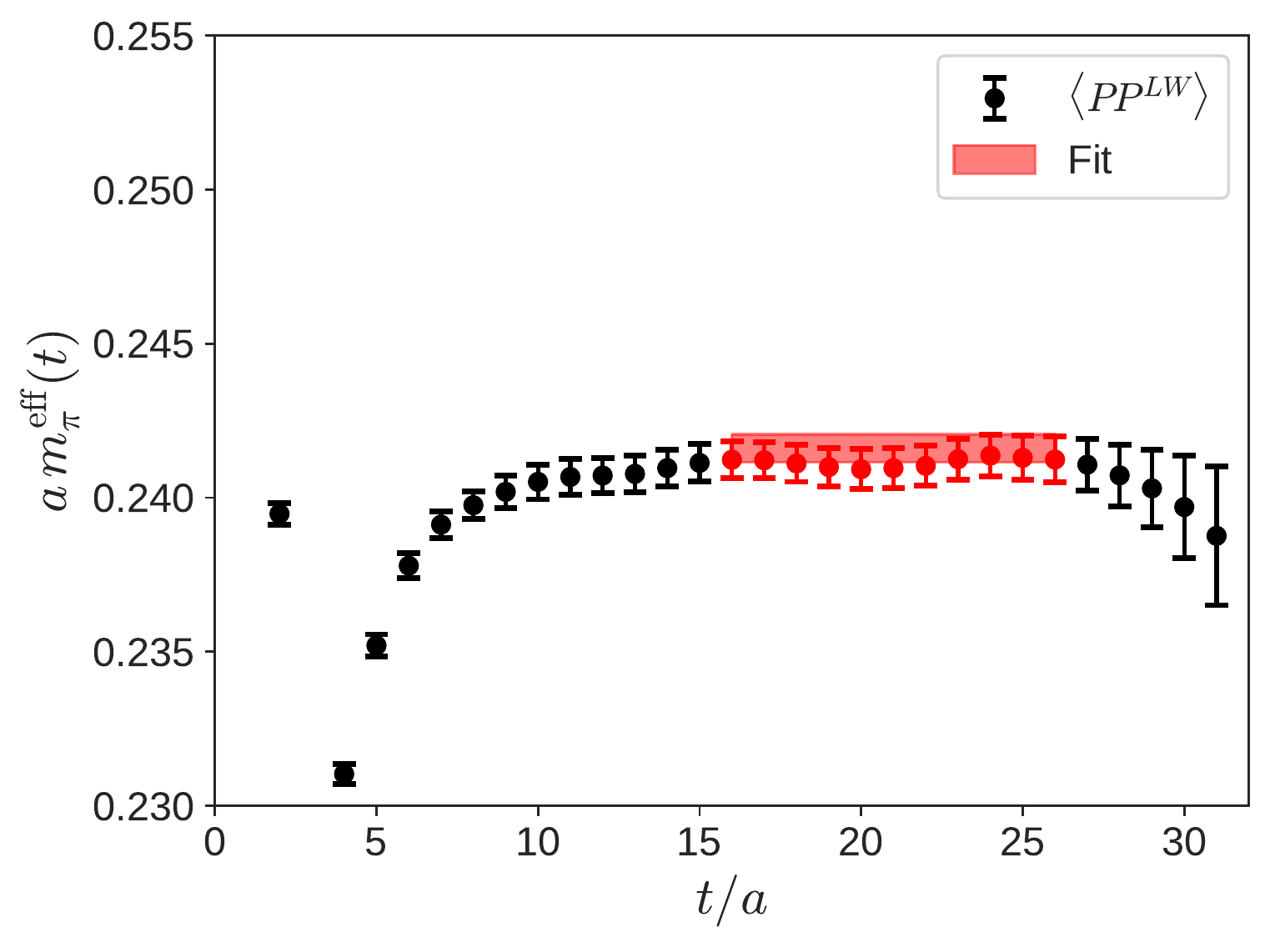}}
\subfloat[24I, \(a m_{l} = 0.005\)]{\includegraphics[width=0.48\linewidth]{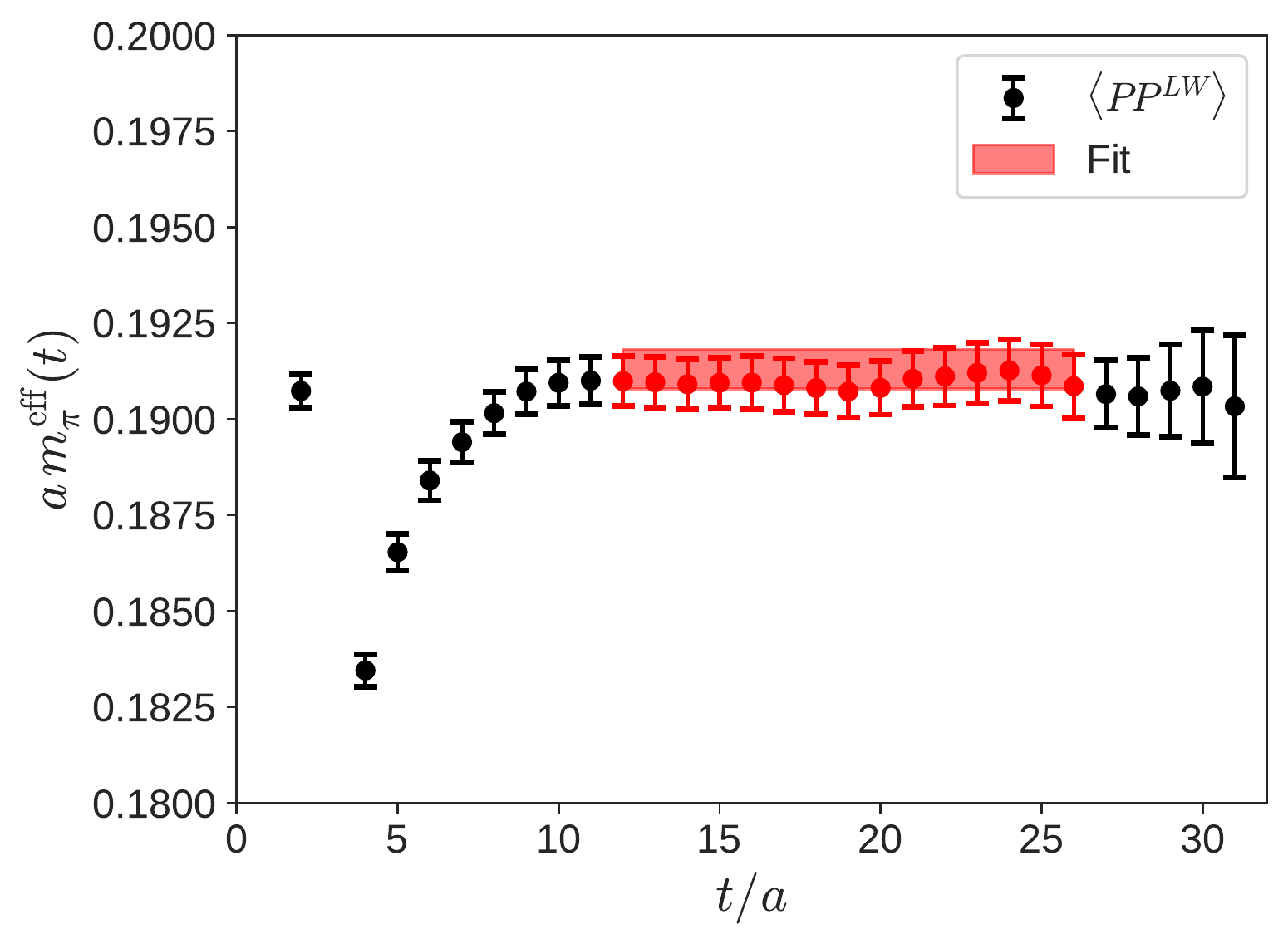}} \\
\subfloat[32I, \(a m_{l} = 0.008\)]{\includegraphics[width=0.48\linewidth]{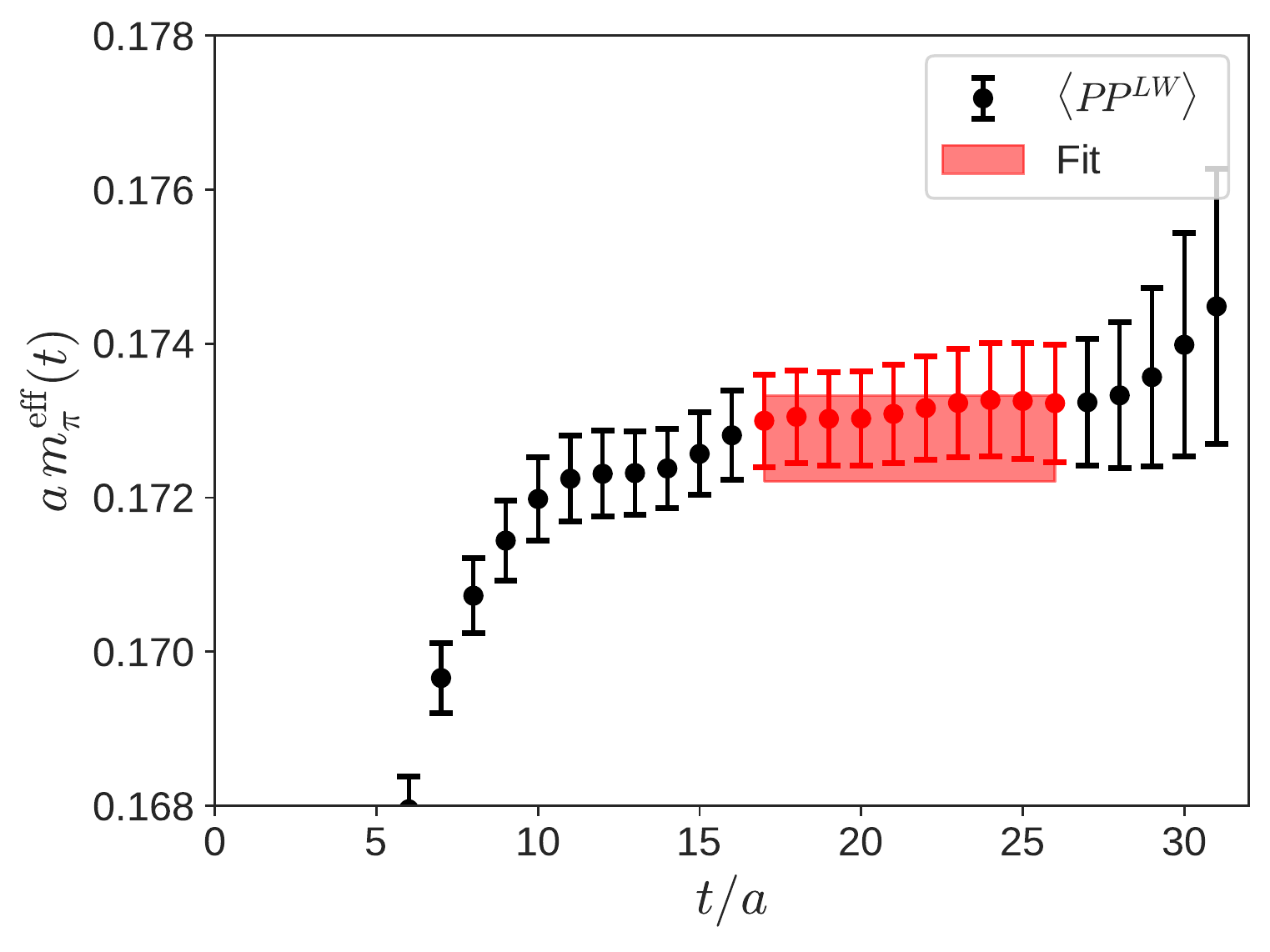}}
\subfloat[32I, \(a m_{l} = 0.006\)]{\includegraphics[width=0.48\linewidth]{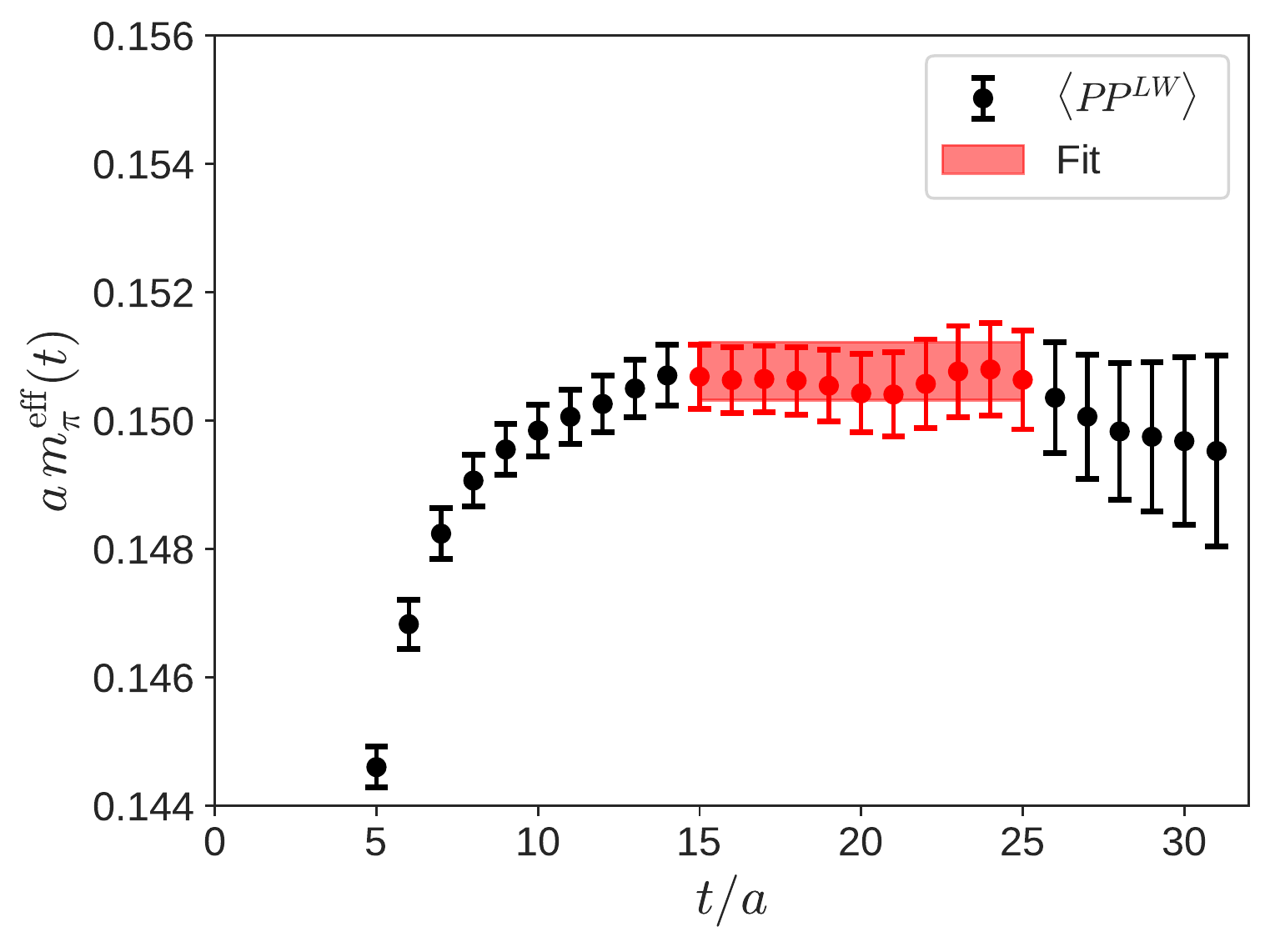}} \\
\subfloat[32I, \(a m_{l} = 0.004\)]{\includegraphics[width=0.48\linewidth]{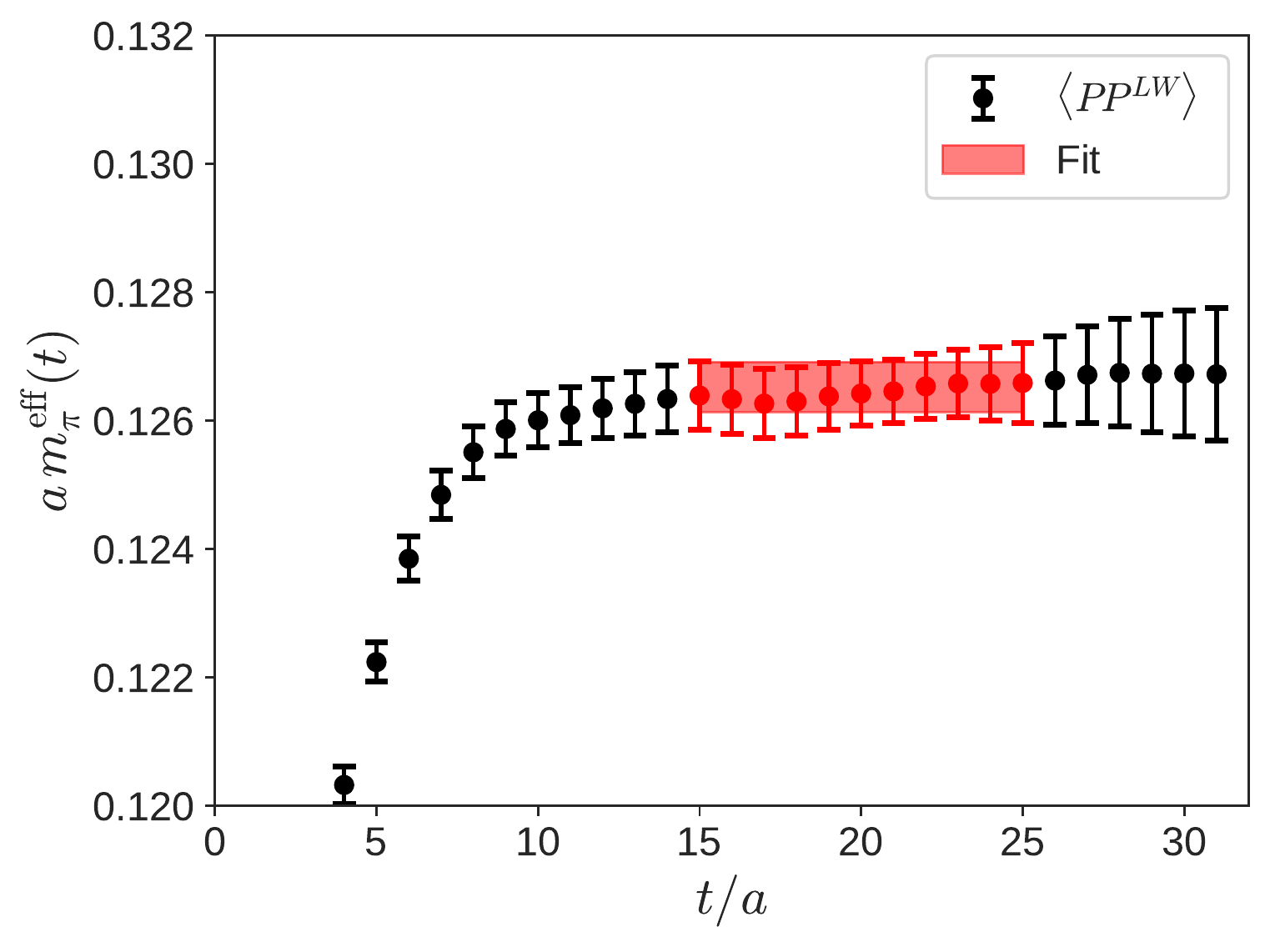}}
\caption{Light quark pseudoscalar-pseudoscalar (PP) two-point functions with wall sources and local sinks.}
\label{fig:2pt_PP_LW}
\end{figure}

\begin{figure}[!ht]
\centering
\subfloat[24I, \(a m_{l} = 0.01\)]{\includegraphics[width=0.48\linewidth]{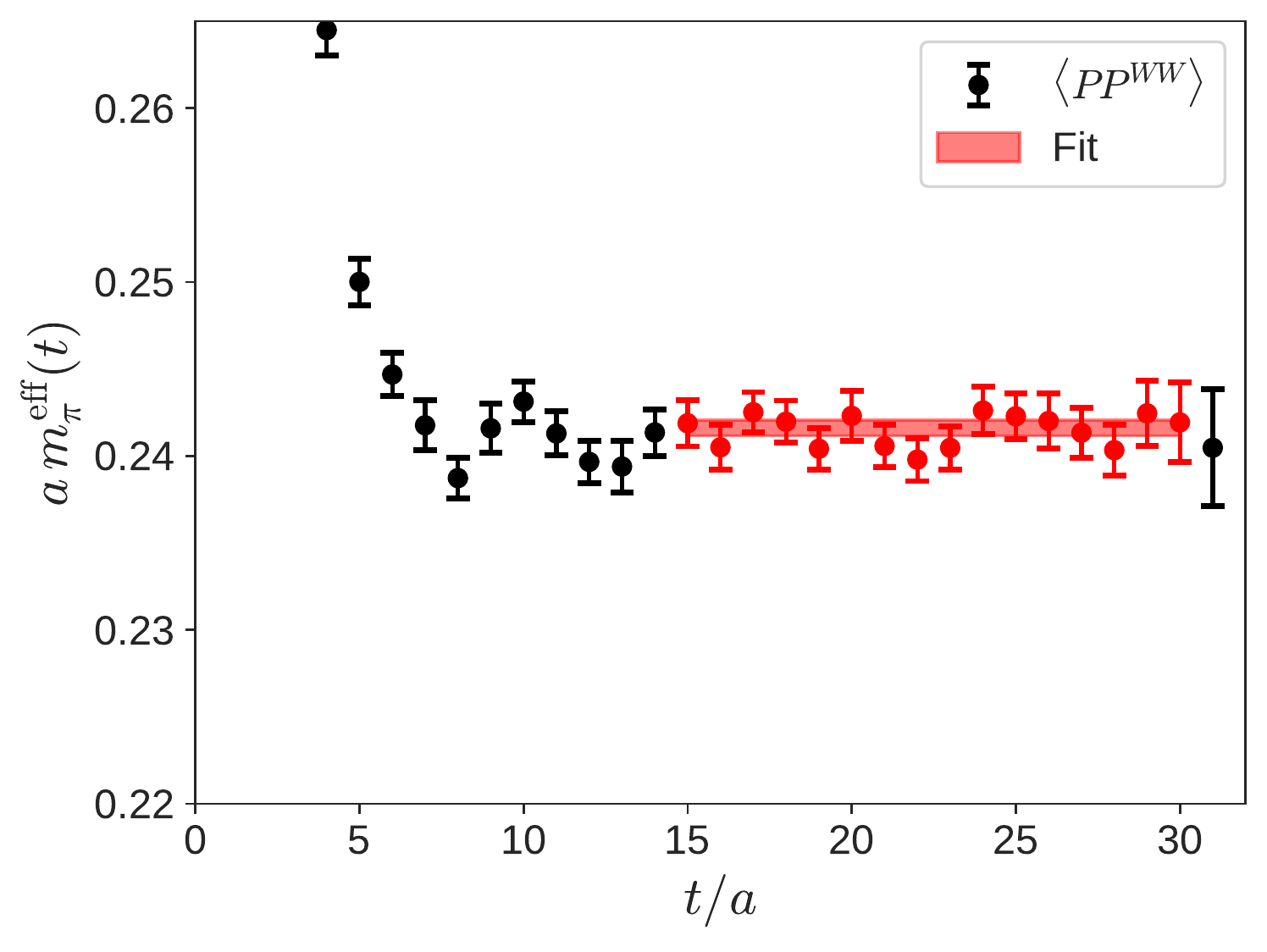}}
\subfloat[24I, \(a m_{l} = 0.005\)]{\includegraphics[width=0.48\linewidth]{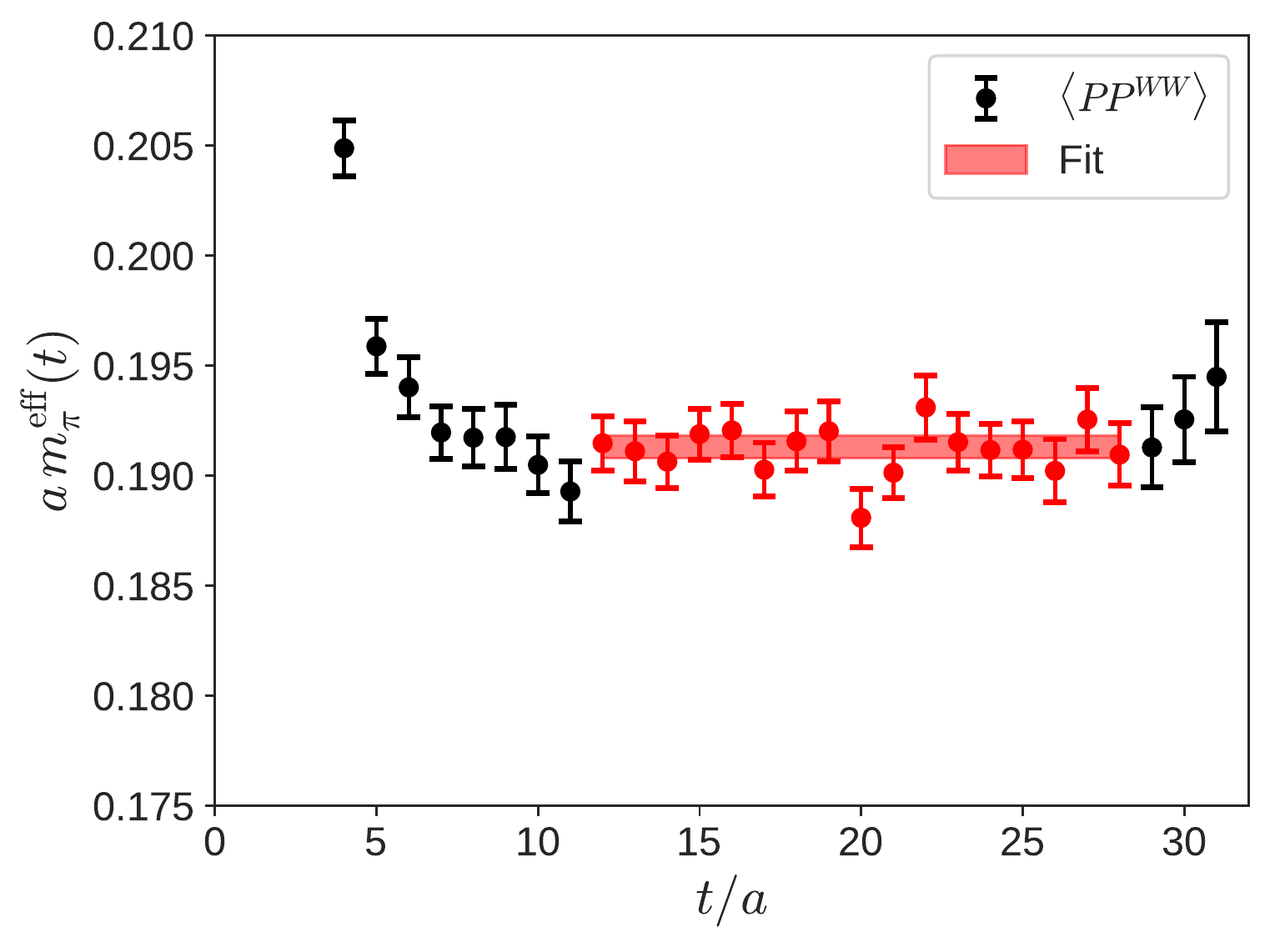}} \\
\subfloat[32I, \(a m_{l} = 0.008\)]{\includegraphics[width=0.48\linewidth]{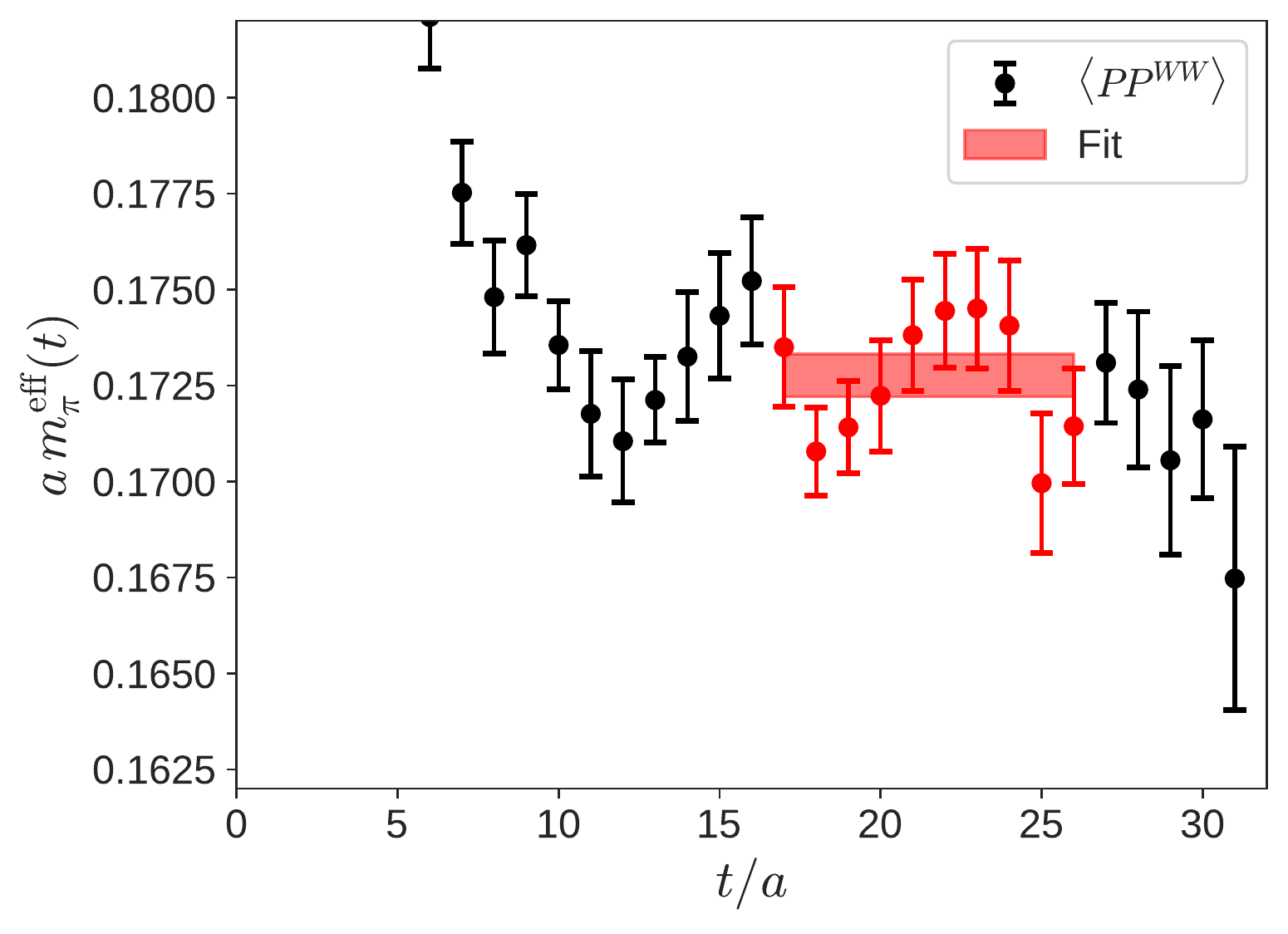}}
\subfloat[32I, \(a m_{l} = 0.006\)]{\includegraphics[width=0.48\linewidth]{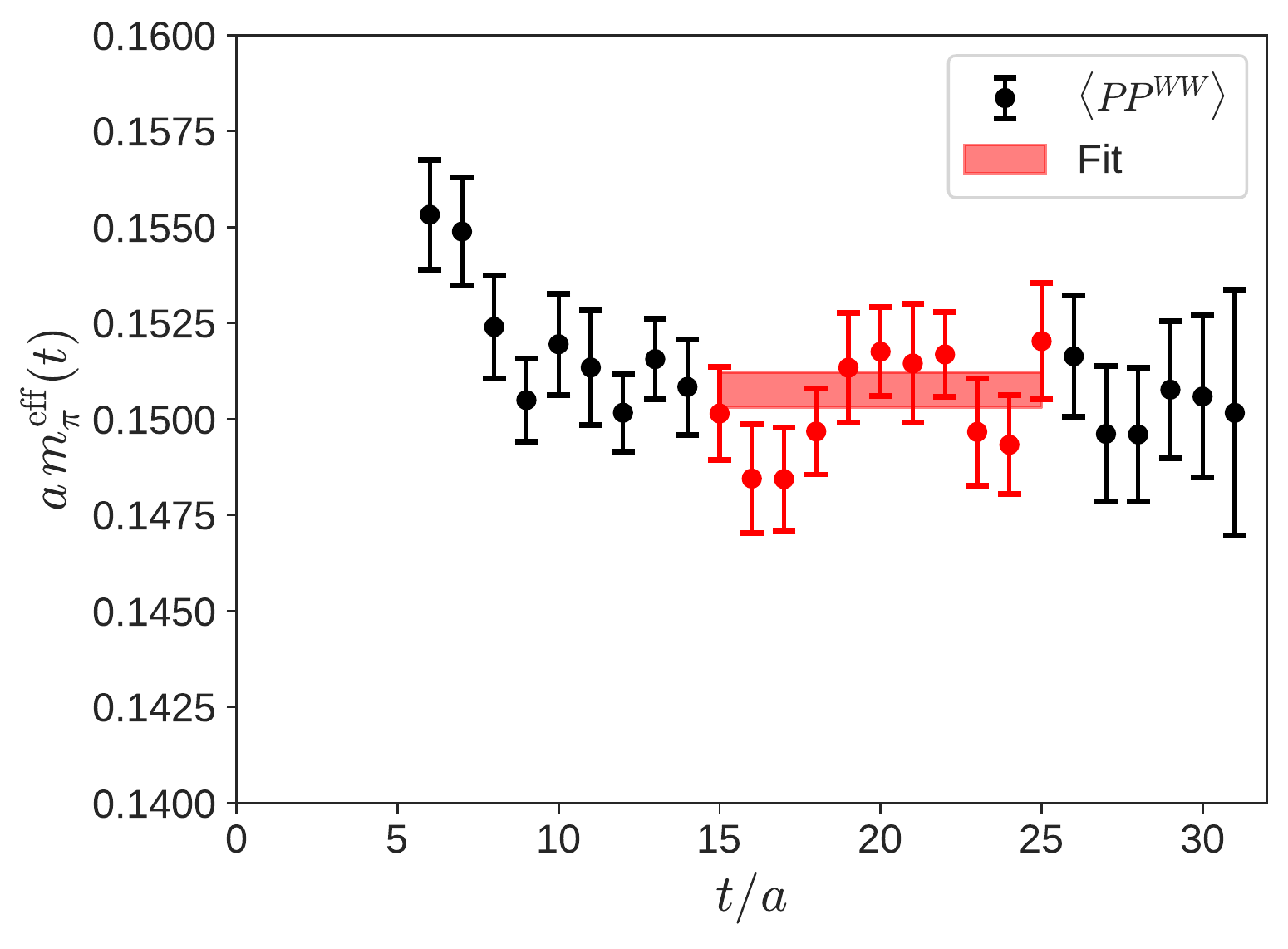}} \\
\subfloat[32I, \(a m_{l} = 0.004\)]{\includegraphics[width=0.48\linewidth]{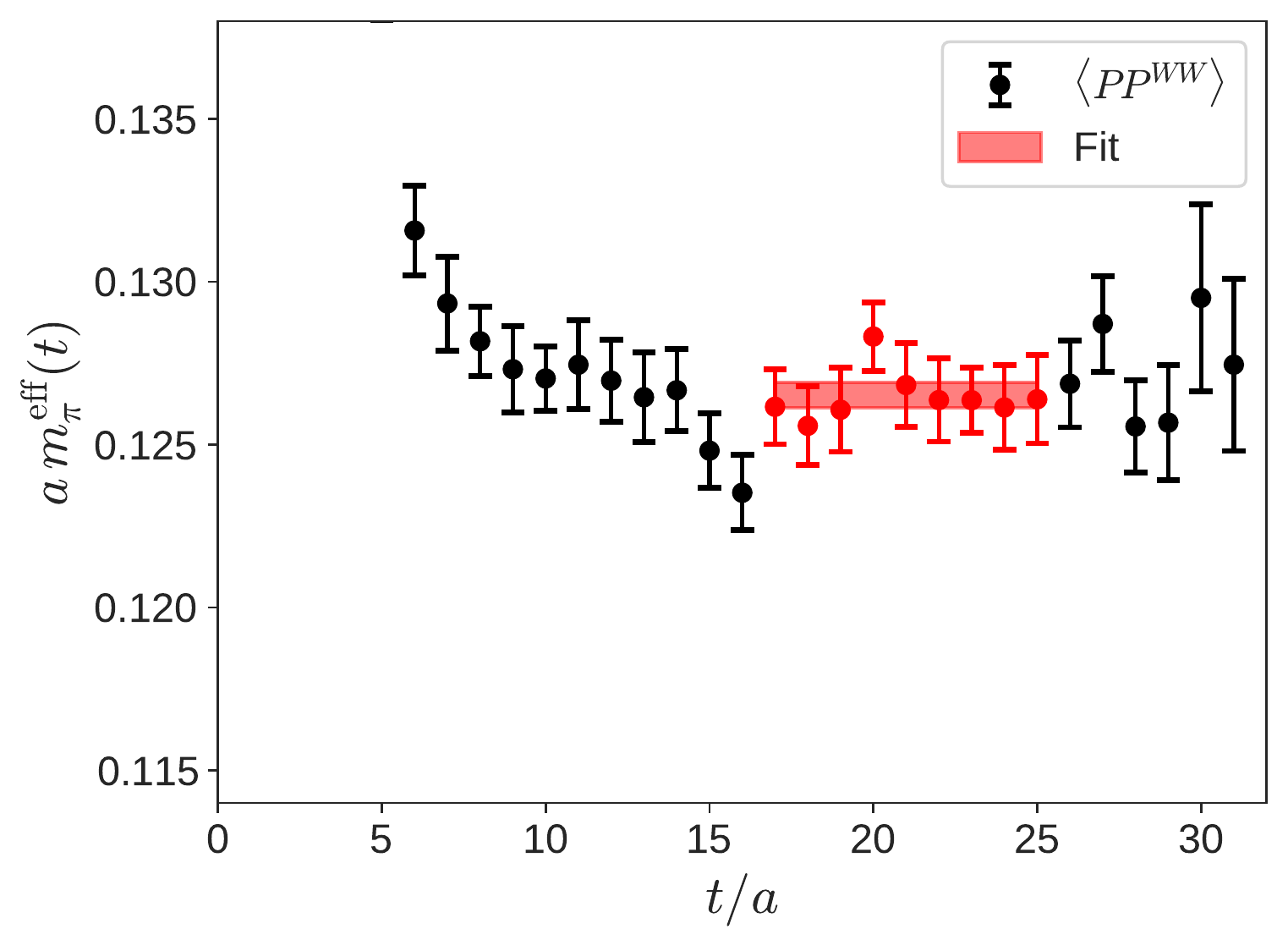}}
\caption{Light quark pseudoscalar-pseudoscalar (PP) two-point functions with wall sources and wall sinks (WW).}
\label{fig:2pt_PP_WW}
\end{figure}

\begin{figure}[!ht]
\centering
\subfloat[24I, \(a m_{l} = 0.01\)]{\includegraphics[width=0.48\linewidth]{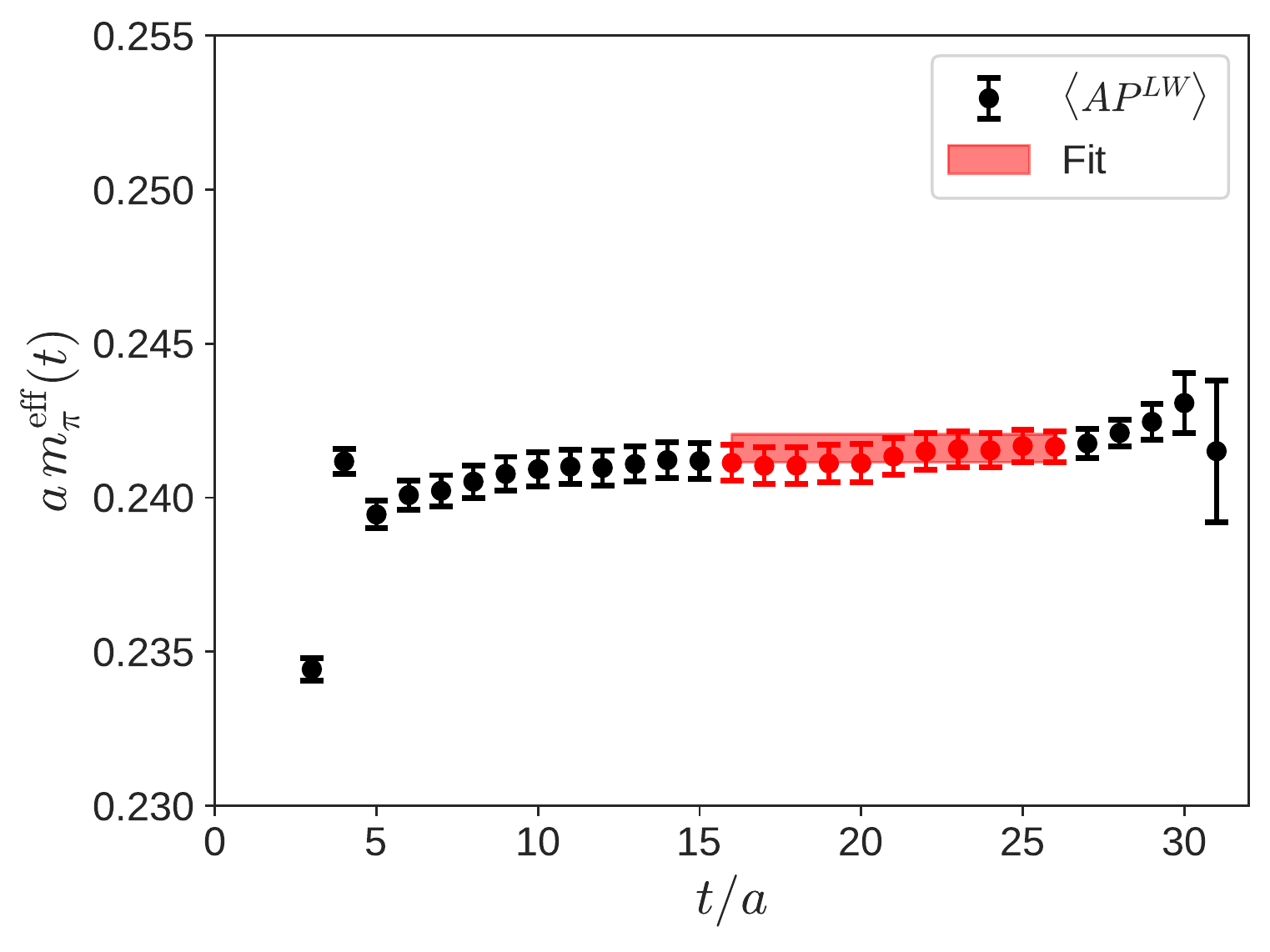}}
\subfloat[24I, \(a m_{l} = 0.005\)]{\includegraphics[width=0.48\linewidth]{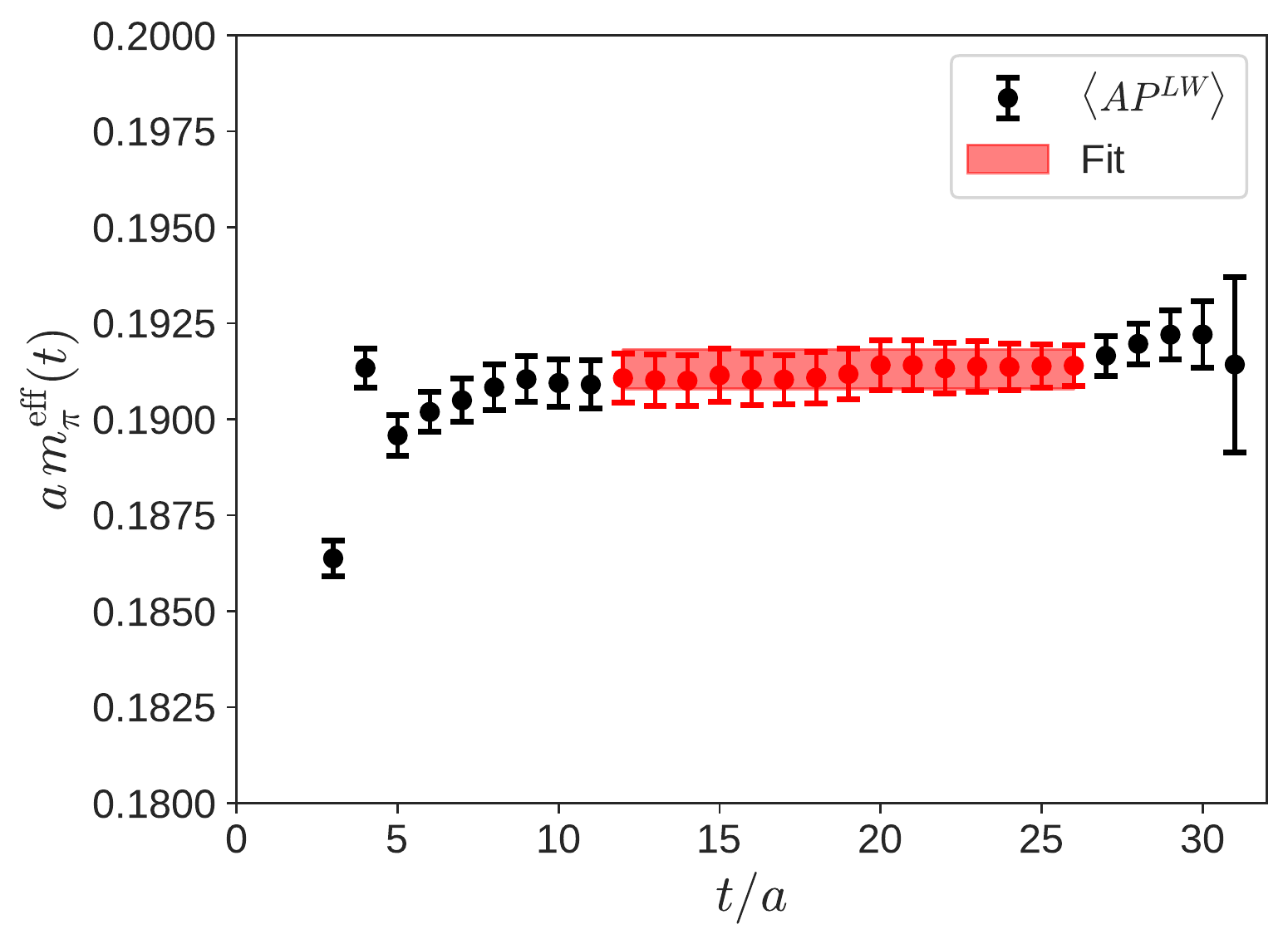}} \\
\subfloat[32I, \(a m_{l} = 0.008\)]{\includegraphics[width=0.48\linewidth]{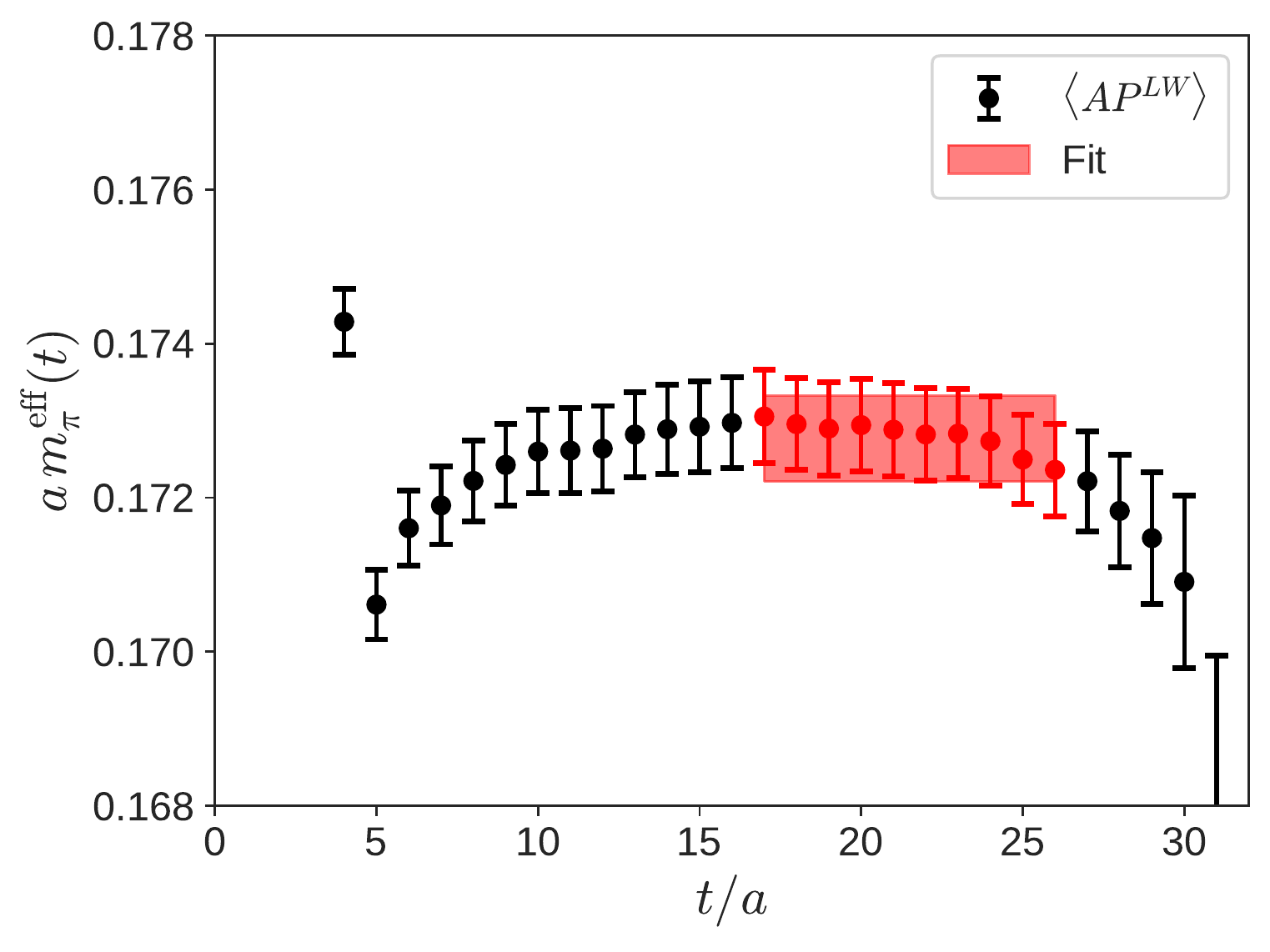}}
\subfloat[32I, \(a m_{l} = 0.006\)]{\includegraphics[width=0.48\linewidth]{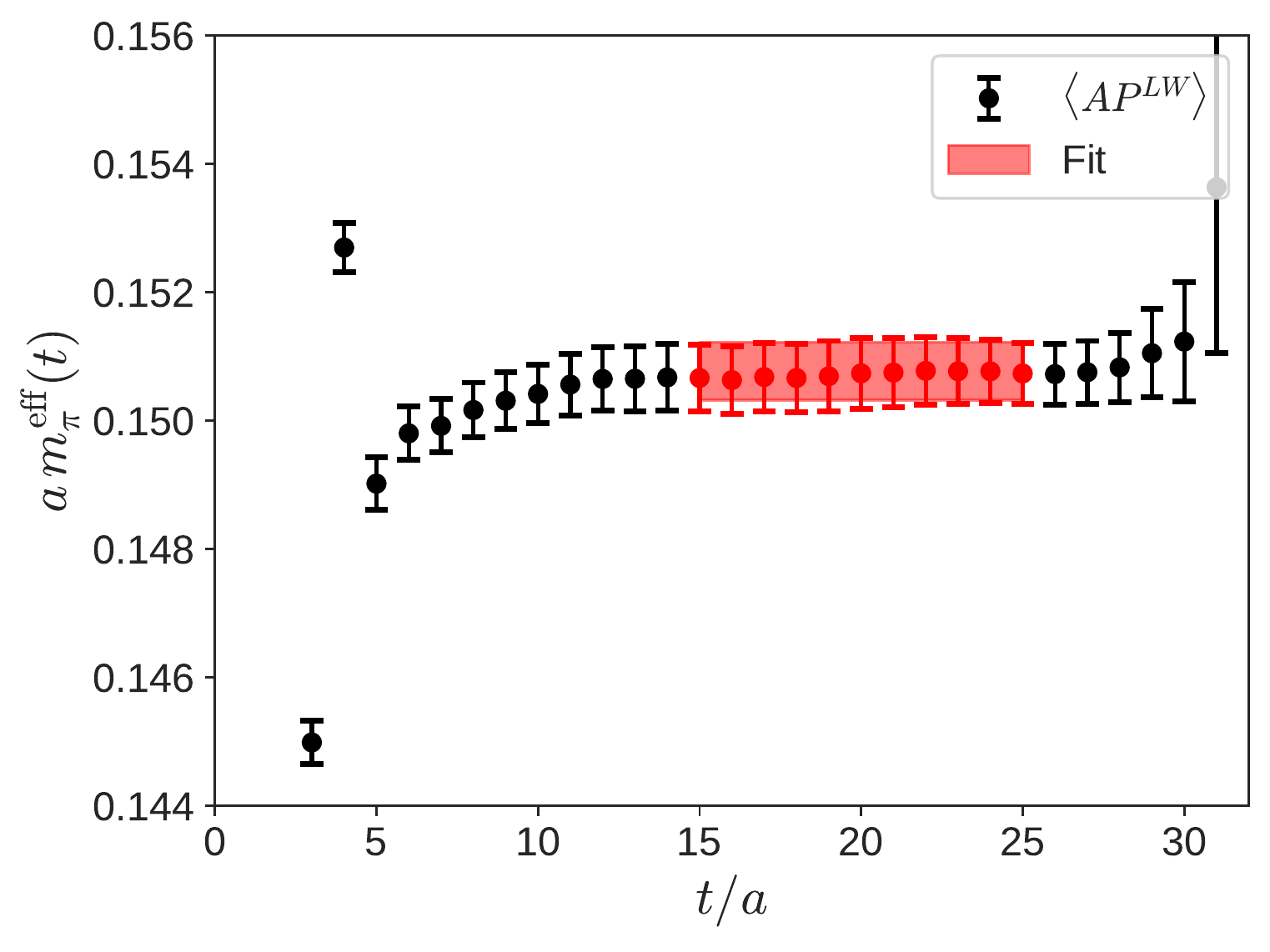}} \\
\subfloat[32I, \(a m_{l} = 0.004\)]{\includegraphics[width=0.48\linewidth]{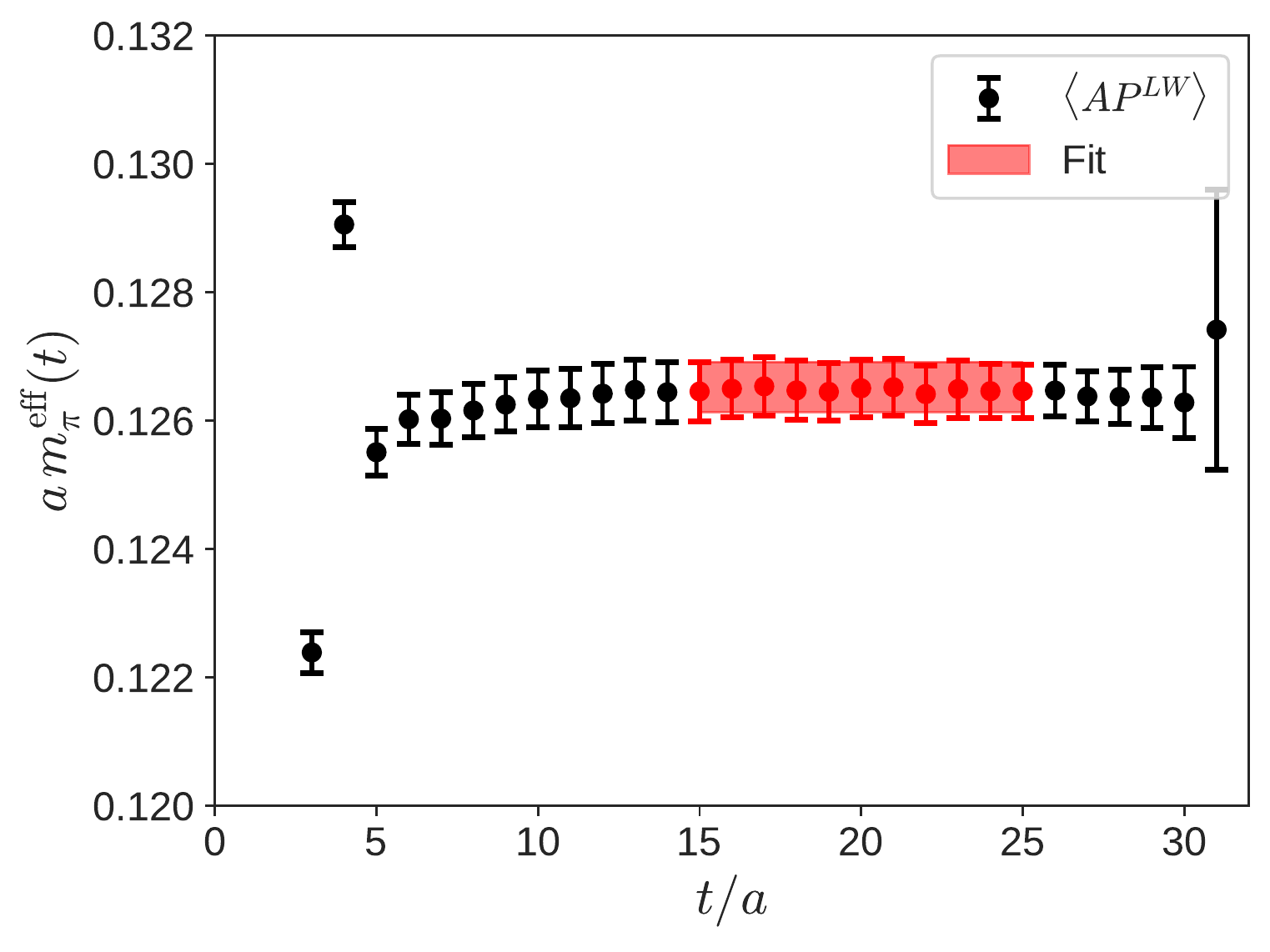}}
\caption{Light quark axial-pseudoscalar (AP) two-point functions with wall sources and local sinks.}
\label{fig:2pt_AP_LW}
\end{figure}

\begin{figure}[!ht]
\centering
\subfloat[24I, \(a m_{l} = 0.01\)]{\includegraphics[width=0.48\linewidth]{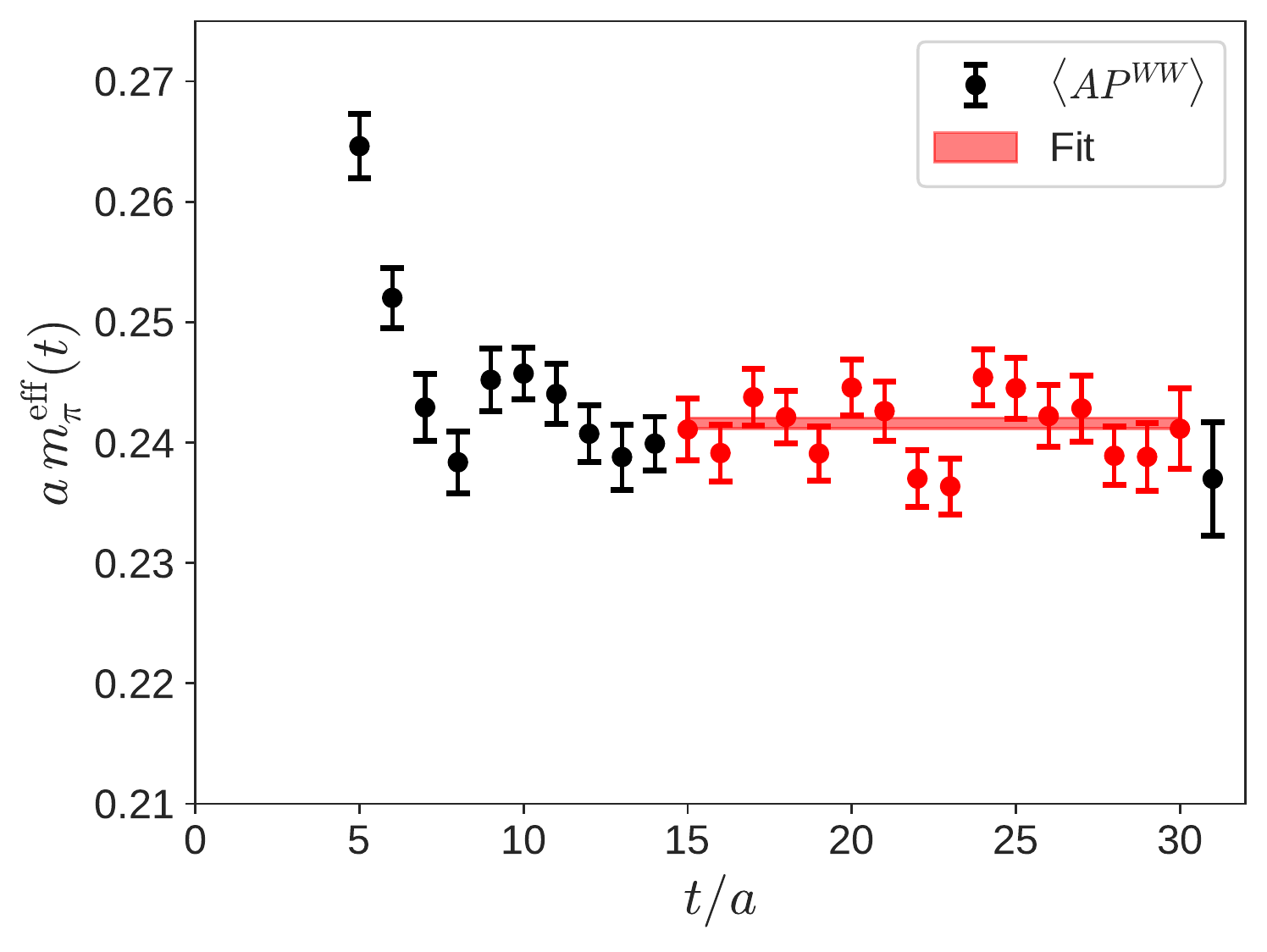}}
\subfloat[24I, \(a m_{l} = 0.005\)]{\includegraphics[width=0.48\linewidth]{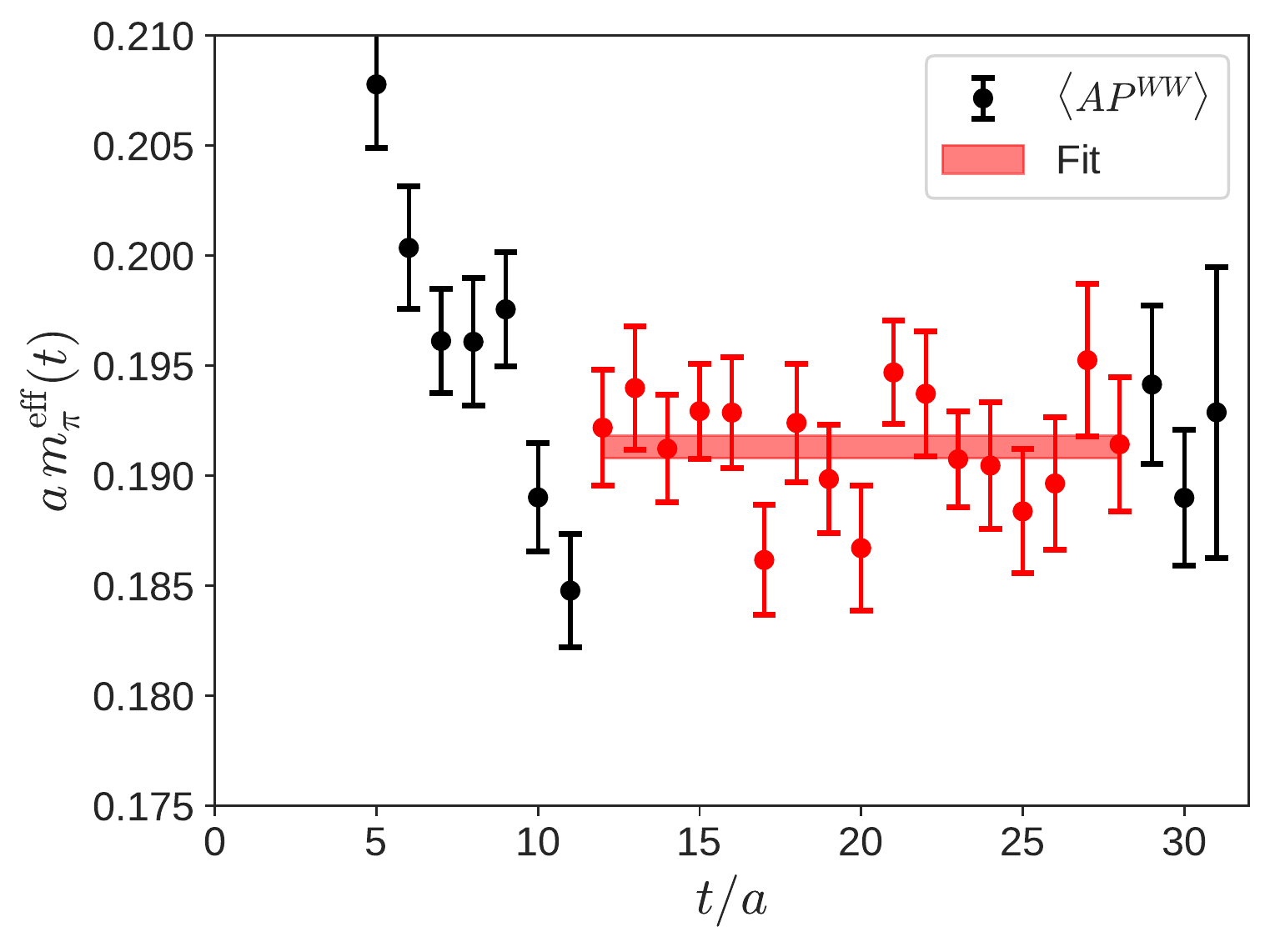}} \\
\subfloat[32I, \(a m_{l} = 0.008\)]{\includegraphics[width=0.48\linewidth]{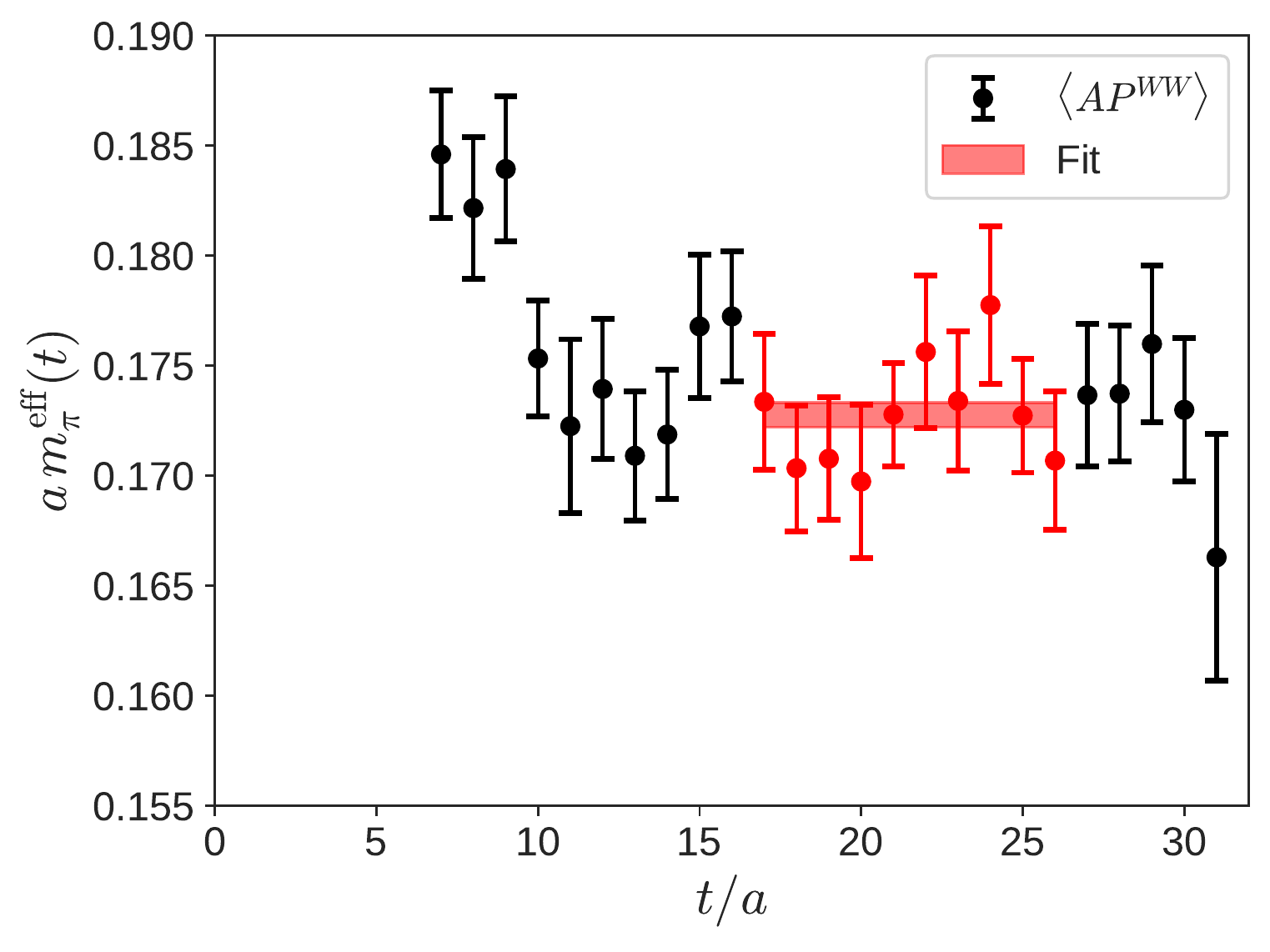}}
\subfloat[32I, \(a m_{l} = 0.006\)]{\includegraphics[width=0.48\linewidth]{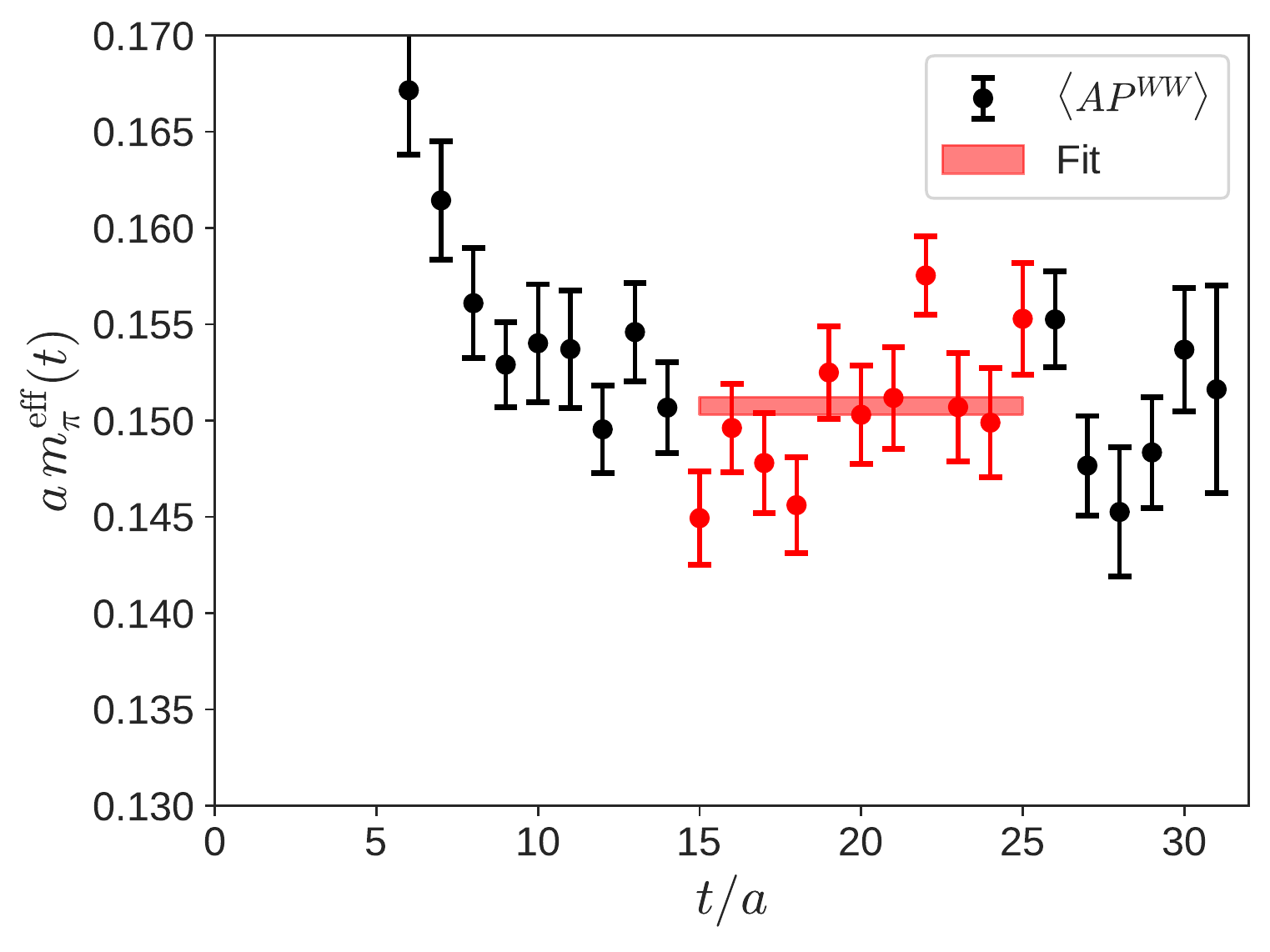}} \\
\subfloat[32I, \(a m_{l} = 0.004\)]{\includegraphics[width=0.48\linewidth]{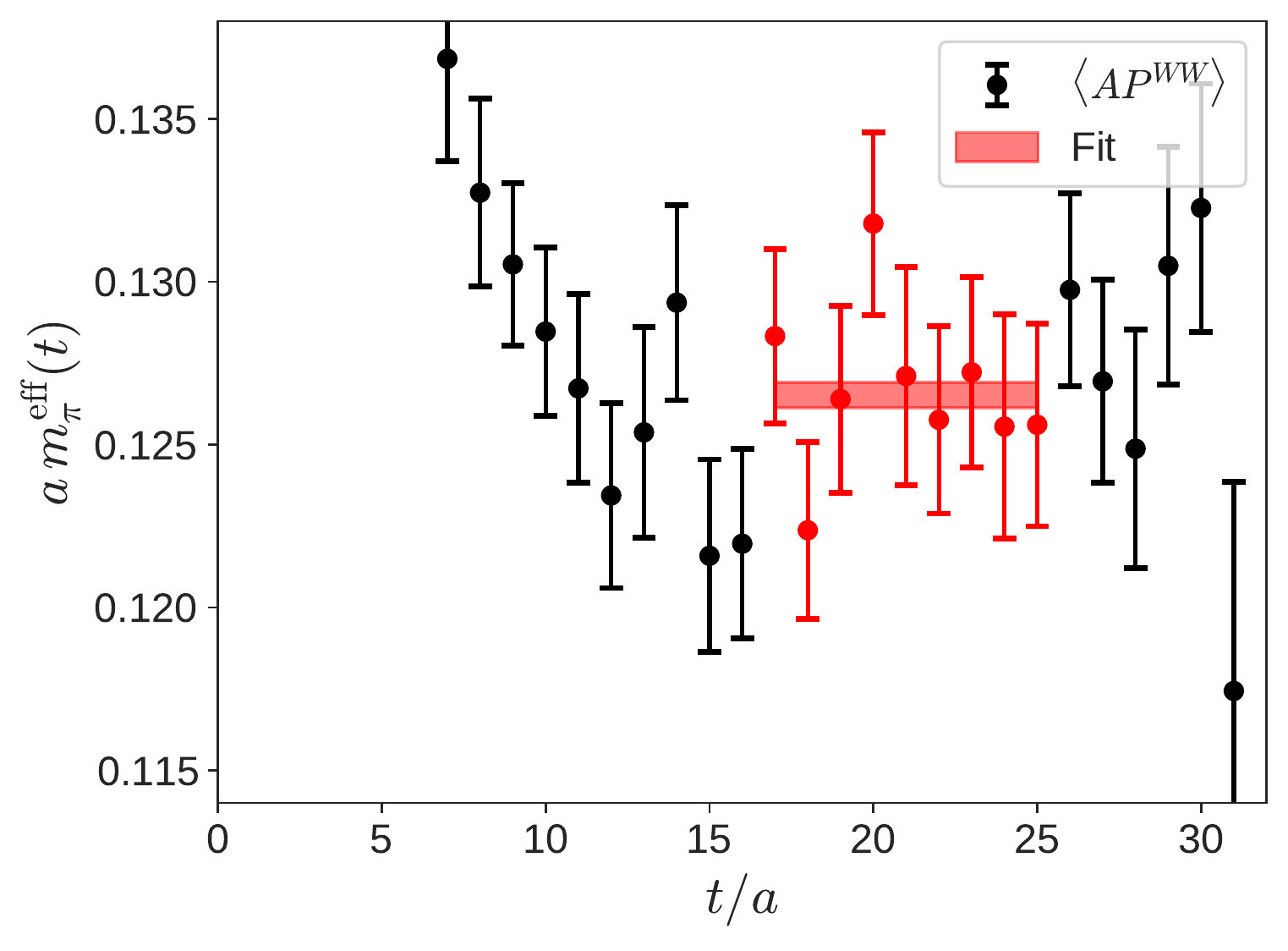}}
\caption{Light quark axial-pseudoscalar (AP) two-point functions with wall sources and wall sinks (WW).}
\label{fig:2pt_AP_WW}
\end{figure}

\begin{figure}[!ht]
\centering
\subfloat[24I, \(a m_{l} = 0.01\)]{\includegraphics[width=0.48\linewidth]{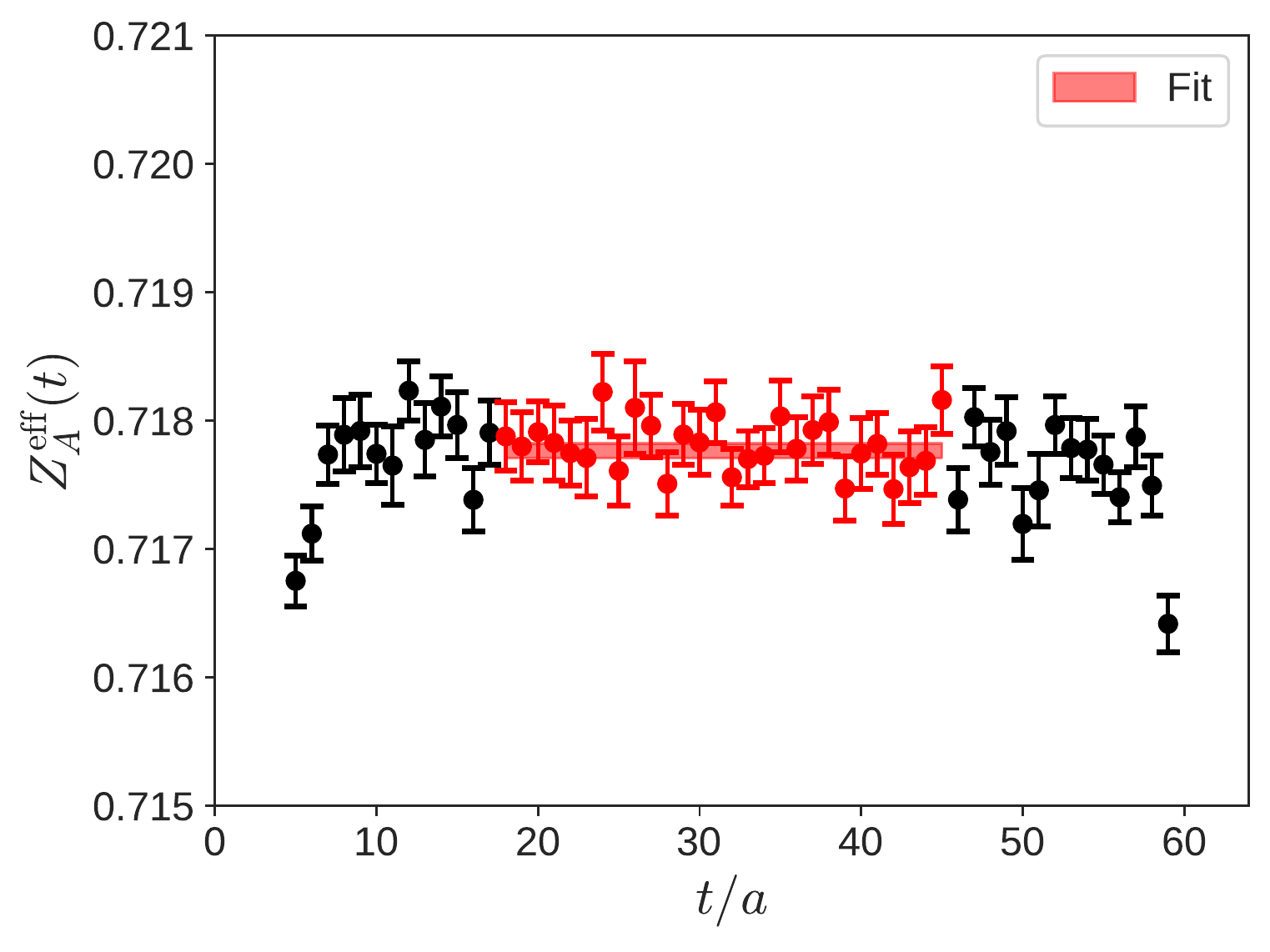}}
\subfloat[24I, \(a m_{l} = 0.005\)]{\includegraphics[width=0.48\linewidth]{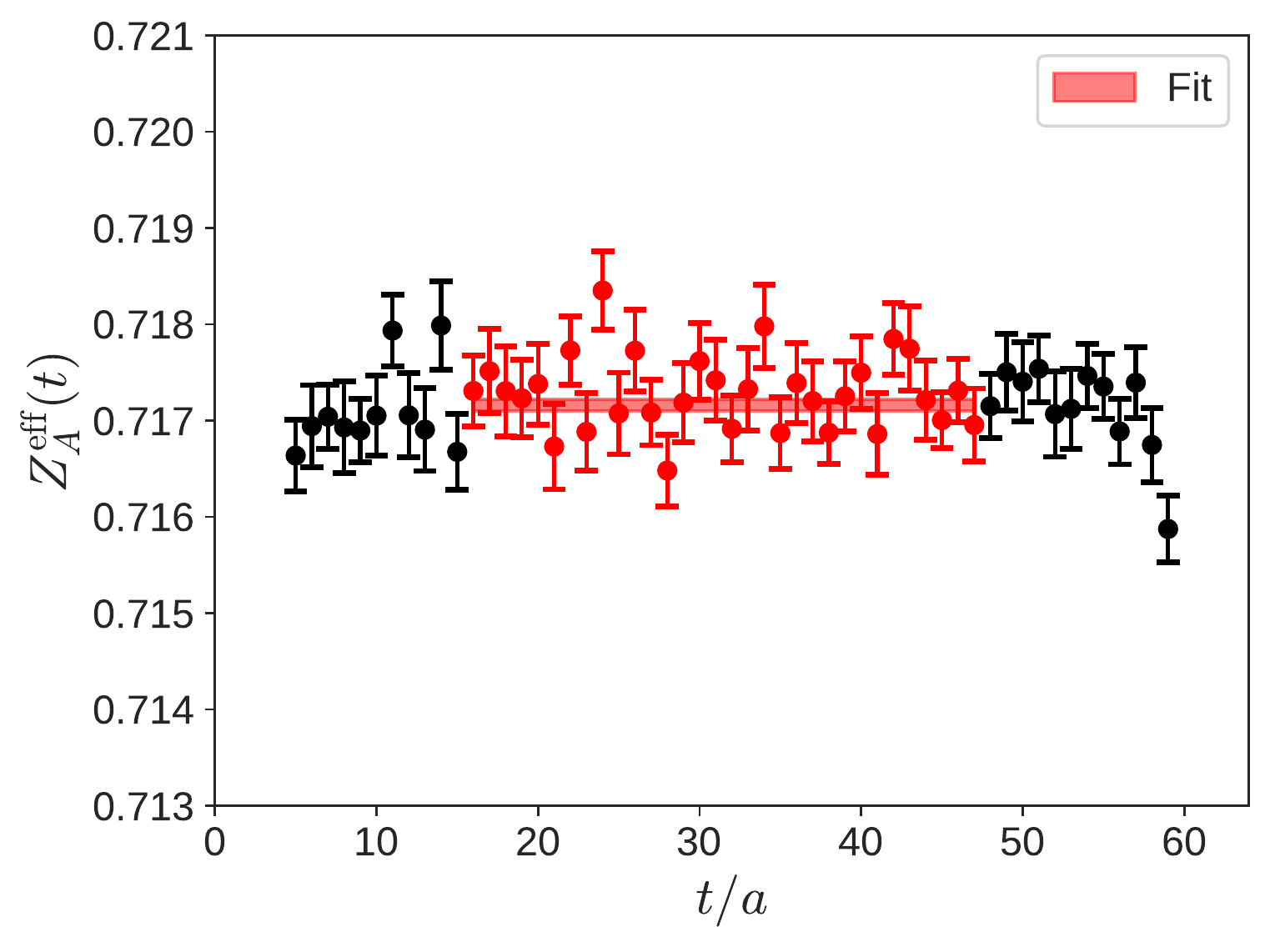}} \\
\subfloat[32I, \(a m_{l} = 0.008\)]{\includegraphics[width=0.48\linewidth]{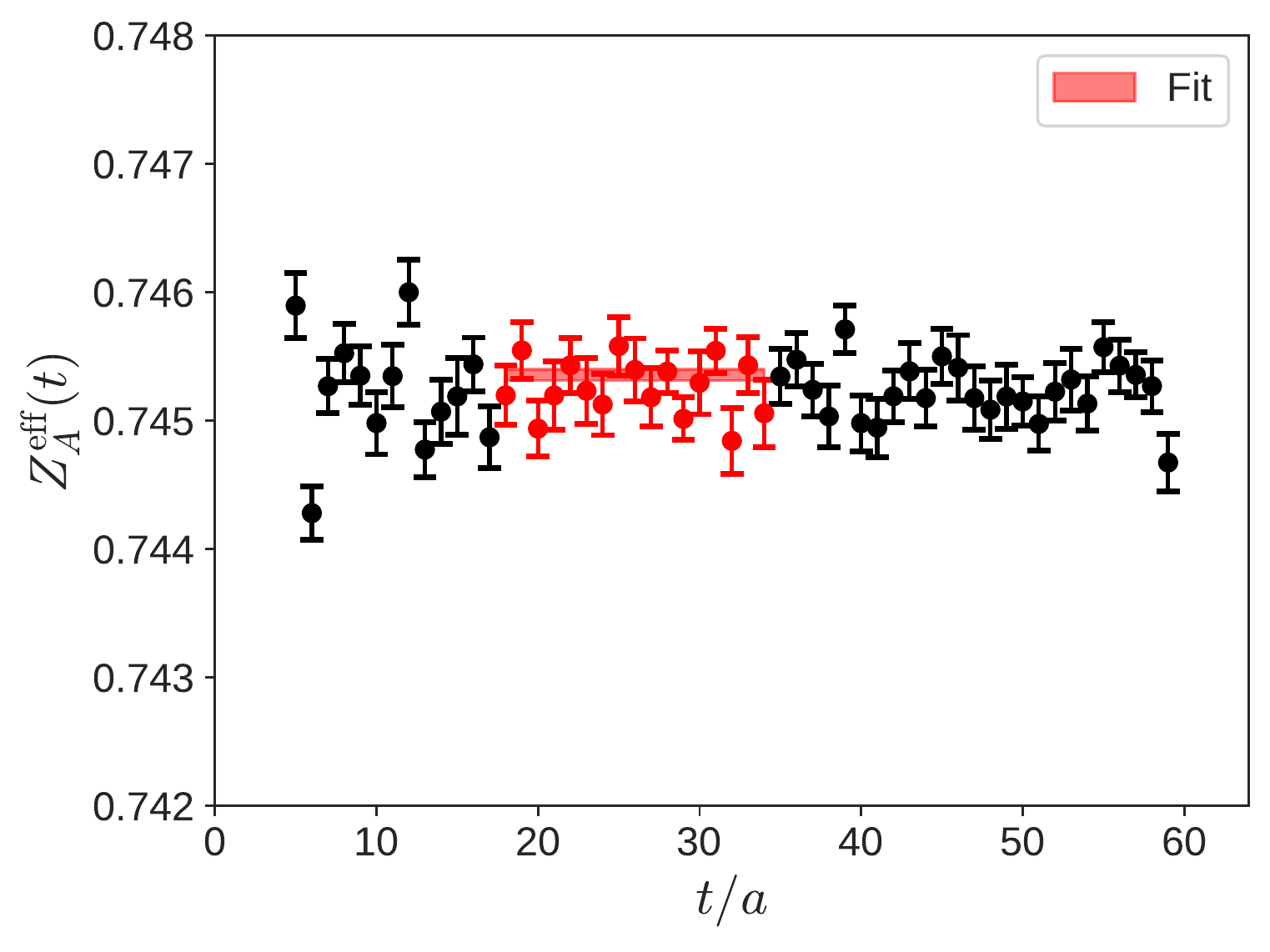}}
\subfloat[32I, \(a m_{l} = 0.006\)]{\includegraphics[width=0.48\linewidth]{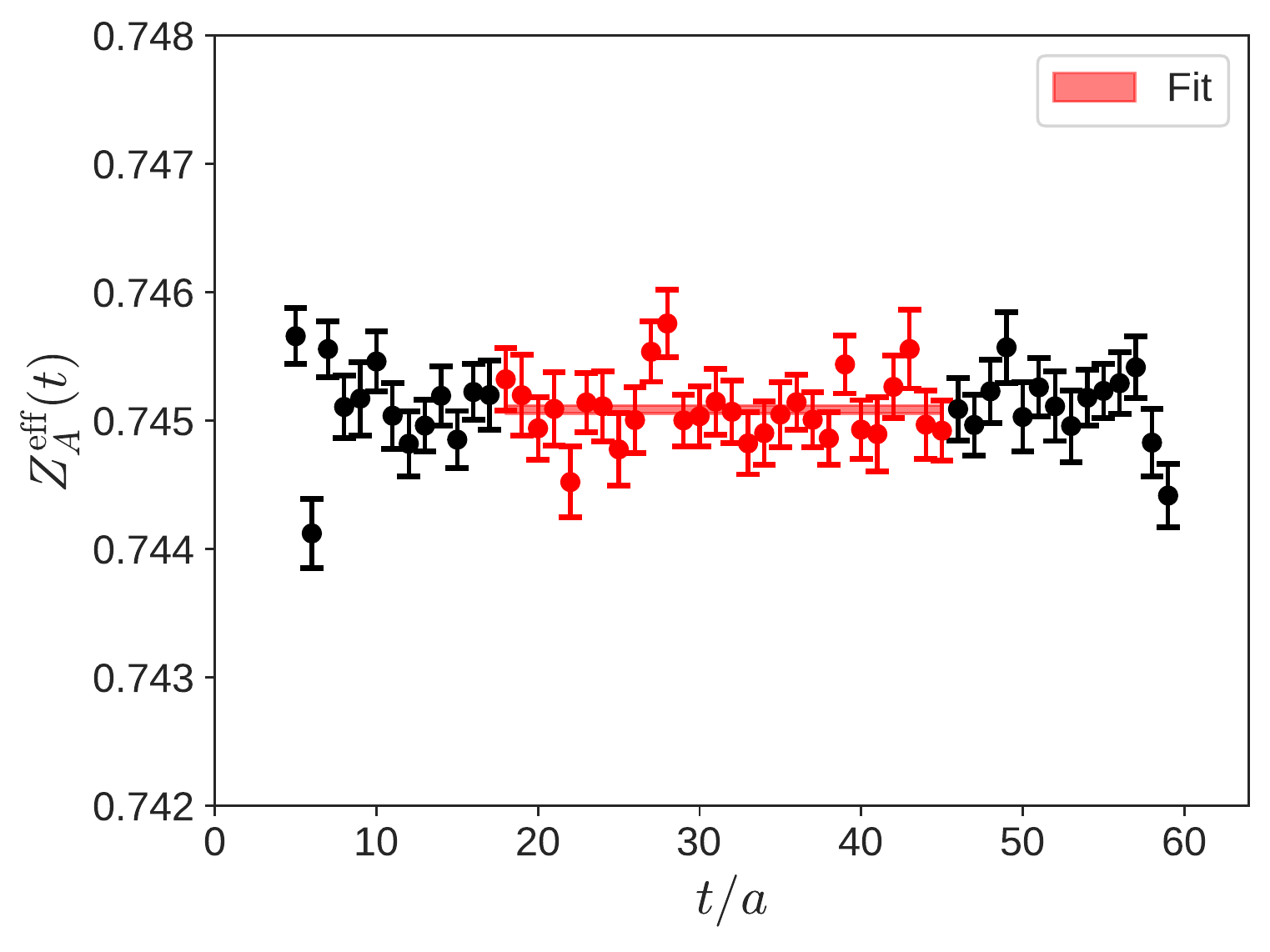}} \\
\subfloat[32I, \(a m_{l} = 0.004\)]{\includegraphics[width=0.48\linewidth]{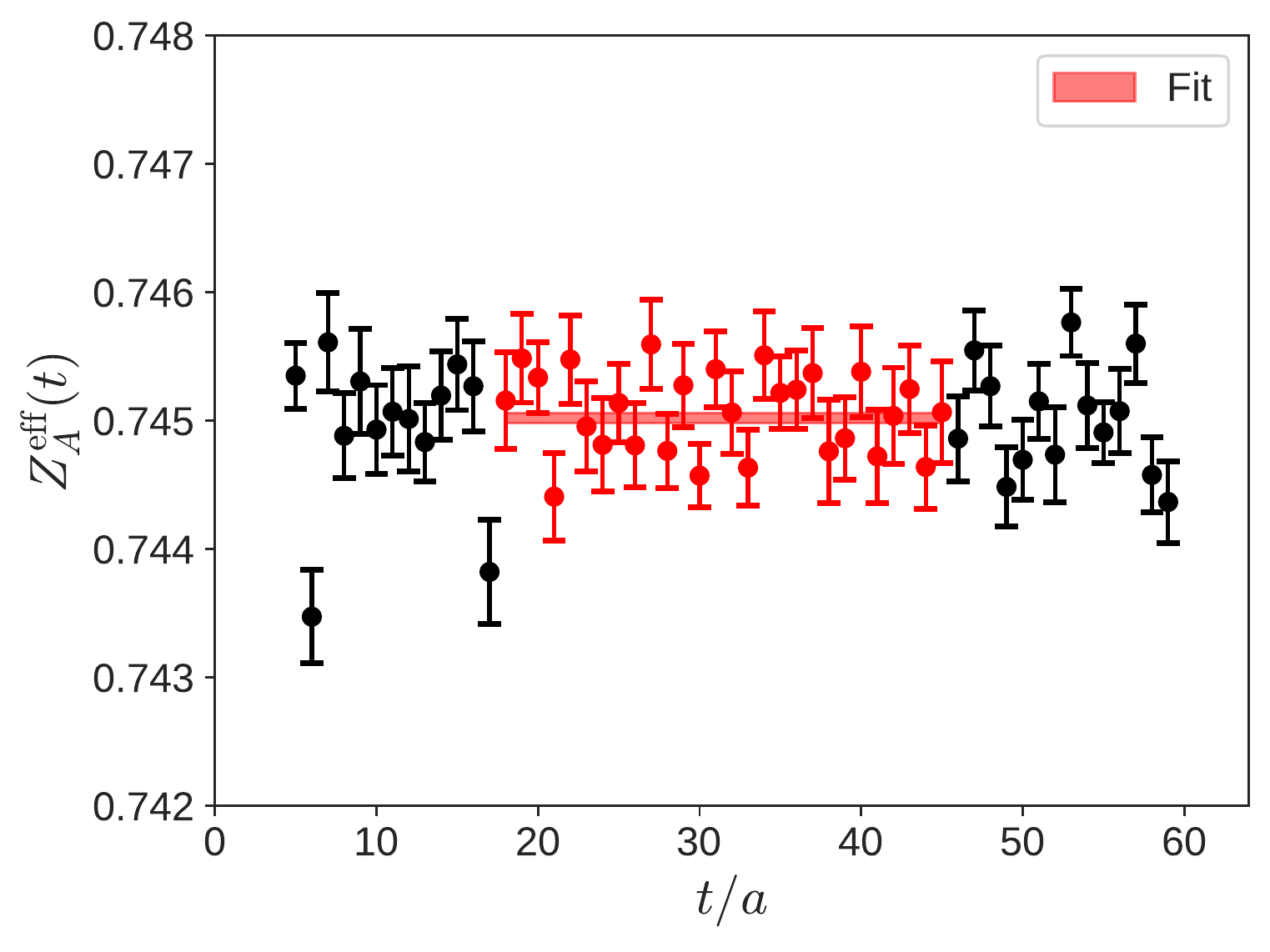}}
\caption{\(Z_{A}\) ratio (Eq.~\eqref{eqn:za_ratio}) two-point functions.}
\label{fig:2pt_za}
\end{figure}

  This section presents Figures \ref{fig:2pt_PP_LW}-\ref{fig:2pt_za}, summarizing the fits to two-point functions performed in Section \ref{subsec:spectrum}. Figures \ref{fig:2pt_PP_LW}-\ref{fig:2pt_AP_WW} show the effective pion masses
\begin{equation}
  a m_{\pi}^{\rm eff} = \cosh^{-1} \left[ \frac{C_{\pi}(t-1) + C_{\pi}(t+1)}{2 C_{\pi}(t)} \right],
\end{equation}
where \(C_{\pi}(t)\) is the pseudoscalar-pseudoscalar two-point function with a local sink (Figure \ref{fig:2pt_PP_LW}), the pseudoscalar-pseudoscalar two-point function with a wall sink (Figure \ref{fig:2pt_PP_WW}), the axial-pseudoscalar two-point function with a local sink (Figure \ref{fig:2pt_AP_LW}), and the axial-pseudoscalar two-point function with a wall sink (Figure \ref{fig:2pt_AP_WW}), respectively. Figure \ref{fig:2pt_za} shows the ratio defined by Eq.~\eqref{eqn:za_ratio}. For each ensemble a single fit is performed to all five quantities simultaneously. In addition, a common value of the pion mass is used for the fits of the pion two-point functions to the ground-state ansatz defined in Eq.~\eqref{eqn:pion_2pt_ansatz}. 

\end{appendices}

\end{document}